\documentclass[11pt,a4paper]{article}
\pdfoutput=1
\usepackage{amsfonts}
\usepackage{amssymb}
\usepackage[centertags]{amsmath}
\usepackage{graphicx}
\usepackage{hyperref}
\usepackage{booktabs}
\usepackage{theorem}
\usepackage{array}
\usepackage{epsfig}
\usepackage{colordvi}
\usepackage{color}
\usepackage{ulem}
\usepackage[small,bf]{caption}
\usepackage{cite}
\usepackage{color}
\usepackage{subfigure}
\usepackage{pdflscape}
\def\1{\mathbf{1}}
\def\3{\mathbf{3}}
\def\2{\mathbf{2}}

\def\gtap{\ \raisebox{-.4ex}{\rlap{$\sim$}} \raisebox{.4ex}{$>$}\ }

\allowdisplaybreaks[1]

\usepackage{a4wide}
\newcommand{\bec}{\begin{cases}}
\newcommand{\eec}{\end{cases}}
\newcommand{\beq}{\begin{equation*}}
\newcommand{\eeq}{\end{equation*}}
\newcommand{\be}{\begin{equation}}
\newcommand{\ee}{\end{equation}}
\newcommand{\ba}{\begin{eqnarray}}
\newcommand{\ea}{\end{eqnarray}}

\DeclareMathOperator{\diag}{diag}

\makeatletter

\newcommand{\Rmnum}[1]{\expandafter\@slowromancap\romannumeral #1@}
\makeatother

\begin{document}

\begin{titlepage}

\vspace*{-15mm}
\begin{flushright}
SISSA 14/2015/FISI\\
IPMU15-0032\\
%arXiv:
\end{flushright}
\vspace*{0.7cm}

\begin{center}
{\bf\Large {Predictions for the Leptonic Dirac CP Violation Phase:}}\\
[4mm]
{\bf\Large {a Systematic Phenomenological Analysis}}\\
[8mm]
\vspace{0.4cm} I. Girardi$\mbox{}^{a)}$, S. T. Petcov$\mbox{}^{a,b)}$
\footnote{Also at: Institute of Nuclear Research and Nuclear Energy,
Bulgarian Academy of Sciences, 1784 Sofia, Bulgaria.}
and A. V. Titov$\mbox{}^{a)}$ 
\\[1mm]
\end{center}
\vspace*{0.50cm}
\centerline{$^{a}$ \it SISSA/INFN, Via Bonomea 265, 34136 Trieste, Italy}
\vspace*{0.2cm}
\centerline{$^{b}$ \it Kavli IPMU (WPI), University of Tokyo,
5-1-5 Kashiwanoha, 277-8583 Kashiwa, Japan}
\vspace*{1.20cm}

\begin{abstract}
\noindent
We derive  predictions for the Dirac phase $\delta$ present
in the $3\times 3$ unitary neutrino mixing
matrix $U = U_e^{\dagger} \, U_{\nu}$, where $U_e$ and $U_{\nu}$ are 
$3\times 3$ unitary matrices which arise from the diagonalisation, 
respectively, of the charged lepton and the neutrino mass matrices.  
We consider forms of $U_e$ and $U_{\nu}$ allowing us to express
$\delta$ as a function of three
neutrino mixing angles,
present in $U$, 
and the angles contained in $U_{\nu}$.
We consider several forms of $U_{\nu}$ 
determined by, or associated with, symmetries,
tri-bimaximal, bimaximal, etc., 
for which the angles in $U_{\nu}$ are 
fixed. For each of these forms and forms of $U_e$ 
allowing one to reproduce the measured values of the neutrino
mixing angles,
we construct the likelihood function
for $\cos \delta$, using i) the latest results of the global
fit analysis of neutrino oscillation data, 
and ii) the prospective sensitivities
on the neutrino mixing angles. 
Our results, in particular, confirm the conclusion, 
reached in earlier similar studies, 
that the measurement of the Dirac phase
in the neutrino mixing matrix, together with an improvement
of the precision on the mixing angles,
can provide unique information as regards the 
possible existence of symmetry 
in the lepton sector. 

\end{abstract}

\vspace{0.5cm}

\end{titlepage}
\setcounter{footnote}{0}

\section{Introduction}

 Understanding the origin of the observed pattern of neutrino mixing, 
establishing the status of the CP symmetry in the lepton sector, 
determining the type of spectrum the neutrino masses obey and 
determining the nature~---~Dirac or Majorana~---~of massive neutrinos 
are among the highest priority goals of the programme of 
future research in neutrino physics (see, e.g., \cite{PDG2014}). 
One of the major experimental efforts within this programme will be dedicated 
to the searches for CP-violating effects in neutrino 
oscillations (see, e.g., \cite{LBLFuture13,deGouvea:2013onf}). 
In the reference three neutrino mixing scheme with three 
light massive neutrinos we are going to consider 
(see, e.g., \cite{PDG2014}), the CP-violating effects in neutrino oscillations 
can be caused, as is well known, by the Dirac CP violation (CPV) 
phase present in the Pontecorvo, Maki, Nakagawa, Sakata (PMNS) 
neutrino mixing matrix. Predictions for the Dirac CPV phase
in the lepton sector can be, and were, obtained, in particular,  
combining the phenomenological approach, developed in 
\cite{GTani02,Xing:2002sw,Frampton:2004ud,Petcov:2004rk,Romanino:2004ww} 
and further exploited in various versions by many authors
with the aim of understanding the pattern of  neutrino mixing 
emerging from the data (see, e.g., \cite{Other,Chao:2011sp,Shimizu:2014ria,Hall:2013yha,Albright:2010ap}), 
with symmetry considerations. In this approach 
one exploits the fact that the PMNS mixing matrix $U$ 
has the form \cite{Frampton:2004ud}:
%%%%%%%%%%%%%%%%%%%%%%%%%%%%%%
\begin{equation}
U = U_e^{\dagger}\, U_{\nu} = 
(\tilde{U}_{e})^\dagger\, \Psi \tilde{U}_{\nu} \, Q_0\,,
\label{Ugeneral}
\end{equation}
%%%%%%%%%%%%%%%%%%%%%%%%%%%%
%
where  $U_e$ and $U_{\nu}$ 
are $3\times 3$ unitary matrices 
originating from the diagonalisation,  
respectively, of the charged lepton 
\footnote{If the charged lepton mass term is 
written in the right-left convention, the matrix $U_e$ 
diagonalises the hermitian matrix $M^\dagger_E M_E$, 
$U_e^{\dagger} M^\dagger_E M_E U_e = {\diag}(m^2_e,m^2_{\mu},m^2_{\tau})$,
$M_E$ being the charged lepton mass matrix.
}
and neutrino mass matrices. 
In eq. (\ref{Ugeneral}) 
$\tilde{U}_e$ and $\tilde{U}_\nu$ 
are CKM-like $3\times 3$ unitary matrices, 
and  $\Psi$ and $Q_0$ are diagonal phase matrices 
each containing in the general case two physical CPV phases 
\footnote{The phases in the matrix $Q_0$ contribute to the Majorana 
phases in the PMNS matrix \cite{Bilenky:1980cx}.}: 
%%%%%%%%%%%%%%%%%%%%%%%%
\begin{equation} 
\Psi = \diag \left(1,e^{-i \psi}, e^{-i \omega} \right)\,,
\quad
Q_0 = {\diag} \left(1,e^{i \frac{\xi_{21}}{2}}, 
e^{i \frac{\xi_{31}}{2}} \right)\,.
\label{PsieQ0}
\end{equation}
%%%%%%%%%%%%%%%%%%%%%%%%%%%%%%
%
It is further assumed that,  
up to subleading perturbative corrections (and phase matrices),   
the PMNS matrix $U$  has a specific known 
form  $\tilde{U}_{\nu}$ that 
is dictated by continuous 
and/or discrete symmetries, or by arguments 
related to symmetries.
This assumption seems very natural 
in view of the observation 
that the measured values 
of the three neutrino mixing angles 
differ from certain possible symmetry values 
by subdominant corrections.
Indeed, the best fit values and the 3$\sigma$ 
allowed ranges of the 
three neutrino mixing parameters
$\sin^2\theta_{12}$, $\sin^2\theta_{23}$ and 
$\sin^2\theta_{13}$ in the standard 
parametrisation of the PMNS matrix (see, e.g., \cite{PDG2014}), 
derived in the global analysis of the neutrino 
oscillation data performed in \cite{Capozzi:2013csa} read
%%%%%%%%%%%%%%%%%%%%%%%%%%%%%%%%%%%%%%%
\begin{eqnarray}
\label{th12values}
(\sin^2 \theta_{12})_{\rm BF} = 0.308\,,~~~~
 0.259 \leq \sin^2 \theta_{12} \leq 0.359\,,\\ [0.30cm]
\label{th23values}
(\sin^2\theta_{23})_{\rm BF} = 0.437~(0.455)\,,~~~~
 0.374~(0.380) \leq \sin^2\theta_{23} \leq 0.626~(0.641)\,,\\[0.30cm]
\label{th13values}
(\sin^2\theta_{13})_{\rm BF} = 0.0234~(0.0240)\,,~~~~
0.0176~(0.0178) \leq \sin^2\theta_{13} \leq 0.0295~(0.0298)\,,
\end{eqnarray}
%%%%%%%%%%
%
where the value (the value in parentheses) 
corresponds to  $\Delta m^2_{31(32)}>0$
($\Delta m^2_{31(32)}<0$), i.e., neutrino mass spectrum 
with normal (inverted) ordering  
\footnote{Similar results were obtained in the global analysis 
of the neutrino oscillation data performed in
\cite{Gonzalez-Garcia:2014bfa}.} (see, e.g., \cite{PDG2014}).
In terms of angles, the best fit values quoted above 
imply: $\theta_{12} \cong \pi/5.34$, 
$\theta_{13} \cong \pi/20$ and $\theta_{23}  \cong \pi/4.35$.
Thus, for instance, $\theta_{12}$ deviates from the possible 
symmetry value $\pi/4$, corresponding to the bimaximal mixing 
\cite{SPPD82,BM}, by approximately 0.2, 
$\theta_{13}$ deviates from 0 (or from 0.32) by approximately  
0.16 and $\theta_{23}$ deviates from the 
symmetry value $\pi/4$ 
by approximately 0.06, where we used $\sin^2\theta_{23} = 0.437$.

  Widely discussed symmetry forms
of $\tilde{U}_{\nu}$ include: 
i) tri-bimaximal (TBM) form \cite{Xing:2002sw,TBM}, 
ii) bimaximal (BM) form, 
or due to a symmetry corresponding
to the conservation of the lepton charge
$L' = L_e - L_{\mu} - L_{\tau}$ (LC) \cite{SPPD82,BM},
iii) golden ratio type A (GRA) form \cite{GRAM,GRAM2}, 
iv) golden ratio type B (GRB) form \cite{GRBM},
and v) hexagonal (HG) form \cite{Albright:2010ap,HGM}.
For all these forms the matrix $\tilde{U}_{\nu}$ 
represents a product of two orthogonal 
matrices describing rotations 
in the 1-2 and 2-3 planes on fixed angles 
$\theta^{\nu}_{12}$ and  $\theta^{\nu}_{23}$: 
%%%%%%%%%%%%%%%%%%%%%%%%%%%%%%%
\begin{equation}
\tilde{U}_\nu = R_{23}(\theta^{\nu}_{23}) \, R_{12}(\theta^{\nu}_{12})\,,
\label{tU2312}
\end{equation}
%%%%%%%%%%%%%%%%%%%%%%%%%%%%%%
%
where 
%%%%%%%%%%%%%%%%%%%%%%%%%%%%%%
\begin{equation}
R_{12}\left( \theta^\nu_{12} \right) = \begin{pmatrix}
\cos \theta^\nu_{12} & \sin \theta^\nu_{12} & 0\\
- \sin \theta^\nu_{12} & \cos \theta^\nu_{12} & 0\\
0 & 0 & 1 \end{pmatrix} \;,
\quad
R_{23}\left( \theta^\nu_{23} \right) = \begin{pmatrix}
1 & 0 & 0\\
0 & \cos \theta^\nu_{23} & \sin \theta^\nu_{23} \\
0 & - \sin \theta^\nu_{23}  & \cos \theta^\nu_{23} \\
\end{pmatrix} \;.
\label{R1223}
\end{equation}
%%%%%%%%%%%%%%%%%%%%%%%%%
%
Thus, $\tilde{U}_\nu$ does not include a rotation in the 1-3 plane, i.e., 
$\theta^{\nu}_{13} = 0$. Moreover, for all the symmetry forms quoted above 
one has also $\theta^{\nu}_{23} = -\,\pi/4$. The forms differ by 
the value of the angle $\theta^{\nu}_{12}$, and, correspondingly, 
of $\sin^2\theta^{\nu}_{12}$:
for the TBM, BM (LC), GRA, GRB and HG forms we have, respectively,
$\sin^2\theta^{\nu}_{12} = 1/3$, $1/2$, $(2 + r)^{-1} \cong 0.276$,
$(3 - r)/4 \cong 0.345$, and $1/4$,
$r$ being the golden ratio, $r = (1 +\sqrt{5})/2$.

 As is clear from the preceding discussion, 
the values of the angles in the matrix $\tilde{U}_{\nu}$, 
which are fixed by symmetry arguments, typically differ from 
the values determined experimentally by relatively small 
perturbative corrections. In the approach we are following, 
the requisite corrections are provided by the angles in the matrix 
$\tilde{U}_e$. The matrix $\tilde{U}_e$ in the general case 
depends on three angles and one phase \cite{Frampton:2004ud}. 
However, in a class of theories of (lepton) flavour 
and neutrino mass generation, 
based on a GUT and/or a discrete symmetry (see, e.g., 
\cite{Gehrlein:2014wda,Meroni:2012ty,Marzocca:2011dh,Antusch:2012fb,Girardi:2013sza,Chen:2009gf}), 
$\tilde{U}_e$ is an orthogonal matrix which describes  
one rotation in the 1-2 plane, 
%%%%%%%%%%%%%%%%%%%%%%%%%%%%%%%%%%%%%%%%%
\begin{align} 
\label{Re12}
\tilde{U}_e &=  R^{-1}_{12}(\theta^{e}_{12})\,,
\end{align}
%%%%%%%%%%%%%%%%%%%%%%%%%%%%%%%%%%
%
or two rotations in 
the planes 1-2 and 2-3,
%%%%%%%%%%%%%%%%%%%%%%%%%%%%%%%%%%%%%%%%%
\begin{align} 
\label{Re12Re23}
\tilde{U}_e &= R^{-1}_{23}(\theta^{e}_{23})\,
R^{-1}_{12}(\theta^{e}_{12})\,,
\end{align}
%%%%%%%%%%%%%%%%%%%%%%%%%%%%%%%%%%%
%
$\theta^{e}_{12}$ 
and $\theta^{e}_{23}$ being the 
 corresponding rotation angles.
Other possibilities include $\tilde{U}_e$ being 
an orthogonal matrix which 
describes  i) one rotation in the 1-3 
plane
\footnote{The case of $\tilde{U}_e$ representing a rotation 
in the 2-3 plane is ruled out for the five symmetry 
forms of $\tilde{U}_{\nu}$ listed above, since 
in this case a realistic value of 
$\theta_{13} \neq 0$ cannot be generated.}, 
%%%%%%%%%%%%%%%%%%%%%%%%%%%%%%%%%%%%%%%%%
\begin{align} 
\label{Re13}
\tilde{U}_e & =  R^{-1}_{13}(\theta^{e}_{13})\,,
\end{align}
%%%%%%%%%%%%%%%%%%%%%%%%%%%%%%%%%%
%
or ii) two rotations in 
any other two of the three planes, e.g., 
%%%%%%%%%%%%%%%%%%%%%%%%%%%%%%%%%%%%%%%%%
\begin{align} 
\label{Re13Re23}
\tilde{U}_e &=  R^{-1}_{23}(\theta^{e}_{23})\, 
R^{-1}_{13}(\theta^{e}_{13})\,,~~{\rm or} \\ %[0.30cm]
\label{Re13Re12}
\tilde{U}_e &= R^{-1}_{13}(\theta^{e}_{13})\,  
R^{-1}_{12}(\theta^{e}_{12})\,. 
\end{align}
%%%%%%%%%%%%%%%%%%%%%%%%%%%%%%%%%%%
%
The use of the inverse matrices in 
eqs. (\ref{Re12})~--~(\ref{Re13Re12})
is a matter of convenience~---~this 
allows us to lighten the notations in 
expressions which will appear further in the text. 

  It was shown in \cite{Petcov:2014laa} 
(see also \cite{Marzocca:2013cr}) 
that for $\tilde{U}_{\nu}$ and $\tilde{U}_e$ given in 
eqs.~(\ref{tU2312}) and (\ref{Re12Re23}), 
the Dirac phase $\delta$ present in the PMNS matrix 
satisfies a sum rule by which it is expressed 
in terms of the three neutrino mixing angles measured  
in the neutrino oscillation experiments and the angle
$\theta^{\nu}_{12}$.
In the standard parametrisation of the PMNS matrix 
(see, e.g., \cite{PDG2014}) 
the sum rule reads \cite{Petcov:2014laa}:
%%%%%%%%%%%%%%%%%%%%%%%%%%%%%%%%%%%%
\begin{equation}
\cos\delta =  \frac{\tan\theta_{23}}{\sin2\theta_{12}\sin\theta_{13}}\,
\left [\cos2\theta^{\nu}_{12} + 
\left (\sin^2\theta_{12} - \cos^2\theta^{\nu}_{12} \right )\,
 \left (1 - \cot^2\theta_{23}\,\sin^2\theta_{13}\right )\right ]\,.
\label{cosdthnu}
\end{equation}
%%%%%%%%%%%%%%%%%%%%%%%%%%%%%%%%%%%%
%
For the specific values of $\theta^{\nu}_{12}= \pi/4$ and 
$\theta^{\nu}_{12} = \sin^{-1} (1/\sqrt{3})$, i.e., 
for the BM (LC) and TBM forms of $\tilde{U}_{\nu}$, 
eq.~(\ref{cosdthnu}) reduces to the expressions for 
$\cos\delta$ derived first in \cite{Marzocca:2013cr}.
On the basis of the analysis performed and the 
results obtained using the best fit values of 
$\sin^2\theta_{12}$, $\sin^2\theta_{13}$ and $\sin^2\theta_{23}$, 
it was concluded in \cite{Petcov:2014laa}, in particular, that  
the measurement of $\cos\delta$ can allow one to distinguish between 
the different symmetry forms of the matrix $\tilde{U}_{\nu}$ considered.

 Within the approach employed, the expression 
for $\cos\delta$ given in eq.~(\ref{cosdthnu}) 
is exact. In \cite{Petcov:2014laa} the correction to the sum rule 
eq.~(\ref{cosdthnu}) due to a non-zero angle $\theta^{e}_{13} \ll 1$ 
in $\tilde{U_e}$, corresponding to 
%%%%%%%%%%%%%%%%%%%%%%%%%%%%%%%%%%
\begin{equation}
\label{Re12Re13Re23}
\tilde{U}_e = R^{-1}_{23}(\theta^{e}_{23})\,R^{-1}_{13}(\theta^{e}_{13})\,
R^{-1}_{12}(\theta^{e}_{12})\,
\end{equation}
%%%%%%%%%%%%%%%%%%%%%%%%%%%%%
%
with $|\sin\theta^e_{13}| \ll 1$, was also derived.

  Using the best fit values of the neutrino mixing parameters 
$\sin^2\theta_{12}$, $\sin^2\theta_{13}$ and $\sin^2\theta_{23}$, 
found in the global analysis in \cite{Capozzi:2013v1}, predictions 
for $\cos\delta$, $\delta$ and the rephasing invariant  
%%%%%%%%%%%%%%%%%%%%%%%%%%%%%%%%
\begin{equation}
J_{\rm CP} = {\rm Im} \left\{ U^*_{e1} U^*_{\mu 3} U_{e3} U_{\mu 1} \right\}
= \frac{1}{8} \sin \delta \sin 2\theta_{13} \sin 2\theta_{23}
\sin 2\theta_{12} \cos \theta_{13} \,,
\label{JCP}
\end{equation}
%%%%%%%%%%%%%%%%%%%%%%%%%%%%%%%
%
which controls the magnitude of CP-violating 
effects in neutrino oscillations \cite{PKSP3nu88},
were presented in \cite{Petcov:2014laa} 
for each of the five symmetry forms of $\tilde{U}_{\nu}$~---~TBM, 
BM (LC), GRA, GRB and HG~---~considered.

 Statistical analysis of the sum rule eq.~(\ref{cosdthnu}) 
predictions for $\delta$ 
and $J_{\rm CP}$ (for $\cos\delta$) 
using the current (the prospective) uncertainties 
in the determination of the three  neutrino mixing 
parameters, $\sin^2\theta_{12}$, $\sin^2\theta_{13}$, $\sin^2\theta_{23}$,   
and $\delta$ 
($\sin^2\theta_{12}$, $\sin^2\theta_{13}$ and $\sin^2\theta_{23}$), 
was performed in \cite{Girardi:2014faa} 
for the five symmetry forms~---~BM (LC), 
TBM, GRA, GRB and HG~---~of $\tilde{U}_{\nu}$. 
Using the current uncertainties in 
the measured values of $\sin^2\theta_{12}$, $\sin^2\theta_{13}$, 
$\sin^2\theta_{23}$ and $\delta$ 
\footnote{We would like to note that the recent statistical 
analyses performed in \cite{Capozzi:2013csa,Gonzalez-Garcia:2014bfa} showed 
indications/hints that $\delta \cong 3\pi/2$. 
As for $\sin^2\theta_{12}$, $\sin^2\theta_{13}$ and  
$\sin^2\theta_{23}$, in the case of $\delta$ we utilise as 
``data'' the results obtained in 
ref. \cite{Capozzi:2013csa}.},
it was found, in particular, that 
for the TBM, GRA, GRB and HG forms, 
$J_{\rm CP}\neq 0$ at $5\sigma$,  
$4\sigma$, $4\sigma$ and $3\sigma$, respectively.
For all these four forms $|J_{\rm CP}|$ is predicted 
at $3\sigma$ to lie in the following narrow interval 
\cite{Girardi:2014faa}: $0.020 \leq |J_{\rm CP}| \leq 0.039$. 
As a consequence, in all these cases the CP-violating 
effects in neutrino oscillations are predicted to be relatively large.
In contrast, for the BM (LC) form, the predicted best fit 
value is  $J_{\rm CP} \cong 0$, and the CP-violating effects 
in neutrino oscillations can be strongly suppressed. 
The statistical analysis of the sum rule predictions for 
$\cos\delta$, performed in \cite{Girardi:2014faa} 
by employing prospective uncertainties of 0.7\%, 3\% and 
5\% in the determination of  $\sin^2\theta_{12}$, 
$\sin^2\theta_{13}$ and $\sin^2\theta_{23}$,  
revealed that with precision in the measurement of 
$\delta$, $\Delta \delta \cong (12^\circ - 16^\circ)$,
which is planned to be achieved in 
the future neutrino experiments like T2HK and ESS$\nu$SB   
\cite{deGouvea:2013onf},  
it will be possible to distinguish at $3\sigma$ 
between the BM (LC), TBM/GRB and GRA/HG forms 
of $\tilde{U}_{\nu}$. Distinguishing between the TBM and GRB forms, 
and between the GRA and HG forms, requires a measurement of 
$\delta$ with an uncertainty of a few degrees.
 
 In the present article we derive new sum rules for 
$\cos\delta$ using the general approach employed, in particular,  
in \cite{Petcov:2014laa,Girardi:2014faa}. 
We perform a systematic 
study of the forms of the matrices $\tilde{U}_e$ and  $\tilde{U}_\nu$, 
for which it is possible to derive sum rules 
for $\cos\delta$ of the type of eq.~(\ref{cosdthnu}), 
but for which the sum rules of interest do not exist in the 
literature. More specifically, we consider the following forms  
of  $\tilde{U}_e$ and  $\tilde{U}_\nu$: 
\begin{itemize}
\item[A.] $\tilde{U}_\nu = R_{23}(\theta^{\nu}_{23})R_{12}(\theta^{\nu}_{12})$ 
with  $\theta^{\nu}_{23} = -\pi/4$ and $\theta^{\nu}_{12}$
corresponding to the TBM, BM (LC), GRA, GRB and HG mixing, and 
i)  $\tilde{U}_e =  R^{-1}_{13}(\theta^{e}_{13})$, 
ii) $\tilde{U}_e =  R^{-1}_{23}(\theta^{e}_{23})R^{-1}_{13}(\theta^{e}_{13})$,
and iii) $\tilde{U}_e = R^{-1}_{13}(\theta^{e}_{13})
R^{-1}_{12}(\theta^{e}_{12})$; 
\item[B.] $\tilde{U}_\nu = R_{23}(\theta^{\nu}_{23})R_{13}(\theta^{\nu}_{13}) 
 R_{12}(\theta^{\nu}_{12})$ with $\theta^{\nu}_{23}$, 
$\theta^{\nu}_{13}$ and $\theta^{\nu}_{12}$ fixed by arguments associated with 
symmetries, and 
iv) $\tilde{U}_e = R^{-1}_{12}(\theta^{e}_{12})$, 
and v) $\tilde{U}_e = R^{-1}_{13}(\theta^{e}_{13})$.
\end{itemize}
 In each of these cases we obtain the respective sum rule for 
$\cos\delta$. This is done first for $\theta^{\nu}_{23} = -\,\pi/4$
in the cases listed in point A, and for the specific 
values of (some of) the angles in  $\tilde U_{\nu}$, characterising 
the cases listed in point B. 
For each of the cases listed in points A and B
we derive also generalised sum rules for 
$\cos\delta$ for arbitrary fixed values of all angles
contained in $\tilde U_{\nu}$ (i.e., without setting 
 $\theta^{\nu}_{23} = -\,\pi/4$ in the cases 
listed in point A, etc.).
Next we derive predictions 
for $\cos\delta$ and $J_{\rm CP}$
($\cos\delta$), performing a statistical 
analysis using the current (the prospective) uncertainties 
in the determination of the neutrino mixing parameters 
$\sin^2\theta_{12}$, $\sin^2\theta_{13}$, $\sin^2\theta_{23}$ and $\delta$ 
($\sin^2\theta_{12}$, $\sin^2\theta_{13}$ and $\sin^2\theta_{23}$). 

It should be noted that the approach to 
understanding the experimentally determined pattern of lepton mixing 
and to obtaining predictions 
for  $\cos\delta$ and $J_{\rm CP}$ employed in the present work and in 
the earlier related studies \cite{Petcov:2014laa} and 
\cite{Girardi:2014faa}, is by no means unique~---~it 
is one of a number of approaches discussed in the 
literature on the problem (see, e.g., \cite{King:2013eh, Winter1,Winter2}).
It is used in a large number of phenomenological studies 
(see, e.g., \cite{Antusch:2005kw,Shimizu:2014ria,GTani02,Frampton:2004ud,Romanino:2004ww,Chao:2011sp,Hall:2013yha}) 
as well as in a class of models 
(see \cite{Girardi:2013sza,Marzocca:2011dh,Antusch:2011qg,Meroni:2012ty,Antusch:2012fb,Chen:2009gf,Gehrlein:2014wda}) 
of neutrino mixing based on discrete symmetries.
However, it should be clear that the conditions  
of the validity of the approach employed in the present work  
are not fulfilled in all theories with discrete flavour symmetries. 
For example, they are not fulfilled in the theories with 
discrete flavour symmetry  $\Delta (6n^2)$ 
studied in \cite{King:2013vna,Hagedorn:2014wha}, 
with the $S_4$ flavour symmetry constructed in \cite{Luhn:2013lkn} 
and in the models discussed in \cite{Altarelli:2012bn}.
Further, the conditions of our analysis are also not fulfilled 
in the phenomenological approach developed and exploited in 
\cite{Winter1,Winter2}. In these articles, in particular,  
the matrices  $U_e$ and $U_{\nu}$ are assumed to have 
specific given fixed forms,  
in which all three mixing angles in each of the two matrices 
are fixed to some numerical 
values, typically, but not only, $\pi/4$, 
or some integer powers $n$ of the parameter 
$\epsilon \cong \theta_C$,
$\theta_C$ being the Cabibbo angle.
The angles  $\theta^{\nu}_{ij} \cong (\theta_C)^n$ with $n > 2$
are set to zero. 
For example, in \cite{Winter2} the following 
sets of values of the angles in 
 $U_e$ and $U_{\nu}$ have been used:
$(\theta^e_{12},\theta^e_{13},\theta^e_{23},\theta^{\nu}_{12},\theta^{\nu}_{13},\theta^{\nu}_{23}) = (*,\pi/4,\pi/4,*,\pi/4,*)$ and 
$(*,*,\pi/4,\pi/4,*,*)$, 
where ``$*$'' means angles not exceeding $\theta_C$.
None of these sets correspond to the cases studied by us.
As a consequence, the sum rules for $\cos\delta$ 
derived in our work and in  \cite{Winter2} are very different.
In  \cite{Winter2} the authors have also considered  
specific textures of the neutrino Majorana mass matrix 
leading to the two sets of values of the angles 
in $U_e$ and $U_{\nu}$ quoted above. 
However, these textures lead 
to values of $\sin^2\theta_{23}$ or of 
$\sin^2\theta_{12}$ which are strongly disfavoured by the
current data. 
Although in \cite{Winter1}
a large variety of forms of $U_e$ and $U_{\nu}$ 
have been investigated, none of them corresponds to 
the forms studied by us, as can be inferred 
from the results on the values of the PMNS angles 
$\theta_{12}$, $\theta_{13}$ and $\theta_{23}$
obtained in \cite{Winter1} and summarised in 
Table 2 in each of the two articles we have cited in 
\cite{Winter1}.

 Our article is organised as follows.
In Section \ref{sec:ije23nu12nu} we consider the 
models which contain one rotation from the charged lepton sector, 
i.e., $\tilde{U}_e = R^{-1}_{12}(\theta^e_{12})$, or 
$\tilde{U}_e = R^{-1}_{13}(\theta^e_{13})$,
and two rotations from the neutrino sector:
$\tilde{U}_\nu = R_{23}(\theta^{\nu}_{23}) \, R_{12}(\theta^{\nu}_{12})$.
In these cases the PMNS matrix reads
%%%%%%%%%%%%%%%%%%%%%%%%%%%%%%%%
\begin{align}
U = R_{ij}(\theta^e_{ij}) \, \Psi \, R_{23}(\theta^{\nu}_{23}) \,
R_{12}(\theta^{\nu}_{12}) \, Q_0 \,, 
\label{eq:Uij}
\end{align}
%%%%%%%%%%%%%%%%%%%%%%%%%%%%%
%
with $(ij) = (12)$, $(13)$. 
The matrix $\tilde U_{\nu}$ is assumed to have the following 
symmetry forms: TBM, BM (LC), GRA, GRB and HG. 
As we have already noted, for all these forms 
$\theta^{\nu}_{23} = -\pi/4$,
but we discuss also the general case
of an arbitrary fixed value of $\theta^{\nu}_{23}$. The forms 
listed above differ 
by the value of the angle $\theta^{\nu}_{12}$, which 
for each of the forms of interest was given earlier.
In Section \ref{sec:ijekle23nu12nu} we analyse the models which
contain two rotations from the charged lepton sector, i.e.,
$\tilde{U}_e = R^{-1}_{23}(\theta^e_{23}) \, R^{-1}_{13}(\theta^e_{13})$,
or $\tilde{U}_e = R^{-1}_{13}(\theta^e_{13}) \, R^{-1}_{12}(\theta^e_{12})$,
and 
\footnote{ We consider only the ``standard'' ordering 
of the two rotations in $\tilde{U}_e$, see 
\cite{Marzocca:2013cr}.  The case with 
$\tilde{U}_e =  R^{-1}_{23}(\theta^e_{23}) \, R^{-1}_{12}(\theta^e_{12})$
has been analysed in detail in 
\cite{Marzocca:2013cr,Petcov:2014laa,Girardi:2014faa} and 
will not be discussed by us.
} 
two rotations from the neutrino sector, i.e.,
%%%%%%%%%%%%%%%%%%%%%%%
\begin{align}
&U = R_{ij}(\theta^e_{ij}) \,  R_{kl}(\theta^e_{kl})  \, \Psi \, 
R_{23}(\theta^{\nu}_{23}) \, R_{12}(\theta^{\nu}_{12}) \, Q_0 \,,
\label{eq:Uijkl} 
\end{align}
%%%%%%%%%%%%%%%%%%%%%%
%
with $(ij)-(kl) = (13)-(23)$, $(12)-(13)$.
First we assume the angle $\theta^{\nu}_{23}$
to correspond to the TBM, BM (LC), GRA, GRB and HG 
symmetry forms of $\tilde{U}_{\nu}$. After that we give
the formulae for an arbitrary fixed value of this angle.
Further, in Section \ref{sec:ije23nu13nu12nu}, 
we generalise the schemes considered in Section \ref{sec:ije23nu12nu} 
by  allowing also a third rotation matrix to be present in 
$\tilde{U}_{\nu}$:
%%%%%%%%%%%%%%%%%%%%%%%
\begin{align}
U = R_{ij}(\theta^e_{ij}) \, \Psi \, R_{23}(\theta^{\nu}_{23})\,
R_{13}(\theta^{\nu}_{13}) \, R_{12}(\theta^{\nu}_{12}) \, Q_0 \,,
\label{eq:Uija} 
\end{align}
%%%%%%%%%%%%%%%%%%%%%%%%%%%
%
with $(ij) = (12)$, $(13)$, $(23)$.
%%%%%%%%%%%%%%%%%%%%%%%

 Using the sum rules for $\cos\delta$ derived in Sections 
\ref{sec:ije23nu12nu}~--~\ref{sec:ije23nu13nu12nu}, 
in Section \ref{sec:predictions} 
we obtain  predictions for $\cos\delta$, 
$\delta$ and 
$J_{\rm CP}$
for each of the models considered in the preceding
sections. Section \ref{sec:summary} contains 
summary of the results of the present study and conclusions.

We note finally that the titles of Sections 2~--~4 and of their subsections 
reflect the rotations
contained in the corresponding parametrisation,
eqs.~(\ref{eq:Uij})~--~(\ref{eq:Uija}).

%%%%%%%%%%%%%%%%%%%%%%%%%%%%%%%%%%%
%
\section{The Cases of $\theta^e_{ij} - (\theta^\nu_{23}, \theta^\nu_{12})$ 
 Rotations }
\label{sec:ije23nu12nu}
%
%%%%%%%%%%%%%%%%%%%%%%%%%%%%%%%%%%%

In this section we  derive the 
sum rules for $\cos \delta$ of interest 
in the case when the matrix $\tilde{U}_\nu = R_{23}(\theta^{\nu}_{23})\,
R_{12}(\theta^{\nu}_{12})$ with fixed (e.g., symmetry) values 
of the angles $\theta^{\nu}_{23}$ and 
$\theta^{\nu}_{12}$, 
gets correction only due to
one rotation from the charged lepton sector.
The neutrino mixing matrix $U$ has the 
form given in eq.~(\ref{eq:Uij}).
We do not consider the cases of eq.~(\ref{eq:Uij}) 
 i) with $(ij) = (23)$,
because the reactor angle
$\theta_{13}$ does not get corrected and remains zero,
and ii) with $(ij) = (12)$ and $\theta^{\nu}_{23} = -\pi/4$,
which has been already analysed in detail
in \cite{Petcov:2014laa,Girardi:2014faa}.

%%%%%%%%%%%%%%%%%%%%%%%
\subsection{The Scheme with $\theta^e_{12} - (\theta^\nu_{23}, \theta^\nu_{12})$ 
 Rotations }
\label{sec:12e23nu12nu}
%%%%%%%%%%%%%%%%%%%%%%%

 For  $\theta^{\nu}_{23} = -\pi/4$ the sum rule for $\cos\delta$ 
in this case was derived in ref.~\cite{Petcov:2014laa} and is 
given in eq.~(50) therein. 
Here we consider the case of an arbitrary fixed value
of the angle $\theta^{\nu}_{23}$. Using eq.~(\ref{eq:Uij}) 
with $(ij) = (12)$, one finds the following expressions 
for the mixing angles 
$\theta_{13}$ and $\theta_{23}$ of the standard 
parametrisation of the PMNS matrix:
%%%%%%%%%%%%%%%%%%%%%%%
\begin{align}
\sin^2 \theta_{13} & = |U_{e3}|^2  = \sin^2 \theta^e_{12} \sin^2 \theta^{\nu}_{23} \label{eq:th13A0}\,,\\
\sin^2 \theta_{23} & = \frac{|U_{\mu3}|^2}{1-|U_{e3}|^2} = \frac{\sin^2 \theta^{\nu}_{23}-\sin^2 \theta_{13}}{1 - \sin^2 \theta_{13}} 
\label{eq:th23A0}\,.
\end{align}
%%%%%%%%%%%%%%%%%%%%%%%
%
Although
eq.~(\ref{cosdthnu}) was derived in \cite{Petcov:2014laa} 
for $\theta^{\nu}_{23} = -\pi/4$ and 
$\tilde{U}_e = R^{-1}_{23}(\theta^{e}_{23})R^{-1}_{12}(\theta^{e}_{12})$,
it is not difficult to convince oneself that 
it holds also in the case under discussion 
for an arbitrary fixed value of $\theta^{\nu}_{23}$. 
The sum rule for $\cos \delta$ of interest,  
expressed in terms of the angles 
$\theta_{12}$, $\theta_{13}$, 
$\theta^{\nu}_{12}$ and $\theta^{\nu}_{23}$,
can be obtained from eq.~(\ref{cosdthnu}) 
by using the expression for $\sin^2 \theta_{23}$ 
given in  eq.~(\ref{eq:th23A0}). 
The result reads:
%%%%%%%%%%%%%%%%%%%%%%%
\begin{align} 
\cos\delta & =
\frac{(\cos 2 \theta_{13} - \cos 2 \theta^{\nu}_{23})^{\frac{1}{2}}}{\sqrt{2}\sin2\theta_{12}\sin\theta_{13} |\cos \theta^{\nu}_{23}|}\,
\bigg[ \cos2\theta^{\nu}_{12} \nonumber \\ 
& + \left (\sin^2\theta_{12} - \cos^2\theta^{\nu}_{12} \right )\,
\frac{2\sin^2 \theta^{\nu}_{23}-(3+\cos 2 \theta^{\nu}_{23}) \sin^2\theta_{13}}{\cos 2 \theta_{13} - \cos 2 \theta^{\nu}_{23}} \bigg]\,.
\label{eq:cosdelta12e23nu12nugen}
\end{align}
%%%%%%%%%%%%%%%%%%%%%%%
%
Setting $\theta^{\nu}_{23} = -\pi/4$ 
in (\ref{eq:cosdelta12e23nu12nugen}), one reproduces 
the sum rule given in 
eq.~(50) in ref.~\cite{Petcov:2014laa}.

%%%%%%%%%%%%%%%%%%%%%%%
\subsection{The Scheme with $\theta^e_{13} - (\theta^\nu_{23}, \theta^\nu_{12})$ 
 Rotations }
\label{sec:13e23nu12nu}
%%%%%%%%%%%%%%%%%%%%%%%

 In the present subsection we consider 
the parametrisation of the neutrino
mixing matrix given in eq.~(\ref{eq:Uij}) with $(ij) = (13)$.
In this set-up the phase $\psi$ 
in the matrix $\Psi$ is unphysical 
(it can be absorbed in the $\mu^\pm$ field) 
and therefore effectively
$\Psi = \diag \left(1, 1, e^{-i \omega}\right)$.
Using eq.~(\ref{eq:Uij}) with $(ij) = (13)$ and 
$\theta^{\nu}_{23} = -\pi/4$
and the standard parametrisation of $U$, we get
%%%%%%%%%%%%%%%%%%%%%%%
\begin{align}
\sin^2 \theta_{13} & = |U_{e3}|^2  = \frac{1}{2} \sin^2 \theta^e_{13} \label{eq:th13B0}\,,\\
\sin^2 \theta_{23} & = \frac{|U_{\mu3}|^2}{1-|U_{e3}|^2} = \frac{1}{2\,(1 - \sin^2 \theta_{13})} \label{eq:th23B0}\,,  \\
\sin^2 \theta_{12} & = \frac{|U_{e2}|^2}{1-|U_{e3}|^2} = \frac{1}{1-\sin^2 \theta_{13}} 
\bigg[ \dfrac{1}{2} \sin^2 \theta^e_{13} \cos^2 \theta^{\nu}_{12}  \nonumber \\
& +  \cos^2 \theta^e_{13} \sin^2 \theta^{\nu}_{12} +  \frac{1}{\sqrt{2}} \sin 2 \theta^e_{13}  \cos \omega \sin \theta^{\nu}_{12} \cos \theta^{\nu}_{12} \bigg]  \label{eq:th12B0}\,.
\end{align}
%%%%%%%%%%%%%%%%%%%%%%%%%%%
%
From eqs.~(\ref{eq:th13B0}) and (\ref{eq:th12B0}) we  
 obtain an expression for 
$\cos \omega$ in terms of
the measured mixing angles 
$\theta_{12}$, $\theta_{13}$  
and the known $\theta^{\nu}_{12}$:
%%%%%%%%%%%%%%%%%%%%%%%
\be
\label{eq:omega1}
\cos \omega = \frac{1-\sin^2 \theta_{13}}{\sin 2\theta^{\nu}_{12}
\sin \theta_{13} (1-2\sin^2 \theta_{13})^{\frac{1}{2}}}
\left[ \sin^2 \theta_{12} - \sin^2 \theta^{\nu}_{12} -
\cos 2\theta^{\nu}_{12} \frac{\sin^2 \theta_{13}}{1-\sin^2 \theta_{13}} \right] \,.
\ee
%%%%%%%%%%%%%%%%%%%%%%%
%
 Further, one can find 
\footnote{We note that the expression for $\cos \omega$ 
we have obtained coincides 
with that  for $\cos \phi$ in the set-up
with the $(ij)= (12)$ rotation in the charged
lepton sector (cf. eq.~(46) in \cite{Petcov:2014laa}).}
a relation between 
$\sin \delta$ ($\cos \delta$) and $\sin\omega$ ($\cos\omega$)
by comparing the imaginary (the real) part
of the quantity $U^*_{e1} U^*_{\mu 3} U_{e3} U_{\mu 1}$,
written by using eq.~(\ref{eq:Uij}) with $(ij) = (13)$
and in the standard parametrisation of $U$.
For the relation between $\sin \delta$ and 
$\sin\omega$ we get
%%%%%%%%%%%%%%%%%%%%%%%%%%%%%%%%%%%%%
\be
\sin \delta =
-\frac{\sin 2 \theta^{\nu}_{12}}{\sin 2 \theta_{12}} \sin \omega \,.
\ee
%%%%%%%%%%%%%%%%%%%%%%%%%%%%%%%%%%
%
The sum rule for  $\cos \delta$ of interest can be derived  
by substituting $\cos \omega$ from eq.~(\ref{eq:omega1}) 
in the relation between $\cos \delta$ and $\cos\omega$ 
(which is not difficult to derive and we do not give).
We obtain
%%%%%%%%%%%%%%%%%%%%%%%
\be
\cos \delta =
-\frac{(1-2\sin^2 \theta_{13})^{\frac{1}{2}}}{\sin 2 \theta_{12} \sin \theta_{13}}
\left[\cos 2 \theta^{\nu}_{12}+(\sin^2 \theta_{12}-\cos^2 \theta^{\nu}_{12})
\frac{1-3\sin^2 \theta_{13}}{1-2\sin^2 \theta_{13}} \right] \,.
\label{eq:cosdelta13e}
\ee
%%%%%%%%%%%%%%%%%%%%%%%
%
We note that the expression for $\cos \delta$ thus found        
differs only by an overall minus sign from the analogous 
expression for $\cos\delta$ 
derived in \cite{Petcov:2014laa}
in the case of $(ij) = (12)$ rotation in the charged lepton 
sector (see eq.~(50) in \cite{Petcov:2014laa}).

 In eq. (\ref{JCP}) we have given the expression for the 
rephasing invariant $J_{\rm CP}$ in the standard 
parametrisation of the PMNS matrix. Below and in the next 
sections we give for completeness also the expressions 
of the $J_{\rm CP}$ factor in terms of the independent
parameters of the set-up considered.
In terms of the parameters $\omega$, 
$\theta^e_{13}$ and $\theta^{\nu}_{12}$ of 
the set-up discussed in the present subsection,
$J_{\rm CP}$ is given by
%%%%%%%%%%%%%%%%%%%%%%%%%%%
\be
J_{\rm CP} = -\frac{1}{8 \sqrt{2}} 
\sin \omega \sin 2\theta^e_{13} \sin 2\theta^{\nu}_{12} \,.
\ee
%%%%%%%%%%%%%%%%%%%%%%%%%%%%
%

 In the case of an arbitrary fixed value of the angle $\theta^{\nu}_{23}$ 
the expressions for the mixing angles $\theta_{13}$ and $\theta_{23}$ 
take the form
%%%%%%%%%%%%%%%%%%%%%%%
\begin{align}
\sin^2 \theta_{13} & = |U_{e3}|^2  = \sin^2 \theta^e_{13} \cos^2 \theta^{\nu}_{23} \label{eq:th13B0gen}\,,\\
\sin^2 \theta_{23} & = \frac{|U_{\mu3}|^2}{1-|U_{e3}|^2} = \frac{\sin^2 \theta^{\nu}_{23}}{1 - \sin^2 \theta_{13}} \label{eq:th23B0gen}\,.
\end{align}
%%%%%%%%%%%%%%%%%%%%%%%
%
The sum rule for $\cos \delta$ in this case can be obtained 
with a simpler procedure, namely,
by using the expressions for the absolute value of the 
element $U_{\mu 1}$ of the PMNS matrix in the 
two parametrisations employed in the present subsection:
%%%%%%%%%%%%%%%%%%%%%%%%
\begin{align}
|U_{\mu 1}| =
|\cos\theta_{23} \sin\theta_{12} + e^{i \delta} \cos\theta_{12} \sin\theta_{13} \sin\theta_{23}|
=  |\cos\theta^{\nu}_{23} \sin\theta^{\nu}_{12}|\,,
\label{eq:Umu113e2rotgen}
\end{align}
%%%%%%%%%%%%%%%%%%%%%%%%%
%
From eq. (\ref{eq:Umu113e2rotgen}) we get
%%%%%%%%%%%%%%%%%%%%%%%
\begin{align} 
\cos\delta & =
-\frac{(\cos 2 \theta_{13} + \cos 2 \theta^{\nu}_{23})^{\frac{1}{2}}}{\sqrt{2}\sin2\theta_{12}\sin\theta_{13}|\sin\theta^{\nu}_{23}|}\,
\bigg[ \cos2\theta^{\nu}_{12} \nonumber \\ 
& + \left (\sin^2\theta_{12} - \cos^2\theta^{\nu}_{12} \right )\,
\frac{2\cos^2 \theta^{\nu}_{23}-(3-\cos 2 \theta^{\nu}_{23}) \sin^2\theta_{13}}{\cos 2 \theta_{13} + \cos 2 \theta^{\nu}_{23}} \bigg]\,.
\label{eq:cosdelta13e23nu12nugen}
\end{align}
%%%%%%%%%%%%%%%%%%%%%%%

{ We will use the sum rules for $\cos\delta$ derived in the present 
and the next two Sections 
to obtain  predictions for $\cos\delta$, 
$\delta$ and for the 
$J_{\rm CP}$ factor in Section \ref{sec:predictions}.

%%%%%%%%%%%%%%%%%%%%%%%%%%%%%%%%%%%
%
\section{The Cases of $(\theta^e_{ij},\theta^e_{kl}) - (\theta^\nu_{23}, \theta^\nu_{12})$ 
 Rotations }
\label{sec:ijekle23nu12nu}
%
%%%%%%%%%%%%%%%%%%%%%%%%%%%%%%%%%%%%
%

As we have seen in the preceding Section,
in the case of one rotation from the charged lepton sector
and for  $\theta^{\nu}_{23} = -\pi/4$, 
the mixing angle $\theta_{23}$ cannot deviate 
significantly from $\pi/4$ due to the smallness of 
the angle $\theta_{13}$.  If the matrix 
$\tilde{U}_{\nu}$ 
has one of the symmetry forms considered in this study,
the matrix $\tilde{U}_e$ has to contain at least
two rotations in order to be possible to 
reproduce the current best fit values
of the neutrino mixing parameters,
quoted in eqs.~(\ref{th12values})~--~(\ref{th13values}). 
This conclusion will remain valid 
if higher precision measurements of 
$\sin^2\theta_{23}$ confirm that 
$\theta_{23}$ deviates significantly from $\pi/4$. 
In what follows we investigate different combinations
of two rotations from the charged lepton sector and 
derive a sum rule for $\cos \delta$ in each set-up.
We will not consider the case 
$(\theta^e_{12},\theta^e_{23}) - (\theta^\nu_{23}, \theta^\nu_{12})$,
because it has been thoroughly analysed in 
refs. \cite{Marzocca:2013cr,Petcov:2014laa,Girardi:2014faa},
and, as we have already noted, the resulting 
sum rule eq.~(\ref{cosdthnu}) derived in \cite{Petcov:2014laa}
holds for an arbitrary fixed value of $\theta^{\nu}_{23}$.

%%%%%%%%%%%%%%%%%%%%%%%
%
\subsection{The Scheme with $(\theta^e_{13},\theta^e_{23}) - 
(\theta^\nu_{23}, \theta^\nu_{12})$ Rotations }
\label{sec:13e23e23nu12nu}
%
%%%%%%%%%%%%%%%%%%%%%%%%%%%%%%%%%%%%%%
%

Following the method used in ref. \cite{Marzocca:2013cr},
the PMNS matrix $U$ 
from eq.~(\ref{eq:Uijkl}) with $(ij) - (kl) = (13) - (23)$,
can be cast in the form:
%%%%%%%%%%%%%%%%%%%%%%%
\be
U = R_{13}(\theta^e_{13}) \, P_1 \, R_{23}(\hat \theta_{23}) \, R_{12}(\theta^{\nu}_{12}) \, \hat Q \,,
\label{eq:U1323b}
\ee
%%%%%%%%%%%%%%%%%%%%%%%
%
where the angle $\hat \theta_{23}$ is determined 
 i) for $\theta^{\nu}_{23} = -\pi/4$ by
%%%%%%%%%%%%%%%%%%%%%%%
\be
\sin^2 \hat \theta_{23} = 
\frac{1}{2}\left(1 - \sin 2 \theta^e_{23} \cos( \omega - \psi)\right) \,,
\label{eq:thetahat}
\ee
%%%%%%%%%%%%%%%%%%%%%%%
%
and ii) for an arbitrary fixed value of $\theta^{\nu}_{23}$ by
%%%%%%%%%%%%%%%%%%%%%%%
\be
\sin^2 \hat \theta_{23} = \sin^2 \theta^e_{23} \cos^2 \theta^{\nu}_{23} + \cos^2 \theta^e_{23} \sin^2 \theta^{\nu}_{23} +
 \frac{1}{2} \sin 2 \theta^e_{23} \sin 2 \theta^{\nu}_{23} 
\cos( \omega - \psi) \,.
\label{eq:hattheta23gen}
\ee
%%%%%%%%%%%%%%%%%%%%%%%%
%
The phase matrices $P_1$ and $\hat Q$ have the form:  
%%%%%%%%%%%%%%%%%%%%%%%
\be
P_1 = \diag (1, 1, e^{-i \alpha}),~~\text{and}~~\hat Q = Q_1 \, Q_0\,, 
~~\text{with}~~Q_1 = \diag (1, 1, e^{i \beta})\,,
\label{eq:P1Q1}
\ee
%%%%%%%%%%%%%%%%%%%%%%%
%
where the phases $\alpha$ and $\beta$ are given by
%%%%%%%%%%%%%%%%%%%%%%%
\be
\alpha = \gamma + \psi + \omega\,,~~\text{with}~~\gamma = 
\arg \left(e^{ -i \psi} \cos \theta^e_{23} \sin \theta^{\nu}_{23} 
+ e^{-i \omega} \sin \theta^e_{23} \cos \theta^{\nu}_{23} \right)\,,
\ee
%%%%%%%%%%%%%%%%%%%%%%%
%%%%%%%%%%%%%%%%%%%%%%%
\be
\beta = \gamma - \phi\,,~~\text{where}~~\phi = 
\arg \left(e^{ -i \psi} \cos \theta^e_{23} \cos \theta^{\nu}_{23} 
- e^{-i \omega} \sin \theta^e_{23} \sin \theta^{\nu}_{23} \right)\,.
\ee
%%%%%%%%%%%%%%%%%%%%%%%

Using eq.~(\ref{eq:U1323b}) and the standard 
parametrisation of $U$, we find:
%%%%%%%%%%%%%%%%%%%%%%%
\begin{align}
\sin^2 \theta_{13} & = |U_{e3}|^2  =  \sin^2 \theta^e_{13} \cos^2 \hat \theta_{23} \,,\\[0.3cm]
\sin^2 \theta_{23} & = \frac{|U_{\mu3}|^2}{1-|U_{e3}|^2} = \dfrac{\sin^2 \hat \theta_{23}}{1 - \sin^2 \theta_{13}} \,,\\[0.3cm]
\sin^2 \theta_{12} & = \frac{|U_{e2}|^2}{1-|U_{e3}|^2} = \dfrac{1}{1 - \sin^2 \theta_{13}}  \bigg[ \cos^2 \theta^e_{13} \sin^2 \theta^{\nu}_{12} \nonumber \\[0.3cm]
& - \dfrac{1}{2} \sin \hat \theta_{23} \sin 2 \theta^e_{13} \sin 2 \theta^{\nu}_{12} \cos \alpha
  + \cos^2 \theta^{\nu}_{12} \sin^2 \theta^e_{13} \sin^2 \hat \theta_{23} \bigg] \,. \label{eq:th12C0}
\end{align}
%%%%%%%%%%%%%%%%%%%%%%%
%
The first two equations allow one to express $\theta^e_{13}$ and $\hat \theta_{23}$ 
in terms of $\theta_{13}$ and  $\theta_{23}$.
Eq.~(\ref{eq:th12C0})
allows us to find $\cos \alpha$ as a function of 
the PMNS mixing angles  $\theta_{12}$,  $\theta_{13}$, $\theta_{23}$
and the angle $\theta^{\nu}_{12}$:
%%%%%%%%%%%%%%%%%%%%%%%
\be
\cos \alpha = 2 \, \frac{\sin^2 \theta^{\nu}_{12} \cos^2 \theta_{23}
+ \cos^2 \theta^{\nu}_{12} \sin^2 \theta_{23} \sin^2 \theta_{13}
- \sin^2 \theta_{12} \left(1 - \sin^2 \theta_{23} \cos^2 \theta_{13}\right)}
{\sin 2 \theta^{\nu}_{12} \sin 2 \theta_{23}  \sin \theta_{13}} \,.
\label{eq:cosalpha13e23e}
\ee
%%%%%%%%%%%%%%%%%%%%%%%
%
The relation 
\footnote{We note that the expression (\ref{eq:cosalpha13e23e}) 
for $\cos \alpha$ can be obtained formally 
from the r.h.s. of the eq. (22) for $\cos\phi$ in \cite{Petcov:2014laa} 
by substituting $\sin \theta_{23}$ with $\cos \theta_{23}$
and vice versa and by changing its overall sign.}
between $\sin \delta$ ($\cos \delta$) 
and $\sin\alpha$ ($\cos\alpha$) 
can be found by comparing
the imaginary (the real) part of the quantity  
$U^*_{e1} U^*_{\mu 3} U_{e3} U_{\mu 1}$,
written using eq.~(\ref{eq:U1323b}) and 
using the standard parametrisation of $U$:
%%%%%%%%%%%%%%%%%%%%%%%
\begin{align}
\sin \delta & = \dfrac{\sin 2 \theta^{\nu}_{12}}{\sin 2 \theta_{12}} \sin \alpha \,, \\[0.3cm]
\cos \delta & = \dfrac{\sin 2 \theta^{\nu}_{12}}{\sin 2 \theta_{12}} \cos \alpha
- \dfrac{\sin \theta_{13}}{\sin 2 \theta_{12}} \tan \theta_{23} \left(\cos 2 \theta_{12} + \cos 2 \theta^{\nu}_{12}\right)\,.
\end{align}
%%%%%%%%%%%%%%%%%%%%%%%
%
The sum rule expression for  $\cos \delta$ as a function 
of the mixing angles $\theta_{12}$, $\theta_{13}$, $\theta_{23}$ 
and $\theta^{\nu}_{12}$, with $\theta^{\nu}_{12}$ 
having an arbitrary fixed value, reads:
%%%%%%%%%%%%%%%%%%%%%%%
\begin{align}
\cos\delta  
=  -\frac{\cot\theta_{23}}{\sin2\theta_{12}\sin\theta_{13}}\,
\left [\cos2\theta^{\nu}_{12} +
\left (\sin^2\theta_{12} - \cos^2\theta^{\nu}_{12} \right )\,
 \left (1 - \tan^2\theta_{23}\,\sin^2\theta_{13}\right )\right ]\,.
 \label{eq:cosdelta13e23e}
\end{align}
%%%%%%%%%%%%%%%%%%%%%%%
%
This sum rule for $\cos\delta$  
can be obtained formally from the 
r.h.s. of eq.~(\ref{cosdthnu}) 
by interchanging $\tan \theta_{23}$ and $\cot \theta_{23}$
and by multiplying it by $(-1)$.
Thus, in the case of $\theta_{23} = \pi/4$, 
the predictions for $\cos \delta$ 
in the case under consideration will differ 
from those obtained using  eq.~(\ref{cosdthnu})  
only by a sign.
We would like to emphasise that, 
as the sum rule in eq.~(\ref{cosdthnu}),
the sum rule in eq.~(\ref{eq:cosdelta13e23e}) is valid for any fixed 
value of $\theta^{\nu}_{23}$.

The $J_{\rm CP}$ factor has the following form in 
the parametrisation of the PMNS matrix 
employed in the present subsection:
%%%%%%%%%%%%%%%%%%%%%%%%%%%%%%%%%%%%%
\be
J_{\rm CP} = \frac{1}{8} \sin 2 \theta^e_{13} \sin 2 \theta^{\nu}_{12} \sin 2 \hat \theta_{23} \cos \hat \theta_{23} \sin \alpha \,.
\ee
%%%%%%%%%%%%%%%%%%%%%%%%%%%%%%%%%%%%%%
%

%%%%%%%%%%%%%%%%%%%%%%%%%%%%%%%
%
\subsection{The Scheme with $(\theta^e_{12},\theta^e_{13}) - 
(\theta^\nu_{23}, \theta^\nu_{12})$ Rotations }
\label{sec:12e13e23nu12nu}
%
%%%%%%%%%%%%%%%%%%%%%%%%%%%
%

 In this subsection we consider the parametrisation 
of the matrix $U$ defined in eq.~(\ref{eq:Uijkl}) 
with $(ij) - (kl) = (12) - (13)$ under the assumption of vanishing $\omega$,
i.e., $\Psi = \diag(1, e^{-i \psi},1)$.
In the case 
of non-fixed $\omega$ it is impossible to express 
$\cos \delta$ {\it only} in terms of the independent 
angles of the scheme. We will comment more on 
this case later.

 Using the parametrisation given in eq.~(\ref{eq:Uijkl}) 
with $\theta^{\nu}_{23} = -\pi/4$ and $\omega = 0$
and the standard one, we find:
%%%%%%%%%%%%%%%%%%%%%%%%%%%%%%%%
\begin{align}
\sin^2 \theta_{13} & = |U_{e3}|^2  = \dfrac{1}{2} \sin^2 \theta^e_{12} + \dfrac{1}{2} \cos^2 \theta^e_{12} \sin^2 \theta^e_{13} - X_{\psi}\,, 
\label{eq:th13omega0}\\
\sin^2 \theta_{23} & = \frac{|U_{\mu3}|^2}{1-|U_{e3}|^2} = \dfrac{1}{\cos^2 \theta_{13}} \bigg[ \dfrac{1}{2} \cos^2 \theta^e_{12} + \dfrac{1}{2} \sin^2 \theta^e_{12} \sin^2 \theta^e_{13} + X_{\psi} \bigg]\,, 
\label{eq:th23omega0}\\
\sin^2 \theta_{12} & = \frac{|U_{e2}|^2}{1-|U_{e3}|^2} = \dfrac{\zeta \sin^2 \theta^e_{12} + \xi}{1-\sin^2 \theta_{13}}\,, 
\label{eq:th12omega0}
\end{align}
%%%%%%%%%%%%%%%%%%%%%%%%%%%
%
where 
%%%%%%%%%%%%%%%%%%%%%%
\begin{align}
X_{\psi} & = \dfrac{1}{2} \sin 2 \theta^e_{12} \sin \theta^e_{13} \cos \psi\,, \\
\zeta & = \cos^2 \theta^e_{13} \cos 2 \theta^{\nu}_{12} + \dfrac{1}{4\sqrt{2}} \sin 2\theta^{\nu}_{12} \cot \theta^e_{13} (3 \cos 2 \theta^e_{13} - 1)\,, \\
\xi & = \cos^2 \theta^e_{13} \sin^2 \theta^{\nu}_{12} + \dfrac{1}{2} (\cos 2 \theta_{13} - \cos 2 \theta^e_{13}) \cos^2 \theta^{\nu}_{12}   \nonumber \\
& + \dfrac{1}{2 \sqrt{2}} \sin 2 \theta^{\nu}_{12} (3 \cos \theta^e_{13} \sin \theta^e_{13}
-2 \cot \theta^e_{13} \sin^2 \theta_{13})\,.
\end{align}
%%%%%%%%%%%%%%%%%%%%%%%%%%%%
%
The dependence on $\cos \psi$ in eq.~(\ref{eq:th12omega0}) 
has been eliminated by solving eq.~(\ref{eq:th13omega0}) for $X_{\psi}$. 
It follows from eqs.~(\ref{eq:th13omega0}) and (\ref{eq:th23omega0}) that 
$\sin^2\theta^e_{13}$ is a function of the known mixing angles 
$\theta_{13}$ and $\theta_{23}$:
%%%%%%%%%%%%%%%%%%%%%%%%%%%
\be
\sin^2 \theta^e_{13} = 1 - 2 \cos^2 \theta_{13} \cos^2 \theta_{23}\, .
\label{eq4:the13}
\ee
%%%%%%%%%%%%%%%%%%%%%%%%%%%%%
%
Inverting the formula for $\sin^2 \theta_{12}$ allows us to find 
$\sin^2 \theta^e_{12}$, which is given by
%%%%%%%%%%%%%%%%%%%%%%%%%
\begin{align}
\sin^2 \theta^e_{12} & = \bigg[ 4 \left[\cos 2 \theta^{\nu}_{12}
(\cos 2 \theta^e_{13} + \sin^2 \theta_{13}) 
-\, \cos 2 \theta_{12} \cos^2 \theta_{13} \right] \tan \theta^e_{13} 
+  \sqrt{2} \sin 2 \theta^{\nu}_{12} \nonumber \\
& \times (3 \cos 2 \theta^e_{13} - 2 \cos 2 \theta_{13} - 1) \bigg] 
\bigg[4 \cos 2 \theta^{\nu}_{12} \sin 2 \theta^e_{13} 
+ \sqrt{2}(3 \cos 2 \theta^e_{13} - 1) \sin 2 \theta^{\nu}_{12} \bigg]^{-1} .
\label{eq4:the12}
\end{align}
%%%%%%%%%%%%%%%%%%%%%%%%%%%%%%%%%%%
%
Using eqs.~(\ref{eq:th13omega0}) and ~(\ref{eq4:the12})
we can write $\cos \psi$ in terms 
of the standard parametrisation mixing angles and the known 
 $\theta^e_{13}$ and $\theta^{\nu}_{12}$:
%%%%%%%%%%%%%%%%%%%%%%%%%%%%
\be
\cos \psi = \frac{\sin^2 \theta^e_{12} + \cos^2 \theta^e_{12} \sin^2 \theta^e_{13}
-2 \sin^2 \theta_{13}}{\sin 2 \theta^e_{12} \sin \theta^e_{13}}\,.
\label{eq4:psi}
\ee
%%%%%%%%%%%%%%%%%%%%%%%%%%%%
%

We find the relation between $\sin \delta$
and $\sin\psi$ by employing  again the standard 
procedure of comparing the expressions of the $J_{\rm CP}$ factor, 
$J_{\rm CP} = {\rm Im}(U^*_{e1} U^*_{\mu 3} U_{e3} U_{\mu 1})$, 
in the two parametrisations~---~the standard one and that defined 
in eq.~(\ref{eq:Uijkl}) (with $\theta^{\nu}_{23} = -\pi/4$ and $\omega = 0$): 
%%%%%%%%%%%%%%%%%%%%%%%%%%
\be
\sin \delta = \frac{\sin 2 \theta^e_{12} \sin \psi }{4 \sin 2 \theta_{12} \sin 2 \theta_{13} \sin \theta_{23}} \bigg[ 2 \sqrt{2} \sin 2 \theta^e_{13} \cos 2 \theta^{\nu}_{12} + (3 \cos 2 \theta^e_{13} - 1) \sin 2 \theta^{\nu}_{12} \bigg]\,,
\ee
%%%%%%%%%%%%%%%%%%%%%%%%%%%%
%
where $\sin2\theta^e_{12}$  ($\sin 2\theta^e_{13}$  and $\cos 2 \theta^e_{13}$) 
can be expressed in terms of  
$\theta_{12}$, $\theta_{13}$, $\theta_{23}$ 
and $\theta^{\nu}_{12}$ ($\theta_{13}$ and $\theta_{23}$) 
using eq. (\ref{eq4:the12}) (eq. (\ref{eq4:the13})).

 We use a much simpler procedure to find $\cos \delta$.
Namely, we compare the expressions for the absolute value of
the element $U_{\tau 1}$ of the PMNS matrix in the standard 
parametrisation and in the symmetry related
one, eq.~(\ref{eq:Uijkl}) 
with $\theta^{\nu}_{23} = -\pi/4$ and $\omega = 0$,  
considered in the present subsection:
%%%%%%%%%%%%%%%%%%%%%%%%%%%%%%%%%
\begin{align}
|U_{\tau 1}| = 
|\sin \theta_{23} \sin \theta_{12} - \sin \theta_{13} \cos \theta_{23} \cos \theta_{12} e^{i \delta}| = 
 |\sin \theta^e_{13} \cos \theta^{\nu}_{12} + \frac{1}{\sqrt{2}} \cos \theta^e_{13} \sin \theta^{\nu}_{12}| \,.
 \label{eq:Utau112e13e}
\end{align}
%%%%%%%%%%%%%%%%%%%%%%%%%%%%
%
From the above equation we get for $\cos \delta$:
%%%%%%%%%%%%%%%%%%%%%%%%%%%%%%%%%%%%%
\begin{align}
\cos\delta & =
-\frac{2}{\sin 2 \theta_{12} \sin 2 \theta_{23} \sin \theta_{13}}
\bigg [ \cos^2 \theta_{23} \sin^2 \theta_{12} \sin^2 \theta_{13} + \cos^2 \theta_{12} \sin^2 \theta_{23} \nonumber \\
& - \bigg( \sqrt{\cos^2 \theta_{13} \cos^2 \theta_{23}} \cos \theta^{\nu}_{12} - \kappa \sqrt{1- 2 \cos^2 \theta_{13} \cos^2 \theta_{23}} \sin \theta^{\nu}_{12} \bigg)^2 \, \bigg] \,,
\label{eq:cosdelta12e13e}
\end{align}
%%%%%%%%%%%%%%%%%%%%%%%%%%%%%
%
where $\kappa = 1$ if $\theta^e_{13}$ belongs to the first or third quadrant, 
and $\kappa = -1$ if $\theta^e_{13}$
is in the second or the fourth one.
%%%%%%%%%%%%%%%%%%%%%%%%%%%%%%%%%
%
In the parametrisation under discussion, 
eq.~(\ref{eq:Uijkl}) with $(ij) - (kl) = (12) - (13)$, 
$\theta^{\nu}_{23} = -\pi/4$
and $\omega = 0$, we have:
%%%%%%%%%%%%%%%%%%%%%%%%%%%%%%%%%
\be
J_{\rm CP} =  \frac{\sqrt{2}}{32} \cos \theta^e_{13} \sin 2 \theta^e_{12} \left( 2 \sqrt{2} \cos 2 \theta^{\nu}_{12} \sin 2 \theta^e_{13} +
 (3 \cos 2 \theta^e_{13} - 1) \sin 2 \theta^{\nu}_{12} \right) \sin \psi \,.
\ee
%%%%%%%%%%%%%%%%%%%%%%%%%%%%%%%%%%
%
In the case of non-vanishing $\omega$, using  
the same method
and eq.~(\ref{eq4:the13}), which also holds for $\omega \neq 0$, 
allows us to show that $\cos \delta$ is a
function of $\cos \omega$ as well: 
%%%%%%%%%%%%%%%%%%%%%%%%%%%%
\begin{align}
\cos \delta  & = -\frac{2 \cos^2 \theta_{23} }{\sin 2 \theta_{12} \sin 2 \theta_{23} \sin \theta_{13}}
\bigg [ (1 - 2 \cos^2 \theta_{13} \cos^2 \theta_{23}) \frac{\cos^2 \theta^{\nu}_{12}}{\cos^2 \theta_{23} } - \sin^2 \theta_{12} \tan^2 \theta_{23}  \nonumber \\
& + ( \cos^2 \theta_{13} \sin^2 \theta^{\nu}_{12} - \cos^2 \theta_{12} \sin^2 \theta_{13}) + \kappa \frac{\cos \theta_{13}}{\cos \theta_{23}} \sqrt{1 - 2 \cos^2 \theta_{13} \cos^2 \theta_{23}} \cos \omega \sin 2 \theta^{\nu}_{12} \bigg] \,.
\label{eq:cosdelta12e13e:omega}
\end{align}
%%%%%%%%%%%%%%%%%%%%%%%%%%
%

Finally, we generalise eq.~(\ref{eq:cosdelta12e13e:omega})
to the case of an arbitrary fixed value of $\theta^{\nu}_{23}$.
In this case 
%%%%%%%%%%%%%%%%%%%%%%%%%%%%%%%%%
\be
\sin^2 \theta^e_{13} = \frac{1 - \cos^2 \theta_{13} \cos^2 \theta_{23} - \sin^2 \theta^{\nu}_{23}}{\cos^2 \theta^{\nu}_{23}} \,,
\ee
%%%%%%%%%%%%%%%%%%%%%%%%%%%%
%
and eqs.~(\ref{eq:Utau112e13e}) and (\ref{eq:cosdelta12e13e:omega}) read: 
%%%%%%%%%%%%%%%%%%%%%%%%%%%%%%%%%%%%%%
\begin{align}
|U_{\tau 1}| = 
|\sin \theta_{23} \sin \theta_{12} - \sin \theta_{13} \cos \theta_{23} \cos \theta_{12} e^{i \delta}| = 
 |\cos \theta^{\nu}_{12} \sin \theta^e_{13} - e^{-i \omega}\cos \theta^e_{13} \sin \theta^{\nu}_{12} \sin \theta^{\nu}_{23}| \,,
 \label{eq:Utau112e13egen}
\end{align}
%%%%%%%%%%%%%%%%%%%%%%%%%%%%

%%%%%%%%%%%%%%%%%%%%%%%%%%%%%%%%%%%%%%
\begin{align}
\cos \delta  & = \dfrac{1}{\sin2 \theta_{12} \sin 2 \theta_{23} \sin \theta_{13}} \bigg [ 
\frac{2 \kappa \cos \omega \sin 2 \theta^{\nu}_{12} \sin \theta^{\nu}_{23} \cos \theta_{13} \cos \theta_{23}}{\cos^2 \theta^{\nu}_{23}} 
(\cos^2 \theta^{\nu}_{23} - \cos^2 \theta_{13} \cos^2 \theta_{23})^{\frac{1}{2}} \nonumber \\
& - \cos 2 \theta^{\nu}_{12} \Big(1 - \frac{\cos^2 \theta_{13} \cos^2 \theta_{23}}{\cos^2 \theta^{\nu}_{23} }(\sin^2 \theta^{\nu}_{23} + 1) \Big)
 +\cos 2 \theta_{12} \big(\cos^2 \theta_{23} \sin^2 \theta_{13} - \sin^2 \theta_{23} \big) \bigg] \,.
\label{eq:cosdelta12e13e:omegagen}
\end{align}
%%%%%%%%%%%%%%%%%%%%%%%%%%%%%%%%%%%%%%
%

It follows from the results for $\cos\delta$ obtained for 
$\cos\omega \neq 0$, eqs. (\ref{eq:cosdelta12e13e:omega}) 
and (\ref{eq:cosdelta12e13e:omegagen}), that in the case 
analysed in the present subsection 
one can obtain predictions for $\cos\delta$ only in theoretical models in 
which the value of the phase $\omega$ is fixed by the model.

%%%%%%%%%%%%%%%%%%%%%%%%%%%%%%%%%%%
%
\section{The Cases of $\theta^e_{ij} - (\theta^\nu_{23}, \theta^\nu_{13}, \theta^\nu_{12})$ 
 Rotations }
\label{sec:ije23nu13nu12nu}
%
%%%%%%%%%%%%%%%%%%%%%%%%%%%%%%%%%%%%%%%%%%%
%

We consider next a generalisation of the cases analysed
in Section \ref{sec:ije23nu12nu} with the presence of 
a third rotation matrix in 
$\tilde{U}_{\nu}$ arising from the neutrino
sector, i.e., we employ the parametrisation of 
$U$ given in eq.~(\ref{eq:Uija}). 
Non-zero values of $\theta^{\nu}_{13}$ are
inspired by certain types of flavour symmetries
(see, e.g., \cite{Bazzocchi:2011ax,Rodejohann:2014xoa,Toorop:2011jn,King:2012in}).
In the case of $\theta^{\nu}_{12} = \theta^{\nu}_{23} = - \pi/4$ 
and $\theta^{\nu}_{13} = \sin^{-1} (1 / 3)$, for instance,
we have the so-called tri-permuting (TP) pattern,
which was proposed and studied in \cite{Bazzocchi:2011ax}.
In the statistical analysis of the predictions 
for $\cos\delta$, $\delta$ and the $J_{\rm CP}$ factor we will perform 
in Section~\ref{sec:predictions}, 
we will consider three representative values of 
$\theta^{\nu}_{13}$ discussed in the literature:
$\theta^{\nu}_{13} = \pi/20,~\pi/10$ and $\sin^{-1} (1 / 3)$. 

 For the parametrisation of the matrix $U$
given in eq.~(\ref{eq:Uija}) with $(ij) = (23)$,
no constraints on the phase $\delta$ can be obtained.
Indeed, after we recast $U$ in the form
%%%%%%%%%%%%%%%%%%%%%%%%%%%%%%%%%%%
\be
U =  R_{23}(\hat \theta_{23}) \, Q_1 \, R_{13}(\theta^{\nu}_{13})\, R_{12}(\theta^{\nu}_{12}) \, Q_0 \,,
\label{eq:parU3rot23e}
\ee
%%%%%%%%%%%%%%%%%%%%%%%%%
%
where $\sin^2 \hat \theta_{23}$ and $Q_1$ are given in eqs.~(\ref{eq:hattheta23gen}) and (\ref{eq:P1Q1}), respectively,
we find employing a similar procedure used in the previous sections:
%%%%%%%%%%%%%%%%%%%%%%%%%%
\be
\sin^2 \theta_{13} = \sin^2 \theta^{\nu}_{13}\,, \quad
\sin^2 \theta_{23}  =  \sin^2 \hat \theta_{23}\,, \quad
\sin^2 \theta_{12}  = \sin^2 \theta^{\nu}_{12}\,, \quad
\sin \delta = \sin \beta\,.
\ee
%%%%%%%%%%%%%%%%%%%%%%%%%%%%%%%%%%
%
 Thus, there is no correlation between the Dirac CPV phase $\delta$
and the mixing angles in this set-up.

%%%%%%%%%%%%%%%%%%%%%%%
%
\subsection{The Scheme with $\theta^e_{12} - 
(\theta^\nu_{23}, \theta^\nu_{13}, \theta^\nu_{12})$ Rotations }
\label{sec:12e23nu13nu12nu} 
%
%%%%%%%%%%%%%%%%%%%%%%%%%%%%
%

In the parametrisation
of the matrix $U$ given in eq.~(\ref{eq:Uija}) with $(ij) = (12)$,
the phase $\omega$ in the matrix $\Psi$ is unphysical
(it ``commutes'' with $R_{12}(\theta^e_{12})$ and can be absorbed 
by the $\mu^\pm$ field). Hence, the matrix $\Psi$ contains 
only one physical phase $\phi$,
$\Psi = \diag \, (1, e ^{i \phi}, 1)$, and $\phi \equiv - \psi$.
Taking this into account and using eq.~(\ref{eq:Uija}) with $(ij) = (12)$ 
and $\theta^{\nu}_{23} = -\pi/4$,
we get the following expressions for 
$\sin^2\theta_{13}$, $\sin^2\theta_{23}$
and $\sin^2\theta_{12}$: 
%%%%%%%%%%%%%%%%%%%%%%%%%%%%%%%%%
\begin{align}
\sin^2 \theta_{13} & = |U_{e3}|^2  = \frac{1}{2} \sin^2 \theta^e_{12} \cos^2 \theta^{\nu}_{13} + \cos^2 \theta^e_{12} \sin^2 \theta^{\nu}_{13} - X_{12} \sin \theta^{\nu}_{13} \,, 
\label{eq:th13}\\
\sin^2 \theta_{23} & = \frac{|U_{\mu3}|^2}{1-|U_{e3}|^2} = 1 - \frac{\cos^2 \theta^{\nu}_{13}}{2\,(1 - \sin^2 \theta_{13})} \,, 
\label{eq:th23} \\
\sin^2 \theta_{12} & = \frac{|U_{e2}|^2}{1-|U_{e3}|^2} = \frac{1}{1-\sin^2 \theta_{13}} 
\bigg[ \dfrac{1}{2} \sin^2 \theta^e_{12} \left( \cos \theta^{\nu}_{12} + \sin \theta^{\nu}_{12} \sin \theta^{\nu}_{13}\right)^2 \nonumber \\
& +  \cos^2 \theta^e_{12} \cos^2 \theta^{\nu}_{13} \sin^2 \theta^{\nu}_{12} + X_{12} \sin \theta^{\nu}_{12} \left( \cos \theta^{\nu}_{12} + \sin \theta^{\nu}_{12} \sin \theta^{\nu}_{13}\right) \bigg] 
\label{eq:th12} \,,
\end{align}
%%%%%%%%%%%%%%%%%%%%%%%%%%
%
where
%%%%%%%%%%%%%%%%%%%%%%%%%%
\be
X_{12} = \frac{1}{\sqrt{2}} \sin 2 \theta^e_{12} \cos \theta^{\nu}_{13} \cos \phi \,.
\ee
%%%%%%%%%%%%%%%%%%%%%%%%

 Solving eq.~(\ref{eq:th13}) for $X_{12}$ and inserting 
the solution in eq.~(\ref{eq:th12}), 
we find $\sin^2 \theta_{12}$
as a function of $\theta_{13}$, $\theta^{\nu}_{12}$, $\theta^{\nu}_{13}$ 
and $\theta^e_{12}$:
%%%%%%%%%%%%%%%%%%%%%%%%%%%
\be
\sin^2 \theta_{12} = \frac{\alpha \sin^2 \theta^e_{12} + \beta}{1-\sin^2 \theta_{13}} \,.
\ee
%%%%%%%%%%%%%%%%%%%%%%
%
Here the parameters $\alpha$ and $\beta$ are given by: 
%%%%%%%%%%%%%%%%%%%%%%
\begin{align}
\alpha & = \frac{1}{4} \left[ 2\cos 2 \theta^{\nu}_{12} + \sin 2 \theta^{\nu}_{12} \frac{\cos^2 \theta^{\nu}_{13} }{\sin \theta^{\nu}_{13} } \right] \,,\\
\beta & = \sin \theta^{\nu}_{12} \left[ \cos^2 \theta_{13} \sin \theta^{\nu}_{12} + \cos \theta^{\nu}_{12} \left( \sin \theta^{\nu}_{13} - \frac{\sin^2 \theta_{13}}{\sin \theta^{\nu}_{13}} \right)\right] \,.
\end{align}
%%%%%%%%%%%%%%%%%%%%%%%
%
Inverting the formula for $\sin^2 \theta_{12}$ allows us to express
$\sin^2 \theta^e_{12}$ in terms of $\theta_{12}$, $\theta_{13}$, 
$\theta^{\nu}_{12}$, $\theta^{\nu}_{13}$:
%%%%%%%%%%%%%%%%%%%%%%%%%%%%%%%%%
\be
\sin^2 \theta^e_{12} = \frac{2 \cos^2 \theta_{13}\sin \theta^{\nu}_{13} (\sin^2 \theta_{12} - \sin^2 \theta^{\nu}_{12}) + \sin 2 \theta^{\nu}_{12} \sin^2 \theta_{13}  - \sin 2\theta^{\nu}_{12} \sin^2 \theta^{\nu}_{13}}
{\cos 2 \theta^{\nu}_{12} \sin \theta^{\nu}_{13} + \cos \theta^{\nu}_{12} \sin \theta^{\nu}_{12} \cos^2 \theta^{\nu}_{13}} .
\label{eq5_1:the12}
\ee
%%%%%%%%%%%%%%%%%%%%%%%%%%%%%%%%%%%
%
In the limit of vanishing $\theta^{\nu}_{13}$
we have $\sin^2 \theta^e_{12} = 2 \sin^2 \theta_{13}$,
which corresponds to the case
of negligible $\theta^e_{23}$  considered in
\cite{Petcov:2014laa}.

  Using eq.~(\ref{eq:th12}), one can express
$\cos \phi$ in terms of the ``standard'' mixing angles
$\theta_{12}$, $\theta_{13}$
and the angles $\theta^e_{12}$, $\theta^{\nu}_{12}$
and $\theta^{\nu}_{13}$ which are assumed to have known values:
%%%%%%%%%%%%%%%%%%%%%%%%%%%%%%%%%
\begin{align}
\label{eq:phi}
\cos \phi  = &  \bigg [2 \cos^2 \theta_{13} (\sin \theta^e_{12})^{-2} (\sin \theta^{\nu}_{12})^{-2} \sin^2 \theta_{12}
- 2 \cos^2 \theta^{\nu}_{13} \cot^2 \theta^e_{12}  
- (\cot \theta^{\nu}_{12} + \sin \theta^{\nu}_{13})^2 \bigg ] \nonumber \\
 \times & (\cos \theta^{\nu}_{13})^{-1} \tan \theta^e_{12} \bigg[ 2 \sqrt{2} (\cot \theta^{\nu}_{12} + \sin \theta^{\nu}_{13})\bigg]^{-1} \,.
\end{align}
%%%%%%%%%%%%%%%%%%%%%%%%%%%%%%%%
%
 We note that from the requirements 
$(0 < \sin^2 \theta^e_{12} < 1) \land (-1 < \cos \phi < 1)$
one can obtain for a given $\theta^{\nu}_{13}$,
each of the symmetry values of $\theta^{\nu}_{12}$ considered 
and $\theta^{\nu}_{23} = -\pi/4$, 
lower and upper bounds on the value of
$\sin^2 \theta_{12}$.
These bounds will be discussed in subsection~\ref{sec:pred12e}.
Comparing the  expressions for 
$J_{\rm CP} = {\rm Im}(U^*_{e1} U^*_{\mu 3} U_{e3} U_{\mu 1})$,
obtained using eq.~(\ref{eq:Uija}) with $(ij) = (12)$ 
and $\theta^{\nu}_{23} = -\pi/4$,
and in the standard parametrisation of $U$,
one gets the relation between $\sin\phi$ and $\sin\delta$:
%%%%%%%%%%%%%%%%%%%%%%%%%%%%
\begin{align}
\sin \delta & = -\frac{\sin 2 \theta^e_{12}}{2 \sin 2 \theta_{12} \sin 2 \theta_{13} \sin \theta_{23}} \bigg[ \cos^2 \theta^{\nu}_{13} \sin 2 \theta^{\nu}_{12} + 2 \cos 2 \theta^{\nu}_{12} \sin \theta^{\nu}_{13} \bigg ] \sin \phi \,.
\end{align}
%%%%%%%%%%%%%%%%%%%%%%%%%%%%%%
%
Similarly to the method employed in the previous Section, 
we use the equality of the 
expressions for $|U_{\tau 1}|$ in the two parametrisations 
in order to derive the sum rule for $\cos \delta$ of interest:
%%%%%%%%%%%%%%%%%%%%%%%
\begin{align}
|U_{\tau 1}| = 
|\sin \theta_{23} \sin \theta_{12} - \sin \theta_{13} \cos \theta_{23} \cos \theta_{12} e^{i \delta}| = 
\frac{1}{\sqrt{2}} |\sin \theta^{\nu}_{12} + \cos \theta^{\nu}_{12} \sin \theta^{\nu}_{13}| \,.
\label{eq:Utau112e}
\end{align}
%%%%%%%%%%%%%%%%%%%%%%%%%%%%
%
From the above equation we find the following sum rule 
for $\cos \delta$:
%%%%%%%%%%%%%%%%%%%%%%%%
\begin{align}
\cos \delta & =
\frac{1}{\sin2\theta_{12} \sin\theta_{13} |\cos\theta^{\nu}_{13}|
(1 - 2\sin^2\theta_{13} + \sin^2\theta^{\nu}_{13})^{\frac{1}{2}}}
\bigg[\left(1 - 2\sin^2\theta_{13} + \sin^2\theta^{\nu}_{13}\right) \sin^2 \theta_{12}
\nonumber \\
& + \cos^2\theta_{12}\sin^2\theta_{13}\cos^2\theta^{\nu}_{13}
- \cos^2\theta_{13} \left(\sin\theta^{\nu}_{12} + \cos\theta^{\nu}_{12} \sin\theta^{\nu}_{13}\right)^2\bigg]\,.
\label{eq:cosdelta12e23nu13nu12nu}
\end{align}
%%%%%%%%%%%%%%%%%%%%%%%%
%
For $\theta^{\nu}_{13} = 0$ this sum rule reduces to the sum rule 
for $\cos\delta$ given in eq. (50) in \cite{Petcov:2014laa}.

 In the parametrisation of the PMNS matrix 
considered in this subsection, 
the rephasing invariant $J_{\rm CP}$ has the form:
%%%%%%%%%%%%%%%%%%%%%%%%%%%%%
\be
J_{\rm CP} = -\frac{1}{8 \sqrt{2}} \sin \phi \cos \theta^{\nu}_{13}  \sin 2\theta^e_{12}
\left[\cos^2 \theta^{\nu}_{13} \sin 2\theta^{\nu}_{12} + 2 \sin \theta^{\nu}_{13} \cos 2\theta^{\nu}_{12}\right] \,.
\label{eq:JCP12e}
\ee
%%%%%%%%%%%%%%%%%%%%%%%%%%%%%%%%%%%%%%%

In the case when $\theta^{\nu}_{23}$ has a fixed value 
which differs from $-\pi/4$,
the expression for $\sin^2 \theta_{23}$, eq. (\ref{eq:th23}),
changes as follows:
%%%%%%%%%%%%%%%%%%%%%%%%%%%%%%%%%%
\begin{align}
\sin^2 \theta_{23} & = \frac{|U_{\mu3}|^2}{1-|U_{e3}|^2} = 
1 - \frac{\cos^2 \theta^{\nu}_{23} \cos^2 \theta^{\nu}_{13}}
{1 - \sin^2 \theta_{13}} \,. 
\label{eq:th23gen}
\end{align}
%%%%%%%%%%%%%%%%%%%%%%%%%%%%%
%
Equations (\ref{eq:Utau112e}) and (\ref{eq:cosdelta12e23nu13nu12nu}) 
are also modified:
%%%%%%%%%%%%%%%%%%%%%%%%%%%%%%%
\begin{align}
|U_{\tau 1}| = 
|\sin \theta_{23} \sin \theta_{12} - \sin \theta_{13} \cos \theta_{23} \cos \theta_{12} e^{i \delta}| = 
 |\sin \theta^{\nu}_{12} \sin \theta^{\nu}_{23} - \cos \theta^{\nu}_{23} \cos \theta^{\nu}_{12} \sin \theta^{\nu}_{13}| \,,
\label{eq:Utau112egen}
\end{align}
%%%%%%%%%%%%%%%%%%%%%%%%%%%%%%%%%%%%%
%
and
%%%%%%%%%%%%%%%%%%%%%%%%%%%%%%%%%%%%%%%%%
\begin{align}
\cos \delta & = \frac{1}{\sin 2 \theta_{12} \sin \theta_{13} | \cos \theta^{\nu}_{13} \cos \theta^{\nu}_{23}| (\cos^2 \theta_{13} - \cos^2 \theta^{\nu}_{13} \cos^2 \theta^{\nu}_{23})^{\frac{1}{2}}}
 \nonumber \\
& \times \bigg[ (\cos^2 \theta_{13} - \cos^2 \theta^{\nu}_{13} \cos^2 \theta^{\nu}_{23}) \sin^2 \theta_{12} + \cos^2 \theta_{12} \sin^2 \theta_{13} \cos^2 \theta^{\nu}_{13} \cos^2 \theta^{\nu}_{23} 
\nonumber \\
& - \cos^2 \theta_{13} ( \cos \theta^{\nu}_{12} \sin \theta^{\nu}_{13} \cos \theta^{\nu}_{23} - \sin \theta^{\nu}_{12} \sin \theta^{\nu}_{23})^2  \bigg] \,.
\label{eq:cosdelta12e23nu13nu12nugen}
\end{align}
%%%%%%%%%%%%%%%%%%%%%%%%%%%%%%%%%%%%%%
%

In the case of bi-trimaximal mixing \cite{Toorop:2011jn}, 
i.e., for  $\theta^\nu_{12} = \theta^{\nu}_{23} = \tan^{-1} (\sqrt{3} - 1)$
and $\theta^\nu_{13} = \sin^{-1} ((3 - \sqrt{3})/6)$,  
the sum rule we have derived reduces to 
the sum rule obtained  in \cite{Ballett:2014dua}.
However, this case is statistically disfavored 
by the current global neutrino oscillation data.

%%%%%%%%%%%%%%%%%%%%%%%%%%%%%%%%
%
\subsection{The Scheme with $\theta^e_{13} - 
(\theta^\nu_{23}, \theta^\nu_{13}, \theta^\nu_{12})$ Rotations }
\label{sec:13e23nu13nu12nu}
%
%%%%%%%%%%%%%%%%%%%%%%%%%%%%%%%%%%

Here we  
switch to the parametrisation 
of the matrix $U$ given in eq.~(\ref{eq:Uija}) with $(ij) = (13)$.
Now the phase $\psi$ in the matrix $\Psi$
is unphysical, and $\Psi = \diag (1, 1, e^{-i \omega})$.
Fixing $\theta^{\nu}_{23} = -\pi/4$ and
using also the standard parametrisation of $U$, we find:
%%%%%%%%%%%%%%%%%%%%%%%%%%%%%%%
\begin{align}
\sin^2 \theta_{13} & = |U_{e3}|^2  = \frac{1}{2} \sin^2 \theta^e_{13} \cos^2 \theta^{\nu}_{13} + \cos^2 \theta^e_{13} \sin^2 \theta^{\nu}_{13} + X_{13}  \sin \theta^{\nu}_{13} \,, 
\label{eq:th13B}\\
\sin^2 \theta_{23} & = \frac{|U_{\mu3}|^2}{1-|U_{e3}|^2} = \frac{\cos^2 \theta^{\nu}_{13}}{2\,(1 - \sin^2 \theta_{13})} \,, 
\label{eq:th23B} \\
\sin^2 \theta_{12} & = \frac{|U_{e2}|^2}{1-|U_{e3}|^2} = \frac{1}{1-\sin^2 \theta_{13}} 
\bigg[ \dfrac{1}{2} \sin^2 \theta^e_{13} \left( \cos \theta^{\nu}_{12} - \sin \theta^{\nu}_{12} \sin \theta^{\nu}_{13}\right)^2 \nonumber \\
& +  \cos^2 \theta^e_{13} \cos^2 \theta^{\nu}_{13} \sin^2 \theta^{\nu}_{12} + X_{13} \sin \theta^{\nu}_{12} \left( \cos \theta^{\nu}_{12} - \sin \theta^{\nu}_{12} \sin \theta^{\nu}_{13}\right) \bigg] \,. 
\label{eq:th12B}
\end{align}
%%%%%%%%%%%%%%%%%%%%%%%%%%%%
%
Here
%%%%%%%%%%%%%%%%%%%%%%%%%%%%%%%%%
\be
X_{13} = \frac{1}{\sqrt{2}} \sin 2 \theta^e_{13} \cos \theta^{\nu}_{13} \cos \omega \,.
\ee
%%%%%%%%%%%%%%%%%%%%%%%%%
%
Solving eq.~(\ref{eq:th13B}) for $X_{13}$
and inserting the solution in eq.~(\ref{eq:th12B}),
it is not dificult to find $\sin^2 \theta_{12}$ as a function of
$\theta_{13}$, $\theta^{\nu}_{12}$, $\theta^{\nu}_{13}$
and $\theta^e_{13}$:
%%%%%%%%%%%%%%%%%%%%%%%%%%%
\be
\sin^2 \theta_{12}  = \frac{\rho \sin^2 \theta^e_{13} 
+ \eta}{1-\sin^2 \theta_{13}} \,,
\label{s2th12rhoeta}
\ee
%%%%%%%%%%%%%%%%%%%%%%%%%%%%
%
where $\rho$ and $\eta$ are given by
%%%%%%%%%%%%%%%%%%%%%%
\begin{align}
\rho & = \frac{1}{4} \left[ 2\cos 2 \theta^{\nu}_{12} - \sin 2 \theta^{\nu}_{12} \frac{\cos^2 \theta^{\nu}_{13} }{\sin \theta^{\nu}_{13} } \right] \,,\\
\eta & = \sin \theta^{\nu}_{12} \left[ \cos^2 \theta_{13} \sin \theta^{\nu}_{12} - \cos \theta^{\nu}_{12} \left( \sin \theta^{\nu}_{13} - \frac{\sin^2 \theta_{13}}{\sin \theta^{\nu}_{13}} \right)\right] \,.
\end{align}
%%%%%%%%%%%%%%%%%%%
%
Using eq. (\ref{s2th12rhoeta}) for $\sin^2 \theta_{12}$
with $\rho$ and $\eta$ as given above,
one can express $\sin^2 \theta^e_{13}$
in terms of $\theta_{12}$, $\theta_{13}$, $\theta^{\nu}_{12}$, $\theta^{\nu}_{13}$:
%%%%%%%%%%%%%%%%%%%%%%%%%%%%
\be
\sin^2 \theta^e_{13} = \frac{2 \cos^2 \theta_{13}\sin \theta^{\nu}_{13} (\sin^2 \theta_{12} - \sin^2 \theta^{\nu}_{12}) - \sin 2 \theta^{\nu}_{12} \sin^2 \theta_{13} + \sin 2\theta^{\nu}_{12} \sin^2 \theta^{\nu}_{13}}
{\cos 2 \theta^{\nu}_{12} \sin \theta^{\nu}_{13} - \cos \theta^{\nu}_{12} \sin \theta^{\nu}_{12} \cos^2 \theta^{\nu}_{13}} \,.
\label{eq:the13}
\ee
%%%%%%%%%%%%%%%%%%%%%%%%%%%%%
%
In the limit of vanishing $\theta^{\nu}_{13}$ we find 
$\sin^2 \theta^e_{13} = 2 \sin^2 \theta_{13}$,
as obtained in subsection \ref{sec:13e23nu12nu}.

Further, using eq.~(\ref{eq:th12B}), we can write $\cos \omega$
in terms of the standard parametrisation mixing angles and the known 
$\theta^e_{13}$, $\theta^{\nu}_{12}$ and $\theta^{\nu}_{13}$:
%%%%%%%%%%%%%%%%%%%%%%%%%%%%
\begin{align}
\label{eq:omega}
\cos \omega  = &  \bigg [ 2 \cos^2 \theta_{13} (\sin \theta^e_{13})^{-2} (\sin \theta^{\nu}_{12})^{-2} \sin^2 \theta_{12}
- 2 \cos^2 \theta^{\nu}_{13} \cot^2 \theta^e_{13}
- (\cot \theta^{\nu}_{12} - \sin \theta^{\nu}_{13})^2 \bigg ] \nonumber \\
 \times & (\cos \theta^{\nu}_{13})^{-1} \tan \theta^e_{13} \bigg[ 2 \sqrt{2} (\cot \theta^{\nu}_{12} - \sin \theta^{\nu}_{13})\bigg]^{-1} .
\end{align}
%%%%%%%%%%%%%%%%%%%%%%%%%%%%%%
%

 Analogously to the case considered in the preceding subsection, 
from the requirements 
$(0 < \sin^2 \theta^e_{13} < 1) \land (-1 < \cos \omega < 1)$
one can obtain for a given $\theta^{\nu}_{13}$,
each of the symmetry values of $\theta^{\nu}_{12}$ considered 
and $\theta^{\nu}_{23} = -\pi/4$
lower and upper bounds on the value of
$\sin^2 \theta_{12}$.
These bounds will be discussed in subsection~\ref{sec:pred13e}.

Comparing again the imaginary parts of
$U^*_{e1} U^*_{\mu 3} U_{e3} U_{\mu 1}$,
obtained using eq.~(\ref{eq:Uija}) with $(ij) = (13)$ 
and $\theta^{\nu}_{23} = -\pi/4$, and
in the standard parametrisation of $U$,
one gets the following relation between $\sin\omega$ and $\sin\delta$
for arbitrarily fixed $\theta^{\nu}_{12}$ and $\theta^{\nu}_{13}$:
%%%%%%%%%%%%%%%%%%%%%%%%%%%%
\begin{align}
\sin \delta & = -\frac{\sin 2 \theta^e_{13}}{2 \sin 2 \theta_{12} \sin 2 \theta_{13} \cos \theta_{23}} \bigg[ \cos^2 \theta^{\nu}_{13} \sin 2 \theta^{\nu}_{12} - 2 \cos 2 \theta^{\nu}_{12} \sin \theta^{\nu}_{13} \bigg ] \sin \omega \,.
\end{align}
%%%%%%%%%%%%%%%%%%%%%%%%%%%
%
 Exploiting the equality of the expressions for $|U_{\mu 1}|$ 
written in the two parametrisations,
%%%%%%%%%%%%%%%%%%%%%%%%%%%
\be
|U_{\mu 1}| =
|\cos\theta_{23} \sin\theta_{12} + e^{i \delta} \cos\theta_{12} \sin\theta_{13} \sin\theta_{23}|
= \frac{1}{\sqrt{2}} |\cos\theta^{\nu}_{12} \sin\theta^{\nu}_{13} - \sin\theta^{\nu}_{12}|\,,
\label{eq:Umu113e}
\ee
%%%%%%%%%%%%%%%%%%%%%%%%%%%
%
we get the following sum rule for $\cos \delta$:
%%%%%%%%%%%%%%%%%%%%%%%%%%%
\begin{align}
\cos \delta & =
-\frac{1}{\sin2\theta_{12} \sin\theta_{13} |\cos\theta^{\nu}_{13}|
(1 - 2\sin^2\theta_{13} + \sin^2\theta^{\nu}_{13})^{\frac{1}{2}}}
\bigg[\left(1 - 2\sin^2\theta_{13} + \sin^2\theta^{\nu}_{13}\right) \sin^2 \theta_{12}
\nonumber \\
& + \cos^2\theta_{12}\sin^2\theta_{13}\cos^2\theta^{\nu}_{13}
- \cos^2\theta_{13} \left(\sin\theta^{\nu}_{12} - \cos\theta^{\nu}_{12} \sin\theta^{\nu}_{13}\right)^2\bigg]\,.
\label{eq:cosdelta13e23nu13nu12nu}
\end{align}
%%%%%%%%%%%%%%%%%%%%%%%%%%%%%%%%%%%%%%%%%%%%
%
For $\theta^{\nu}_{13} = 0$ this sum rule reduces to the sum rule 
for $\cos\delta$ given in eq.~(\ref{eq:cosdelta13e}).

 In the parametrisation of the PMNS matrix considered 
in this subsection, the $J_{\rm CP}$ factor reads:
%%%%%%%%%%%%%%%%%%%%%%%%%%%%%%%%%%%%%%%%
\be
J_{\rm CP} = 
-\,\frac{1}{8 \sqrt{2}} \sin \omega \cos \theta^{\nu}_{13}  \sin 2\theta^e_{13}
\left[\cos^2 \theta^{\nu}_{13} \sin 2\theta^{\nu}_{12} 
-\, 2 \sin \theta^{\nu}_{13} \cos 2\theta^{\nu}_{12}\right] \,.
\ee
%%%%%%%%%%%%%%%%%%%%%%%%%%%%%%%%
%
 
In the case of an arbitrary fixed value of $\theta^{\nu}_{23}$,
as it is not difficult to show, we have:
%%%%%%%%%%%%%%%%%%%%%%%%%%%%%%%%
\begin{align}
\sin^2 \theta_{23} & = \frac{|U_{\mu3}|^2}{1-|U_{e3}|^2} = \frac{\sin^2 \theta^{\nu}_{23} \cos^2 \theta^{\nu}_{13}}{1 - \sin^2 \theta_{13}} \,, 
\label{eq:th23Bgen}
\end{align}
%%%%%%%%%%%%%%%%%%%%%%%%%%%%%%%%%% 
%
and
%%%%%%%%%%%%%%%%%%%%%%%%%%%
\be
|U_{\mu 1}| =
|\cos\theta_{23} \sin\theta_{12} + e^{i \delta} \cos\theta_{12} \sin\theta_{13} \sin\theta_{23}|
=  |\cos\theta^{\nu}_{12} \sin\theta^{\nu}_{13} \sin \theta^{\nu}_{23} + \sin\theta^{\nu}_{12} \cos \theta^{\nu}_{23}|\,.
\label{eq:Umu113egen}
\ee

%%%%%%%%%%%%%%%%%%%%%%%%%%%
%
Using eqs.~(\ref{eq:th23Bgen}) and (\ref{eq:Umu113egen}), 
we obtain in this case
%%%%%%%%%%%%%%%%%%%%%%%%%%%
\begin{align}
\cos \delta & = -\frac{1}{\sin 2 \theta_{12} \sin \theta_{13} | \cos \theta^{\nu}_{13} \sin \theta^{\nu}_{23}| (\cos^2 \theta_{13} - \cos^2 \theta^{\nu}_{13} \sin^2 \theta^{\nu}_{23})^{\frac{1}{2}}}
 \nonumber \\
& \times \bigg[ (\cos^2 \theta_{13} - \cos^2 \theta^{\nu}_{13} \sin^2 \theta^{\nu}_{23}) \sin^2 \theta_{12} + \cos^2 \theta_{12} \sin^2 \theta_{13} \cos^2 \theta^{\nu}_{13} \sin^2 \theta^{\nu}_{23} 
\nonumber \\
& - \cos^2 \theta_{13} ( \cos \theta^{\nu}_{12} \sin \theta^{\nu}_{13} \sin \theta^{\nu}_{23} + \sin \theta^{\nu}_{12} \cos \theta^{\nu}_{23})^2  \bigg] \,.
\label{eq:cosdelta13e23nu13nu12nugen}
\end{align}
%%%%%%%%%%%%%%%%%%%%%%%%%%%
%

The sum rules derived in 
Sections \ref{sec:ije23nu12nu}~--~\ref{sec:ije23nu13nu12nu}
and corresponding to arbitrary fixed values of the angles 
contained in the matrix $\tilde U_{\nu}$,
eqs.~(\ref{cosdthnu}), (\ref{eq:cosdelta12e23nu12nugen}), 
(\ref{eq:cosdelta13e23nu12nugen}), 
(\ref{eq:cosdelta13e23e}), (\ref{eq:cosdelta12e13e:omegagen}),
(\ref{eq:cosdelta12e23nu13nu12nugen}) and (\ref{eq:cosdelta13e23nu13nu12nugen}),
are summarised in Table~\ref{tab:summarysumrules}.
In Table~\ref{tab:summarysin2th23} we give
the corresponding formulae for $\sin^2 \theta_{23}$.

\begin{landscape}
\pagestyle{empty}
%%%%%%%%%%%%%%%%%%%%%%%%%%
\begin{table}[h]
%\centering
\renewcommand*{\arraystretch}{1.2}
\hspace{-0.7cm}
\begin{tabular}{|c|c|}
\hline
& \\ [-10pt]
Parametrisation of $U$ & $\cos \delta$ \\ [8pt]
\hline 
& \\ [-10pt]
$R_{12}(\theta^e_{12}) \, \Psi \, R_{23}(\theta^{\nu}_{23}) \, R_{12}(\theta^{\nu}_{12}) \, Q_0 $ & 
$\phantom{-}\dfrac{(\cos 2 \theta_{13} - \cos 2 \theta^{\nu}_{23})^{\frac{1}{2}}}{\sqrt{2}\sin2\theta_{12}\sin\theta_{13}|\cos \theta^{\nu}_{23}|}\,
\bigg[ \cos2\theta^{\nu}_{12} + \left (\sin^2\theta_{12} - \cos^2\theta^{\nu}_{12} \right )\,
\dfrac{2\sin^2 \theta^{\nu}_{23}-(3+\cos 2 \theta^{\nu}_{23}) \sin^2\theta_{13}}{\cos 2 \theta_{13} - \cos 2 \theta^{\nu}_{23}} \bigg]$ \\ [15pt]
\hline
& \\ [-10pt]
$R_{13}(\theta^e_{13}) \, \Psi \, R_{23}(\theta^{\nu}_{23}) \, R_{12}(\theta^{\nu}_{12}) \, Q_0 $ & 
$-\dfrac{(\cos 2 \theta_{13} + \cos 2 \theta^{\nu}_{23})^{\frac{1}{2}}}{\sqrt{2}\sin2\theta_{12}\sin\theta_{13}|\sin\theta^{\nu}_{23}|}\,
\bigg[ \cos2\theta^{\nu}_{12} + \left (\sin^2\theta_{12} - \cos^2\theta^{\nu}_{12} \right )\,
\dfrac{2\cos^2 \theta^{\nu}_{23}-(3-\cos 2 \theta^{\nu}_{23}) \sin^2\theta_{13}}{\cos 2 \theta_{13} + \cos 2 \theta^{\nu}_{23}} \bigg]$ \\ [15pt]
\hline
& \\ [-10pt]
$R_{12}(\theta^e_{12}) \,  R_{23}(\theta^e_{23})  \, \Psi \, 
R_{23}(\theta^{\nu}_{23}) \, R_{12}(\theta^{\nu}_{12}) \, Q_0$ & 
$\phantom{-}\dfrac{\tan\theta_{23}}{\sin2\theta_{12}\sin\theta_{13}}\,
\left [\cos2\theta^{\nu}_{12} + 
\left (\sin^2\theta_{12} - \cos^2\theta^{\nu}_{12} \right )\,
 \left (1 - \cot^2\theta_{23}\,\sin^2\theta_{13}\right )\right ]$ \\ [15pt]
\hline
& \\ [-10pt]
$R_{13}(\theta^e_{13}) \,  R_{23}(\theta^e_{23})  \, \Psi \, 
R_{23}(\theta^{\nu}_{23}) \, R_{12}(\theta^{\nu}_{12}) \, Q_0$ & 
$ -\dfrac{\cot\theta_{23}}{\sin2\theta_{12}\sin\theta_{13}}\,
\left [\cos2\theta^{\nu}_{12} +
\left (\sin^2\theta_{12} - \cos^2\theta^{\nu}_{12} \right )\,
 \left (1 - \tan^2\theta_{23}\,\sin^2\theta_{13}\right )\right ]$ \\ [15pt]
 \hline
 & \\ [-10pt]
 $R_{12}(\theta^e_{12}) \,  R_{13}(\theta^e_{13})  \, \Psi \, 
R_{23}(\theta^{\nu}_{23}) \, R_{12}(\theta^{\nu}_{12}) \, Q_0$ & 
$\dfrac{1}{\sin2 \theta_{12} \sin 2 \theta_{23} \sin \theta_{13}} \bigg [ 
\dfrac{2 \kappa \cos \omega \sin 2 \theta^{\nu}_{12} \sin \theta^{\nu}_{23} \cos \theta_{13} \cos \theta_{23}}{\cos^2 \theta^{\nu}_{23}} 
(\cos^2 \theta^{\nu}_{23} - \cos^2 \theta_{13} \cos^2 \theta_{23})^{\frac{1}{2}}$ \\
& $ - \cos 2 \theta^{\nu}_{12} \Big(1 - \dfrac{\cos^2 \theta_{13} \cos^2 \theta_{23}}{\cos^2 \theta^{\nu}_{23} }(\sin^2 \theta^{\nu}_{23} + 1) \Big) +\cos 2 \theta_{12} \big(\cos^2 \theta_{23} \sin^2 \theta_{13} - \sin^2 \theta_{23} \big) \bigg]$ 
%
%
%
%$\frac{1}{2} \csc \theta_{12} \csc \theta_{13} \csc \theta_{23} \sec \theta_{12} \sec \theta_{23} \bigg \{
%\sin^2 \theta_{12} \sin^2 \theta_{23} + \cos^2 \theta_{23} \big( \cos^2 \theta_{12} \sin^2 \theta_{13} 
%- \cos^2 \theta_{13} \sin^2 \theta^{\nu}_{12} \tan^2 \theta^{\nu}_{23} \big)$ \\
%& $ + \cos^2 \theta^{\nu}_{12} \big[ \big( - 1 + \cos^2 \theta_{13} \cos^2 \theta_{23} \big) \sec^2 \theta^{\nu}_{23} + \tan^2 \theta^{\nu}_{23} \big]$ \\
%& $ + \kappa \cos \omega \sin 2 \theta^{\nu}_{12} \sin \theta^{\nu}_{23} \sqrt{\cos^2 \theta_{13} \cos^2 \theta_{23} \sec^2 \theta^{\nu}_{23} 
%\big[ \big( 1- \cos^2 \theta_{13} \cos^2 \theta_{23} \big) \sec^2 \theta^{\nu}_{23} - \tan^2 \theta^{\nu}_{23} \big]} \bigg \} $ 
\\ [15pt]
\hline
 & \\ [-10pt]
$R_{12}(\theta^e_{12}) \, \Psi \, R_{23}(\theta^{\nu}_{23})\,
R_{13}(\theta^{\nu}_{13}) \, R_{12}(\theta^{\nu}_{12}) \, Q_0$ & 
 $\phantom{-}\dfrac{1}{\sin 2 \theta_{12} \sin \theta_{13} | \cos \theta^{\nu}_{13} \cos \theta^{\nu}_{23}| (\cos^2 \theta_{13} - \cos^2 \theta^{\nu}_{13} \cos^2 \theta^{\nu}_{23})^{\frac{1}{2}}} \bigg[ (\cos^2 \theta_{13} - \cos^2 \theta^{\nu}_{13} \cos^2 \theta^{\nu}_{23}) \sin^2 \theta_{12}$ \\
 &  $+ \cos^2 \theta_{12} \sin^2 \theta_{13} \cos^2 \theta^{\nu}_{13} \cos^2 \theta^{\nu}_{23} 
 - \cos^2 \theta_{13} ( \cos \theta^{\nu}_{12} \sin \theta^{\nu}_{13} \cos \theta^{\nu}_{23} - \sin \theta^{\nu}_{12} \sin \theta^{\nu}_{23})^2  \bigg] $
%
%$\dfrac{\csc \theta_{12} \sec \theta_{12} \cot \theta_{13}}
%{2 \sqrt{\cos^2 \theta^{\nu}_{13} \cos^2 \theta^{\nu}_{23} - \cos^4 \theta^{\nu}_{13} \cos^4 \theta^{\nu}_{23} \sec^2 \theta_{13}}}
% \bigg\{  (1 - \cos^2 \theta^{\nu}_{13} \cos^2 \theta^{\nu}_{23} \sec^2 \theta_{13}) \sin^2 \theta_{12}$ \\
%& $ + \cos^2 \theta_{12} \cos^2 \theta^{\nu}_{13} \cos^2 \theta^{\nu}_{23} \tan^2 \theta_{13}
%- (\cos \theta^{\nu}_{12} \cos \theta^{\nu}_{23} \sin \theta^{\nu}_{13} - \sin \theta^{\nu}_{12} \sin \theta^{\nu}_{23})^2 \bigg \}$ 
\\ [15pt]
\hline
 & \\ [-10pt]
$R_{13}(\theta^e_{13}) \, \Psi \, R_{23}(\theta^{\nu}_{23})\,
R_{13}(\theta^{\nu}_{13}) \, R_{12}(\theta^{\nu}_{12}) \, Q_0$ & 
$-\dfrac{1}{\sin 2 \theta_{12} \sin \theta_{13} | \cos \theta^{\nu}_{13} \sin \theta^{\nu}_{23}| (\cos^2 \theta_{13} - \cos^2 \theta^{\nu}_{13} \sin^2 \theta^{\nu}_{23})^{\frac{1}{2}}}
\bigg[ (\cos^2 \theta_{13} - \cos^2 \theta^{\nu}_{13} \sin^2 \theta^{\nu}_{23}) \sin^2 \theta_{12}$ \\
&  $+ \cos^2 \theta_{12} \sin^2 \theta_{13} \cos^2 \theta^{\nu}_{13} \sin^2 \theta^{\nu}_{23}  - \cos^2 \theta_{13} ( \cos \theta^{\nu}_{12} \sin \theta^{\nu}_{13} \sin \theta^{\nu}_{23} + \sin \theta^{\nu}_{12} \cos \theta^{\nu}_{23})^2  \bigg]$
%$\dfrac{1}{2 \sqrt{\cos^2 \theta^{\nu}_{13} \sin^2 \theta^{\nu}_{23} \sec^2 \theta_{13} - \cos^4 \theta^{\nu}_{13} \sin^4 \theta^{\nu}_{23} \sec^4 \theta_{13}}} 
% \bigg \{ \csc \theta_{13} \big[ \tan \theta_{12}(-1 + \cos^2 \theta^{\nu}_{13} \sin^2 \theta^{\nu}_{23} \sec^2 \theta_{13})$\\
% & $+ \csc_{12} \sec_{12} \big( \cos \theta^{\nu}_{23} \sin \theta^{\nu}_{12} 
%+ \cos \theta^{\nu}_{12} \sin \theta^{\nu}_{13} \sin \theta^{\nu}_{23} \big)^2 \big] 
%- \tan \theta_{13} \cos^2 \theta^{\nu}_{13} \sin^2 \theta^{\nu}_{23} \cot \theta_{12} \sec \theta_{13} \bigg \}$ 
\\ [15pt]
\hline
\end{tabular}
\caption{Summary of the sum rules for $\cos\delta$.
The parameter $\kappa$ is defined in Section \ref{sec:12e13e23nu12nu}
after eq.~(\ref{eq:cosdelta12e13e}). 
The sum rule corresponding to the 
parametrisation of $U$, $R_{12}(\theta^e_{12})R_{23}(\theta^e_{23})\Psi
R_{23}(\theta^{\nu}_{23})R_{12}(\theta^{\nu}_{12})Q_0$, 
is the one quoted in eq. (\ref{cosdthnu}) and was derived in 
\cite{Petcov:2014laa}.
}
\label{tab:summarysumrules}
\end{table}
%%%%%%%%%%%%%%%%%%%%%%%%%%
\end{landscape}
%\pagestyle{plain}

%%%%%%%%%%%%%%%%%%%%%%%%%%
\begin{table}[h]
\centering
\renewcommand*{\arraystretch}{1.2}
%\hspace{-0.2cm}
\begin{tabular}{|c|c|}
\hline
& \\ [-10pt]
Parametrisation of $U$ & $\sin^2 \theta_{23} $ \\ [8pt]
\hline 
& \\ [-10pt]
$R_{12}(\theta^e_{12}) \, \Psi \, R_{23}(\theta^{\nu}_{23}) \, R_{12}(\theta^{\nu}_{12}) \, Q_0 $ & 
$\dfrac{\sin^2 \theta^{\nu}_{23}-\sin^2 \theta_{13}}{1 - \sin^2 \theta_{13}}$  \\ [15pt]
\hline
& \\ [-10pt]
$R_{13}(\theta^e_{13}) \, \Psi \, R_{23}(\theta^{\nu}_{23}) \, R_{12}(\theta^{\nu}_{12}) \, Q_0 $ & 
$\dfrac{\sin^2 \theta^{\nu}_{23}}{1 - \sin^2 \theta_{13}}$  \\ [15pt]
\hline
& \\ [-10pt]
$R_{12}(\theta^e_{12}) \,  R_{23}(\theta^e_{23})  \, \Psi \, 
R_{23}(\theta^{\nu}_{23}) \, R_{12}(\theta^{\nu}_{12}) \, Q_0$ & $\dfrac{\sin^2 \hat \theta_{23} - \sin^2 \theta_{13}}{1 - \sin^2 \theta_{13}}$  \\ [15pt]
\hline
& \\ [-10pt]
$R_{13}(\theta^e_{13}) \,  R_{23}(\theta^e_{23})  \, \Psi \, 
R_{23}(\theta^{\nu}_{23}) \, R_{12}(\theta^{\nu}_{12}) \, Q_0$ & 
$\dfrac{\sin^2 \hat \theta_{23}}{1 - \sin^2 \theta_{13}}$  \\ [15pt]
 \hline
 & \\ [-10pt]
 $R_{12}(\theta^e_{12}) \,  R_{13}(\theta^e_{13})  \, \Psi \, 
R_{23}(\theta^{\nu}_{23}) \, R_{12}(\theta^{\nu}_{12}) \, Q_0$ & 
$\dfrac{\sin^2 \theta^{\nu}_{23} - \sin^2 \theta_{13} + \sin^2 \theta^e_{13} \cos^2 \theta^{\nu}_{23}}{1 - \sin^2 \theta_{13}}$  \\ [15pt]
\hline
 & \\ [-10pt]
$R_{12}(\theta^e_{12}) \, \Psi \, R_{23}(\theta^{\nu}_{23})\,
R_{13}(\theta^{\nu}_{13}) \, R_{12}(\theta^{\nu}_{12}) \, Q_0$ & 
$ 1 - \dfrac{\cos^2 \theta^{\nu}_{23} \cos^2 \theta^{\nu}_{13}}{1 - \sin^2 \theta_{13}} $  \\ [15pt]
\hline
 & \\ [-10pt]
$R_{13}(\theta^e_{13}) \, \Psi \, R_{23}(\theta^{\nu}_{23})\,
R_{13}(\theta^{\nu}_{13}) \, R_{12}(\theta^{\nu}_{12}) \, Q_0$ & 
$\dfrac{\sin^2 \theta^{\nu}_{23} \cos^2 \theta^{\nu}_{13}}{1 - \sin^2 \theta_{13}}$ \\ [15pt]
\hline
\end{tabular}
\caption{Summary of the formulae for $\sin^2 \theta_{23}$. The formula for $\sin^2 \hat \theta_{23}$
is given in eq.~(\ref{eq:hattheta23gen}).}
\label{tab:summarysin2th23}
\end{table}
%%%%%%%%%%%%%%%%%%%%%%%%%%

%%%%%%%%%%%%%%%%%%%%%%
%
\section{Predictions}
\label{sec:predictions}
%
%%%%%%%%%%%%%%%%%%%%%%%%%%%%%
%

In this Section we present results 
of a statistical analysis, performed using the procedure
described in Appendix~\ref{app:statdetails} (see also \cite{Girardi:2014faa}), 
which allows us to get the dependence of
the $\chi^2$ function on the value of $\delta$ and on the value 
of the  $J_{\rm CP}$ factor. 
In what follows we always assume
that $\theta^{\nu}_{23} = -\pi/4$.
We find that in 
the case corresponding to eq.~(\ref{eq:Uij}) with $(ij) = (12)$, 
analysed in \cite{Petcov:2014laa}, 
the results for $\chi^2$ as a function of $\delta$ or $J_{\rm CP}$ 
are rather similar to those obtained in \cite{Girardi:2014faa} 
in the case of the parametrisation defined by 
eq.~(\ref{eq:Uijkl}) with $(ij) - (kl) = (12) - (23)$. The main difference
between these two cases is the predictions for $\sin^2 \theta_{23}$, which 
can deviate only by approximately $0.5 \sin^2 \theta_{13}$
from $0.5$ in the first case and by a significantly larger amount in the 
second. As a consequence, the predictions in the first case
are somewhat less favoured by the current
data than in the second case, which is reflected
in the higher value of $\chi^2$ 
at the minimum, $\chi^2_{\rm min}$. 
Similar conclusions hold comparing the results in the case 
of $\theta^e_{13} - (\theta^\nu_{23},\theta^\nu_{12})$ 
rotations, described in Section \ref{sec:13e23nu12nu}, 
and in the corresponding case defined by  
eq.~(\ref{eq:Uijkl}) with $(ij) - (kl) = (13) - (23)$ and
discussed in Section \ref{sec:13e23e23nu12nu}.
Therefore, in what concerns these four schemes, 
 in what follows we will present results of the statistical 
analysis of the predictions for $\delta$ and the $J_{\rm CP}$ factor 
only for the scheme with 
 $(\theta^e_{13},\theta^e_{23}) - (\theta^\nu_{23},\theta^\nu_{12})$ rotations, 
considered in Section \ref{sec:13e23e23nu12nu}.

  We show in Tables~\ref{tab:12} and \ref{tab:1} the predictions 
for $\cos\delta$ and $\delta$ for all the schemes considered 
in the present study using the current best fit values 
of the neutrino mixing parameters 
$\sin^2\theta_{12}$, $\sin^2\theta_{23}$ and $\sin^2\theta_{13}$, 
quoted in eqs.~(\ref{th12values})~--~(\ref{th13values}),
which enter into the sum rule expressions for $\cos \delta$, 
eqs.~(\ref{cosdthnu}), (\ref{eq:cosdelta13e}), (\ref{eq:cosdelta13e23e}),
(\ref{eq:cosdelta12e13e}), (\ref{eq:cosdelta12e23nu13nu12nu}), 
(\ref{eq:cosdelta13e23nu13nu12nu}) and eq.~(50) in ref.~\cite{Petcov:2014laa},
unless other values of the indicated mixing parameters
are explicitly specified.
We present results  only for the NO neutrino mass spectrum, 
since the results for the IO spectrum differ insignificantly. 
Several comments are in order.
%%%%%%%%%%%%%%%%%%%%%%%%%%
\begin{table}[h!]
\centering
\renewcommand*{\arraystretch}{1.3}
\begin{tabular}{|c|c|c|c|c|c|}
\hline
 Scheme  & TBM & GRA & GRB & HG &  BM (LC) \\
\hline
  $\theta^e_{12} - (\theta^\nu_{23}, \theta^\nu_{12})$ &$-0.114$ & $\phantom{-}0.289$ & $-0.200$ & $\phantom{-}0.476$ & | \\
 $\theta^e_{13} - (\theta^\nu_{23}, \theta^\nu_{12})$ & $\phantom{-}0.114$ & $-0.289$ & $\phantom{-}0.200$ & $-0.476$ & | \\
\hline
 $(\theta^e_{12},\theta^e_{23}) - (\theta^\nu_{23}, \theta^\nu_{12})$ & $-0.091$ & $\phantom{-}0.275$ & $-0.169$ & $\phantom{-}0.445$ & | \\
 $(\theta^e_{13},\theta^e_{23}) - (\theta^\nu_{23}, \theta^\nu_{12})$ & $\phantom{-}0.151$ & $-0.315$ & $\phantom{-}0.251$ & $-0.531$ & | \\
 $(\theta^e_{12},\theta^e_{13}) - (\theta^\nu_{23}, \theta^\nu_{12})$ & $-\,0.122$ & $\phantom{-}0.282$ & $-0.208$ & $\phantom{-}0.469$ & | \\
\hline
\hline
 Scheme  &  $[\pi/20,-\pi/4]$ &  $[\pi/10,-\pi/4]$ & $[a,-\pi/4]$ & $[\pi/20,b]$ & $[\pi/20,\pi/6]$ \\
\hline
 $\theta^e_{12} - (\theta^\nu_{23}, \theta^\nu_{13}, \theta^\nu_{12})$ 
& $-\,0.222$  &  $\phantom{-}0.760$ & $\phantom{-}0.911$ & $-0.775$ & $-0.562$ \\
\hline
\hline
 Scheme & $[\pi/20,c]$ & $[\pi/20,\pi/4]$ & $[\pi/10,\pi/4]$ & $[a,\pi/4]$ & $[\pi/20,d]$ \\
\hline
 $\theta^e_{13} - (\theta^\nu_{23}, \theta^\nu_{13}, \theta^\nu_{12})$ 
& $-0.866$ & $\phantom{-}0.222$ & $-0.760$ & $-0.911$ & $-0.791$ \\ 
\hline
  \end{tabular}
%%%%%%%%%%%%%%%%%%%%%%%%%%%
\caption{
The predicted values of $\cos\delta$ 
using the current best fit values 
 of the mixing angles, quoted in eqs.~(\ref{th12values})~--~(\ref{th13values})
 and corresponding to neutrino mass spectrum with NO,
 except for the case 
 $(\theta^e_{12},\theta^e_{13}) - (\theta^\nu_{23}, \theta^\nu_{12})$ 
 with $\omega = 0$ and $\kappa = 1$, in 
which $\sin^2 \theta_{23} = 0.48802$ is used. 
 We have defined $a = \sin^{-1} (1 / 3)$, $b = \sin^{-1} (1 / \sqrt{2 + r})$,
 $c = \sin^{-1} (1 / \sqrt{3})$ and $d = \sin^{-1} (\sqrt{3 - r}/2)$. 
 For the last two schemes we give in square brackets 
  the values of $[\theta^\nu_{13}, \theta^\nu_{12}]$.
 TBM, GRA, GRB, HG and BM (LC) refer, in particular, 
 to the different fixed values 
 of $\theta^{\nu}_{12} = c, b, d, \pi/6$ and $\pi/4$, 
 respectively. 
See text for further details.
}
\label{tab:12}
\end{table}
%%%%%%%%%%%%%%%%%%%%%%%%%%
%
%%%%%%%%%%%%%%%%%%%%%%%%%%
\begin{table}[h!]
\centering
\renewcommand*{\arraystretch}{1.3}
\begin{tabular}{|c|c|c|c|c|c|}
\hline
 Scheme  & TBM & GRA & GRB & HG &  BM (LC)\\
\hline
  $\theta^e_{12} - (\theta^\nu_{23}, \theta^\nu_{12})$ & $\phantom{0}97 \lor 263$ & $\phantom{0}73 \lor 287$ & $102 \lor 258$ & $\phantom{0}62 \lor 298$ & | \\
 $\theta^e_{13} - (\theta^\nu_{23}, \theta^\nu_{12})$  & $\phantom{0}83 \lor 277$ & $107 \lor 253$ & $\phantom{0}78 \lor 282$ & $118 \lor 242$ & | \\
\hline
 $(\theta^e_{12},\theta^e_{23}) - (\theta^\nu_{23}, \theta^\nu_{12})$ & $\phantom{0}95 \lor 265$ & $\phantom{0}74 \lor 286$ & $100 \lor 260$ & $\phantom{0}64 \lor 296$ & | \\
 $(\theta^e_{13},\theta^e_{23}) - (\theta^\nu_{23}, \theta^\nu_{12})$ & $\phantom{0}81 \lor 279$ & $108 \lor 252$ & $\phantom{0}75 \lor 285$ & $122 \lor 238$ & | \\
 $(\theta^e_{12},\theta^e_{13}) - (\theta^\nu_{23}, \theta^\nu_{12})$ & $\phantom{0}97 \lor 263$ & $\phantom{0}74 \lor 286$ & $102 \lor 258$ & $\phantom{0}62 \lor 298$ & | \\
\hline
\hline
 Scheme  &  $[\pi/20,-\pi/4]$ &  $[\pi/10,-\pi/4]$ & $[a,-\pi/4]$ & $[\pi/20,b]$ & $[\pi/20,\pi/6]$ \\
\hline
 $\theta^e_{12} - (\theta^\nu_{23}, \theta^\nu_{13}, \theta^\nu_{12})$ & $103 \lor 257$ & $\phantom{0}41 \lor 319$ & $\phantom{0}24 \lor 336$ & $141 \lor 219$ & $124 \lor 236$ \\
\hline
\hline
 Scheme & $[\pi/20,c]$ & $[\pi/20,\pi/4]$ & $[\pi/10,\pi/4]$ & $[a,\pi/4]$ & $[\pi/20,d]$ \\
\hline
 $\theta^e_{13} - (\theta^\nu_{23}, \theta^\nu_{13}, \theta^\nu_{12})$ & $150 \lor 210$ & $\phantom{0}77 \lor 283$ & $139 \lor 221$ & $156 \lor 204$ & $142 \lor 218$ \\ 
\hline
  \end{tabular}
\caption{ The same as in Table \ref{tab:12}, but for $\delta$ given 
in degrees (see text for further details).
}
\label{tab:1}
\end{table}
%%%%%%%%%%%%%%%%%%%%%%%%%%
%

 We do not present predictions for the BM (LC) symmetry form 
of $\tilde{U}_{\nu}$ in Tables \ref{tab:12} and \ref{tab:1},
because for the current best fit values of 
$\sin^2 \theta_{12}$, $\sin^2 \theta_{23}$, $\sin^2 \theta_{13}$ 
the corresponding sum rules give unphysical values of 
$\cos\delta$ (see, however, refs. \cite{Petcov:2014laa,Girardi:2014faa}).
Using the best fit value of $\sin^2 \theta_{13}$, 
we get physical values of $\cos \delta$ in the BM case for
the following minimal values of $\sin^2 \theta_{12}$:
\begin{align*}
& \mbox{$\cos \delta = -0.993$ ($\delta \cong \pi$) for $\sin^2 \theta_{12} = 0.348$ in the scheme $\theta^e_{12} - (\theta^\nu_{23}, \theta^\nu_{12})$,} \\
& \mbox{$\cos \delta = +0.993$  ($\delta \cong 0$)
for $\sin^2 \theta_{12} = 0.348$ in the scheme $\theta^e_{13} - (\theta^\nu_{23}, \theta^\nu_{12})$,} \\
& \mbox{$\cos \delta = -0.994$  ($\delta \cong \pi$) 
for $\sin^2 \theta_{12} = 0.349$ in the scheme $(\theta^e_{12},\theta^e_{13}) - (\theta^\nu_{23}, \theta^\nu_{12})$,}\\
% \begin{align*}
& \mbox{$\cos \delta = -0.996$  ($\delta \cong \pi$) 
for $\sin^2 \theta_{12} = 0.332$ in the scheme $(\theta^e_{12},\theta^e_{23}) - (\theta^\nu_{23}, \theta^\nu_{12})$,} \\
& \mbox{$\cos \delta = +0.997$ ($\delta \cong 0$) 
for $\sin^2 \theta_{12} = 0.368$ in the scheme $(\theta^e_{13},\theta^e_{23}) - (\theta^\nu_{23}, \theta^\nu_{12})$,}
% \end{align*}
\end{align*}
%%%%%%%%%%%%%%%%%%%%%%%%
%
where in the case of the scheme 
$(\theta^e_{12},\theta^e_{13}) - (\theta^\nu_{23}, \theta^\nu_{12})$
we fixed $\sin^2 \theta_{23} = 0.48802$ (we will comment later on this choice), 
while $\sin^2 \theta_{23}$ was set to its best fit value 
for the last two set-ups.

Results for the scheme 
$(\theta^e_{12},\theta^e_{23}) - (\theta^\nu_{23}, \theta^\nu_{12})$ 
in the cases of the TBM and BM symmetry forms of the matrix 
$\tilde{U}_{\nu}$ were presented first in \cite{Marzocca:2013cr}, 
while results for the same scheme and 
the GRA, GRB and HG symmetry forms 
of $\tilde{U}_{\nu}$, as well as 
for the scheme $\theta^e_{12} - (\theta^\nu_{23}, \theta^\nu_{12})$
for all symmetry forms considered,
were obtained first in \cite{Petcov:2014laa}. 
The predictions for $\cos\delta$ and $\delta$ 
were derived in \cite{Petcov:2014laa} and \cite{Marzocca:2013cr} 
for the best fit values of the relevant 
neutrino mixing parameters found in an earlier 
global analysis performed in \cite{Capozzi:2013v1} 
and differ somewhat (albeit not much) from those 
quoted in Tables \ref{tab:12} and \ref{tab:1}.
The values under discussion given in these tables 
are from \cite{Girardi:2014faa}
and correspond to the best fit values quoted in 
eqs. (\ref{th12values})~--~(\ref{th13values}).

The predictions for $\cos\delta$ of the 
 $\theta^e_{12} - (\theta^\nu_{23}, \theta^\nu_{12})$
and  $\theta^e_{13} - (\theta^\nu_{23}, \theta^\nu_{12})$
schemes for each of the symmetry forms of $\tilde{U}_{\nu}$ 
considered differ only by sign. The 
$\theta^e_{12} - (\theta^\nu_{23}, \theta^\nu_{12})$ 
scheme and the  $(\theta^e_{12},\theta^e_{13}) - (\theta^\nu_{23}, \theta^\nu_{12})$ 
scheme with $\omega = 0$ provide 
very similar  predictions for $\cos\delta$.

 In the schemes with three rotations in $\tilde{U}_{\nu}$ 
we consider, $\cos\delta$ has values 
which differ significantly (being larger in absolute value) from 
the values predicted by the schemes 
with two rotations in  $\tilde{U}_{\nu}$ discussed by us,
 the only exceptions being 
i) the 
 $\theta^e_{12(13)} - (\theta^\nu_{23}, \theta^\nu_{13}, \theta^\nu_{12})$ 
scheme with  $[\theta^\nu_{13},\theta^\nu_{12}] =[\pi/20,^{~-~}_{(+)}\pi/4]$,
for which $|\cos\delta| = 0.222$, 
and ii)  $\theta^e_{12} - (\theta^\nu_{23}, \theta^\nu_{13}, \theta^\nu_{12})$ 
scheme with  $[\theta^\nu_{13},\theta^\nu_{12}] =[\pi/20,\pi/6]$ 
in which $\cos\delta = -\,0.562$. 

The predictions for $\cos\delta$ of the schemes 
denoted as $(\theta^e_{12},\theta^e_{23}) - (\theta^\nu_{23}, \theta^\nu_{12})$ 
and $(\theta^e_{13},\theta^e_{23}) - (\theta^\nu_{23}, \theta^\nu_{12})$
differ for each of the symmetry forms of  $\tilde{U}_{\nu}$
considered both by sign and magnitude. 
If the best fit value of $\theta_{23}$ were $\pi/4$, these 
predictions would differ only by sign.

 In the case of the 
 $(\theta^e_{12},\theta^e_{13}) - (\theta^\nu_{23}, \theta^\nu_{12})$ 
scheme with $\omega = 0$, the predictions for $\cos \delta$ are very sensitive 
to the value of $\sin^2 \theta_{23}$. Using the best fit values of
$\sin^2 \theta_{12}$ and $\sin^2 \theta_{13}$ for the NO 
neutrino mass spectrum, 
quoted in eqs.~(\ref{th12values}) and (\ref{th13values}),
we find from the constraints  $(-1 < \cos \psi < 1)$ 
and $(0 < \sin^2 \theta^e_{13} < 1) \land (0 < \sin^2 \theta^e_{12} < 1)$,
where $\sin^2 \theta^e_{13}$, $\sin^2 \theta^e_{12}$ and $\cos \psi$
are given in eqs.~(\ref{eq4:the13})~--~(\ref{eq4:psi}),
that $\sin^2 \theta_{23}$ should lie in the following intervals:
%%%%%%%%%%%%%%%%%%%%%%%%%%%%%%%%%%%%%%%
\begin{align*}
&(0.488,0.496) \cup (0.847,0.909)  \mbox{ for TBM}; \\
&(0.488,0.519) \cup (0.948,0.971)  \mbox{ for BM}; \\
&(0.488,0.497) \cup (0.807,0.880)  \mbox{ for GRA}; \\
&(0.488,0.498) \cup (0.856,0.914)  \mbox{ for GRB}; \\
&(0.488,0.500) \cup (0.787,0.866)  \mbox{ for HG}.
\end{align*}
%%%%%%%%%%%%%%%%%%%%%%%%%%%%%%%%%%%%%%
%
Obviously, the quoted intervals with 
$\sin^2 \theta_{23} \geq 0.78$ are ruled out by the current data. 
We observe that a small increase of $\sin^2 \theta_{23}$ 
from the value $0.48802$~
\footnote{For $\sin^2 \theta_{23} < 0.48802$, 
 $\cos\delta$ has an unphysical (complex) value.
}
produces a relatively large variation of $\cos \delta$.
The strong dependence of $\cos \delta$ on $\sin^2 \theta_{23}$
takes place for values of $\omega$
satisfying roughly $\cos \omega \gtap 0.01$. 
In contrast, for $\cos \omega = 0$, $\cos \delta$ exhibits a 
relatively weak dependence on $\sin^2 \theta_{23}$.
For the reasons related to the dependence of 
$\cos\delta$ on $\omega$ we are not going to present results 
of the statistical analysis in this case. 
This can be done in specific models of neutrino mixing, 
in which the value of the phase $\omega$ is fixed by the model.

%%%%%%%%%%%%%%%%
%
\subsection{The Scheme with 
$(\theta^e_{13},\theta^e_{23}) - (\theta^\nu_{23}, \theta^\nu_{12})$ Rotations }
\label{sec:pred13e23e}
%
%%%%%%%%%%%%%%%%%%

In the left panel of Fig.~\ref{Fig:cosdeltaNO_13e23e} we show 
the likelihood function, defined as
%%%%%%%%%%%%%%%%%%%%%%%%%%%%%%
\be
L(\cos \delta) \propto \exp \left(- \frac{\chi^2 (\cos \delta)}{2} \right) \,,
\ee
%%%%%%%%%%%%%%%%%%%%%%%%%%%%%%%%%%%%%
%
versus $\cos\delta$ for the NO neutrino mass spectrum for the scheme 
with $(\theta^e_{13},\theta^e_{23}) - (\theta^\nu_{23}, \theta^\nu_{12})$ 
rotations 
\footnote{The corresponding results for the IO neutrino 
mass spectrum differ little from those shown
in the left panel of Fig.~\ref{Fig:cosdeltaNO_13e23e}.
}. 
This function represents the most probable 
values of $\cos\delta$ for each of the symmetry forms
considered. In the analysis performed by us we use as input the
current global neutrino oscillation data 
on $\sin^2 \theta_{12}$, $\sin^2 \theta_{23}$, $\sin^2 \theta_{13}$ 
and $\delta$ \cite{Capozzi:2013csa}.
The maxima of $L(\cos \delta)$, $L(\chi^2 = \chi^2_{\rm min})$, 
for the different symmetry forms of $\tilde{U}_{\nu}$ considered, 
correspond to the values of $\cos\delta$ given in Table \ref{tab:12}.  
The results shown are obtained
by marginalising over $\sin^2 \theta_{13}$ and 
$\sin^2 \theta_{23}$ for a fixed value
of $\delta$ (for details of the statistical analysis
see Appendix~\ref{app:statdetails} and \cite{Girardi:2014faa}).
The $n \sigma$ confidence level (C.L.) region corresponds to the
interval of values of $\cos \delta$ for which
$L(\cos \delta) \geq L(\chi^2 = \chi^2_{\rm min}) \cdot L(\chi^2 = n^2)$.
Here $\chi^2_{\rm min}$ is the value of $\chi^2$
in the minimum.

As can be observed from the left panel of 
Fig.~\ref{Fig:cosdeltaNO_13e23e}, 
for the TBM and GRB forms there is a substantial 
overlap of the corresponding likelihood functions.
The same observation 
holds also for the GRA and HG forms.
However,  the likelihood functions of these
two sets of symmetry forms overlap 
only at $3\sigma$ and 
in a small interval of values of $\cos \delta$.
Thus, the TBM/GRB, GRA/HG and BM (LC) symmetry forms 
might be distinguished with 
a not very demanding (in terms of 
precision) measurement  of $\cos \delta$.
At the maximum, the non-normalised likelihood
function equals $\exp( - \chi^2_{\rm min}/2)$, 
and this value allows one to 
judge quantitatively 
about the compatibility of a given symmetry form
with the global neutrino oscillation data,
as we have pointed out.

In the right panel of Fig.~\ref{Fig:cosdeltaNO_13e23e} we present $L$ 
versus $\cos \delta$ within the Gaussian approximation
(see \cite{Girardi:2014faa} for details), using the current best fit values 
of $\sin^2 \theta_{12}$, $\sin^2 \theta_{23}$, $\sin^2 \theta_{13}$ 
for NO spectrum, given in 
eqs.~(\ref{th12values})~--~(\ref{th13values}), 
 and the prospective 
$1\sigma$ uncertainties 
in the measurement of these mixing parameters.
More specifically, we use as $1\sigma$ 
uncertainties i) 0.7\% for $\sin^2 \theta_{12}$, 
which is the prospective sensitivity 
of the JUNO experiment \cite{Wang:2014iod}, 
ii) 5\% for $\sin^2 \theta_{23}$~ 
\footnote{This sensitivity
is planned to be achieved
in future neutrino facilities \cite{Coloma:2014kca}.}, 
obtained from the prospective uncertainty 
of 2\% \cite{deGouvea:2013onf}
on $\sin^2 2 \theta_{23}$ expected to be reached in 
the NOvA and T2K experiments, and
iii) 3\% for $\sin^2 \theta_{13}$, deduced from the error of 3\% on 
$\sin^2 2 \theta_{13}$ planned to be reached in the  Daya Bay 
experiment \cite{deGouvea:2013onf,Zhang:2015fya}.
The BM (LC) case is  quite sensitive to the 
values of $\sin^2 \theta_{12}$ and $\sin^2 \theta_{23}$ 
and for the current best fit values
is disfavoured at more than $2\sigma$.
%
%%%%%%%%%%%%%%%%%%%%%%%%%%%%%%%%%
\begin{figure}[h!]
  \begin{center}
   \subfigure
 {\includegraphics[width=7cm]{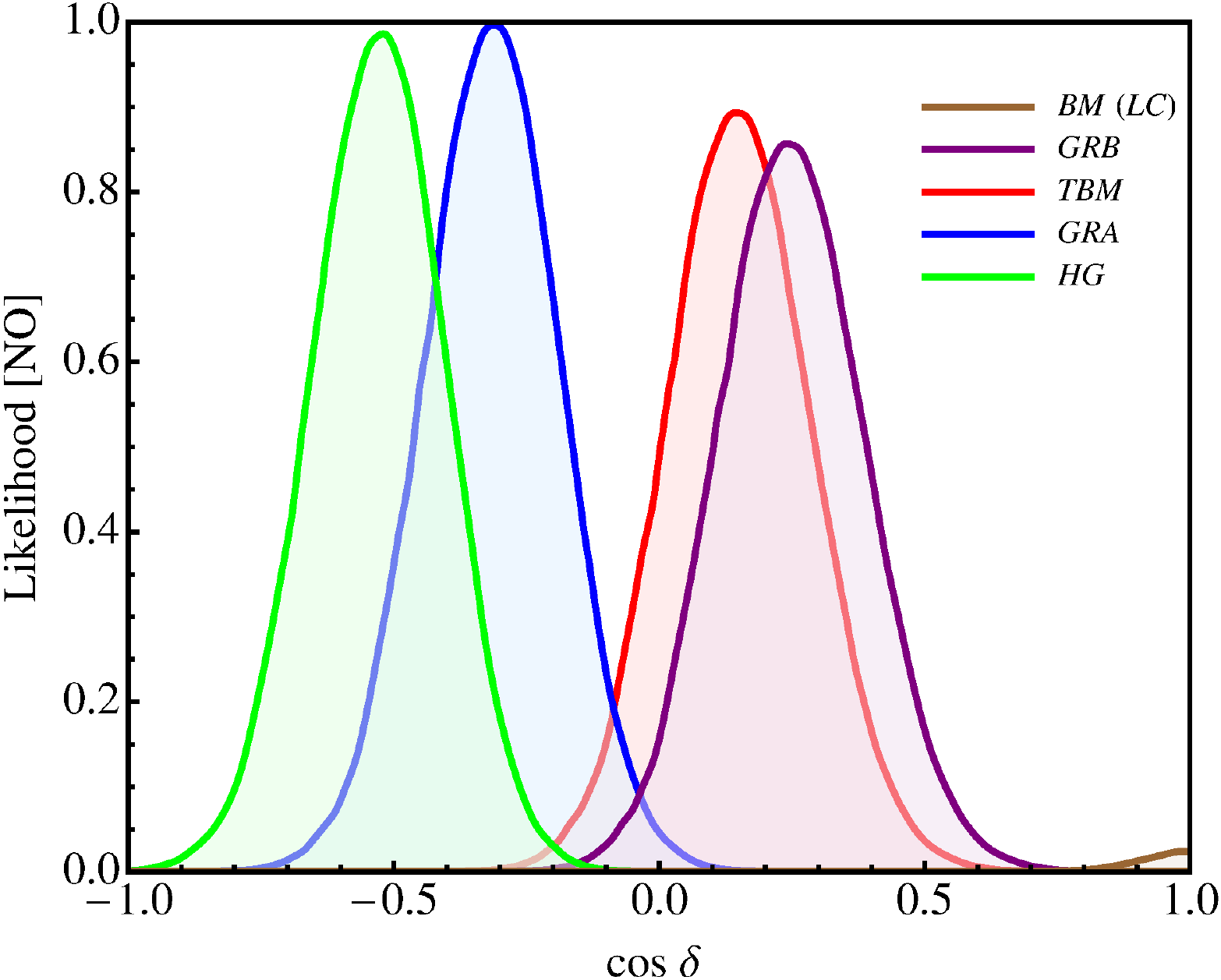}}
  {\includegraphics[width=7cm]{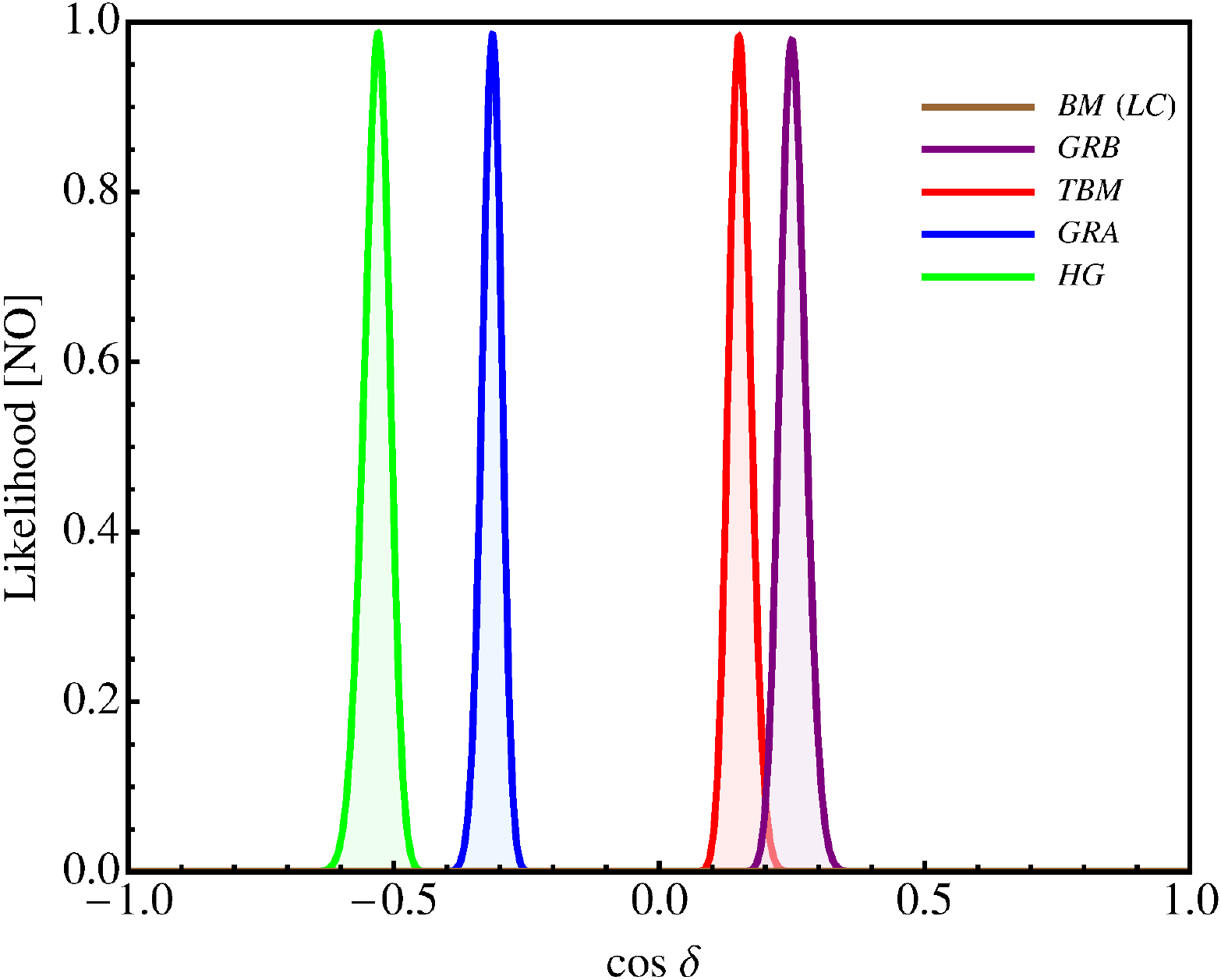}}
   \vspace{5mm}
     \end{center}
\vspace{-1.0cm} \caption{\label{Fig:cosdeltaNO_13e23e}
The likelihood function versus $\cos \delta$ for the NO
neutrino mass spectrum after marginalising over
$\sin^2\theta_{13}$ and $\sin^2 \theta_{23}$
for the TBM, BM (LC), GRA, GRB and HG symmetry forms
of the matrix $\tilde{U}_{\nu}$ 
in the $(\theta^e_{13},\theta^e_{23}) - (\theta^\nu_{23}, \theta^\nu_{12})$ 
set-up.
The results shown are obtained 
using eq.~(\ref{eq:cosdelta13e23e})
and i) the latest results on the mixing parameters
$\sin^2\theta_{12}$, $\sin^2\theta_{13}$, $\sin^2\theta_{23}$
and $\delta$ 
found in the global analysis of the neutrino oscillation data 
\cite{Capozzi:2013csa} 
(left panel), and ii) the prospective $1\sigma$ uncertainties on 
$\sin^2\theta_{12}$, $\sin^2\theta_{13}$, $\sin^2\theta_{23}$
and the Gaussian approximation 
for the likelihood function (right panel) 
(see text for further details).
}
\end{figure}
%%%%%%%%%%%%%%%%%%%%%%%%%
%
That the BM (LC) case is disfavoured 
by the current data can be understood, in particular, 
from the following observation.
Using the best fit values of $\sin^2 \theta_{13}$ 
and $\sin^2 \theta_{12}$ as well as the constraint 
$-1 \leq \cos \alpha \leq 1$, where $\cos \alpha$ is
defined in eq.~(\ref{eq:cosalpha13e23e}),
one finds that $\sin^2 \theta_{23}$ 
should satisfy $\sin^2 \theta_{23} \geq 0.63$, 
which practically coincides with 
the currently allowed maximal value of 
$\sin^2 \theta_{23}$ at $3\sigma$
 (see eq.~(\ref{th23values})).

 It is interesting to compare the results described above 
and obtained in the scheme denoted by  
$(\theta^e_{13},\theta^e_{23}) - (\theta^\nu_{23}, \theta^\nu_{12})$
with those obtained in \cite{Girardi:2014faa} in the  
$(\theta^e_{12},\theta^e_{23}) - (\theta^\nu_{23}, \theta^\nu_{12})$ set-up.
We recall that for each of the symmetry forms 
we have considered~---~TBM, BM, GRA, GRB and 
HG~---~$\theta^\nu_{12}$ has a specific fixed value and                 
$\theta^\nu_{23} = -\pi/4$.
The first thing to note is that for a given 
symmetry form,  $\cos\delta$ is predicted to have opposite 
signs in the two schemes. 
In the scheme
$(\theta^e_{13},\theta^e_{23}) - (\theta^\nu_{23}, \theta^\nu_{12})$  
analysed in the present article, one has $\cos\delta > 0$ in the
TBM, GRB and BM (LC) cases, while $\cos\delta <0$ in the
cases of the GRA and HG symmetry forms. 
As in the $(\theta^e_{12},\theta^e_{23}) - (\theta^\nu_{23}, \theta^\nu_{12})$
set-up, there are significant overlaps between the TBM/GRB
and GRA/HG forms of $\tilde U_{\nu}$, respectively.
The BM (LC) case is disfavoured at more than $2\sigma$ confidence level.
It is also important to notice that due to the
fact that the best fit value of $\sin^2\theta_{23} < 0.5$, 
the predictions for $\cos\delta$ for
each symmetry form, obtained in the two set-ups
differ not only by sign but also in absolute value, 
as was already pointed out in 
Section~\ref{sec:13e23e23nu12nu}. Thus, a
precise measurement of $\cos \delta$ would 
allow one to distinguish not only between the symmetry forms
of $\tilde U_{\nu}$, but also could provide an indication about
the structure of the matrix $\tilde U_e$.

 We note that the predictions for $\sin^2 \theta_{23}$ are rather similar 
in the cases of the two schemes discussed, 
$(\theta^e_{13},\theta^e_{23}) - (\theta^\nu_{23}, \theta^\nu_{12})$
and $(\theta^e_{12},\theta^e_{23}) - (\theta^\nu_{23}, \theta^\nu_{12})$.
We give for completeness 
$N_{\sigma} \equiv \sqrt{\chi^2}$ as a function of $\sin^2 \theta_{23}$ in
Appendix \ref{app:th2313e23e}.

For the rephasing invariant $J_{\rm CP}$, using the current global 
neutrino oscillation data, we find for the symmetry forms considered the 
following best fit values and the $3\sigma$ ranges
for the NO neutrino mass spectrum:
 
%%%%%%%%%%%%%%%%%%%%%%%%%%%%%%%
\begin{align}
&\mbox{$J_{\rm CP}=-\,0.033\,,\phantom{0}$ $-0.039 \leq J_{\rm CP}\leq  -0.026$,~ 
$0.030 \leq J_{\rm CP}\leq 0.036$ for TBM;}\\
&\mbox{$J_{\rm CP}= -\,0.004\,,\phantom{0}$ $-0.026  \leq J_{\rm CP}\leq  0.023$ for BM (LC) ;}\\
&\mbox{$J_{\rm CP}=-\,0.032\,,\phantom{0}$ $-0.037 \leq J_{\rm CP}\leq  -0.024$,~ 
$0.029  \leq J_{\rm CP}\leq 0.035$ for  GRA;}\\
&\mbox{$J_{\rm CP} =-\,0.033\,,\phantom{0}$ $-0.039  \leq J_{\rm CP}\leq  -0.023$,~ 
$0.028  \leq J_{\rm CP}\leq 0.036$ for GRB;}\\
&\mbox{$J_{\rm CP}=-\,0.028\,,\phantom{0}$ $-0.035 \leq J_{\rm CP}\leq -0.014$,~
$0.021  \leq J_{\rm CP}\leq 0.032$ for HG.}
\end{align}
%%%%%%%%%%%%%%%%%%%%%%%%%%%%%%%%
%
Thus, relatively large 
CP-violating effects in neutrino oscillations are predicted 
for all symmetry forms considered, the only 
exception being the case of the BM symmetry form.
\\

%%%%%%%%%%%%%%%%%%%%%%%%
%
\subsection{The Scheme with 
$\theta^e_{12} - (\theta^\nu_{23}, \theta^\nu_{13}, \theta^\nu_{12})$ Rotations }
\label{sec:pred12e}
%
%%%%%%%%%%%%%%%%%%%%%%%%%%%%
%
 For the scheme with
$\theta^e_{12} - (\theta^\nu_{23}, \theta^\nu_{13}, \theta^\nu_{12})$
rotations we find that only for particular values
of $\theta^{\nu}_{12}$ and $\theta^{\nu}_{13}$, 
among those considered by us,
the allowed intervals of values of $\sin^2\theta_{12}$ 
satisfy the requirement that they contain 
in addition to the best fit value of $\sin^2 \theta_{12}$ 
also the $1.5\sigma$ experimentally allowed range of 
$\sin^2 \theta_{12}$.
Indeed, combining the 
conditions $0 < \sin^2 \theta^e_{12} < 1$
and $|\cos \phi| < 1$,
where $\sin^2 \theta^e_{12}$ and $\cos \phi$ are given in 
eqs.~(\ref{eq5_1:the12}) and (\ref{eq:phi}), respectively,
and allowing $\sin^2 \theta_{13}$ to vary in the $3\sigma$ range
for the NO spectrum, we get restrictions on
the value of $\sin^2 \theta_{12}$, presented in Table~\ref{tab:mod1cosphi}.
We see from the Table that only five out of 18 combinations
of the angles $\theta^{\nu}_{12}$ and $\theta^{\nu}_{13}$ 
considered by us
satisfy the requirement formulated above.
In Table~\ref{tab:mod1cosphi} these cases are marked
with the subscripts I, II, III, IV, V,
while the ones marked 
with an asterisk contain values of $\sin^2 \theta_{12}$
allowed at $2\sigma$ \cite{Capozzi:2013csa}.
%%%%%%%%%%%%%%%%%%%%%%%%%%
\begin{table}[h]
\centering
\renewcommand*{\arraystretch}{1.3}
\begin{tabular}{|c|l|l|l|l|}
\hline
$\theta^{\nu}_{12}$ & $\theta^{\nu}_{13} = \pi/20$ & $\theta^{\nu}_{13} = \pi/10$ & $\theta^{\nu}_{13} =  \sin^{-1} (1 / 3)$ \\
\hline
$\sin^{-1} (1 / \sqrt{3})$ & $(0.319,0.654)^*$ & $(0.471,0.773)$ & $(0.495,0.789)$ \\
$\pi/4$ & $(0.484,0.803)$ & $(0.639,0.897)$& $(0.662,0.909)$ \\
$-\pi/4$ & $(0.197,0.516)_{\rm III}$ & $(0.103,0.361)_{\rm I}$& $(0.091,0.338)_{\rm IV}$ \\
$\sin^{-1} (1 / \sqrt{2 + r})$ & $(0.262,0.594)_{\rm II}$ & $(0.409,0.719)$ & $(0.434,0.737)$ \\
$\sin^{-1} (\sqrt{3 - r} / 2)$ & $(0.331,0.666)^*$ & $(0.484,0.784)$ & $(0.508,0.800)$ \\
$\pi/6$ & $(0.236,0.564)_{\rm V}$ & $(0.380,0.692)$ & $(0.404,0.710)$ \\
\hline
\end{tabular}
\caption{Ranges of $\sin^2 \theta_{12}$ obtained from the requirements $(0 < \sin^2 \theta^e_{12} < 1) \land (-1 < \cos \phi < 1)$ 
allowing $\sin^2 \theta_{13}$ to vary in the 3$\sigma$ allowed range
for the NO neutrino mass spectrum, quoted in eq.~(\ref{th13values}).
The cases for which the best fit value of $\sin^2 \theta_{12} = 0.308$ 
is within the corresponding allowed ranges are marked with 
the subscripts I, II, III, IV, V.
The cases marked 
with an asterisk contain values of $\sin^2 \theta_{12}$
allowed at $2\sigma$ \cite{Capozzi:2013csa}.
}
\label{tab:mod1cosphi}
\end{table}
%%%%%%%%%%%%%%%%%%%%%%%%%%

 Equation~(\ref{eq:th23}) implies that
$\sin^2 \theta_{23}$ is fixed by the value
of $\theta^{\nu}_{13}$, and for the best fit value
of $\sin^2 \theta_{13}$ 
 and the values of 
$\theta^{\nu}_{13} = 0$, $\pi/20$, $\pi/10$, $\sin^{-1}(1/3)$,
considered by us, we get, respectively:
$\sin^2 \theta_{23} = 0.488$, $0.501$, $0.537$, $0.545$.
Therefore a 
measurement of $\sin^2 \theta_{23}$ with 
a sufficiently high precision would
rule out at least some of the cases with fixed 
values of $\theta^{\nu}_{13}$ considered in the literature.

  We will perform a statistical analysis of the predictions 
for $\cos\delta$ in the five cases~---~I, II, III, IV, V~---~
listed above.
The analysis is similar to the one 
discussed in Section \ref{sec:pred13e23e}. 
The only difference is that when we
consider the prospective sensitivities on the PMNS mixing angles 
we will assume $\sin^2 \theta_{23}$ to have 
the following potential best fit values: 
$\sin^2 \theta_{23} = 0.488$, $0.501$, $0.537$, $0.545$. 
 Note that for the best fit value of 
$\sin^2\theta_{13}$, $\sin^2 \theta_{23} = 0.488$ 
does not correspond to 
any of the values of $\theta^{\nu}_{13}$  
in the five cases~---~I, II, III, IV, V~---~of interest. 
Thus, $\sin^2 \theta_{23} = 0.488$ is not 
the most probable value in any of the five cases 
considered: depending on the case, 
the most probable value is one of the other three 
values of $\sin^2 \theta_{23}$ listed above.
We include results for  
$\sin^2 \theta_{23} = 0.488$ to illustrate 
how the likelihood function 
changes when the best fit value of 
$\sin^2 \theta_{23}$, determined in a 
global analysis, differs from the 
value of $\sin^2 \theta_{23}$ 
predicted in a given case.

In Fig.~\ref{Fig:cosdeltaNO_12e} we show
the likelihood function versus $\cos \delta$ for
all the cases marked with the subscripts in 
Table~\ref{tab:mod1cosphi}.
The maxima of the likelihood function in the five cases considered 
take place at the corresponding 
values of $\cos\delta$ cited in Table \ref{tab:12}.
As Fig.~\ref{Fig:cosdeltaNO_12e} clearly indicates,
the cases differ not only in the predictions
for $\sin^2 \theta_{23}$, which in the considered set-up
is a function of  $\sin^2 \theta^{\nu}_{13}$ and $\sin^2 \theta_{13}$,
but also in the predictions for $\cos \delta$.
Given the values of $\theta_{12}$ and $\theta_{13}$, 
the positions of the peaks are determined by the values 
of  $\theta^{\nu}_{12}$ and $\theta^{\nu}_{13}$.

The Cases I and IV are disfavoured by the current data because 
the corresponding values of $\sin^2\theta_{23} = 0.537$ and 
0.545 are disfavoured. The Cases II, III and V 
are less favoured for the NO neutrino mass spectrum than for 
the IO spectrum since  $\sin^2\theta_{23} = 0.501$ 
is less favoured for the first than for the second spectrum.

In Fig.~\ref{Fig:cosdeltaNO_12e_fut}
we show the predictions for $\cos \delta$
using the prospective precision in the measurement 
of  $\sin^2\theta_{12}$, $\sin^2\theta_{13}$, 
$\sin^2\theta_{23}$,
the best fit values for $\sin^2 \theta_{12}$ and $\sin^2 \theta_{13}$
as in eqs.~(\ref{th12values}) and (\ref{th13values})
and the potential best fit values of
$\sin^2 \theta_{23} = 0.488$, $0.501$, $0.537$, $0.545$.
The values of $\sin^2 \theta_{23}$
correspond in the scheme discussed 
to the best fit value of $\sin^2 \theta_{13}$ 
in the cases which are compatible with the 
current $1.5\sigma$
range of allowed values of 
$\sin^2\theta_{12}$.
The position of the peaks, obviously,  
does not  depend explicitly on the  assumed 
experimentally determined 
best fit value of $\sin^2 \theta_{23}$.
For the best fit value of 
$\sin^2\theta_{13}$ used, 
the corresponding sum rule for $\cos \delta$ 
depends on the given fixed value of $\theta^{\nu}_{13}$, 
and via it, on the predicted value of $\sin^2\theta_{23}$ 
(see eqs.~(\ref{eq:th23}) and (\ref{eq:cosdelta12e23nu13nu12nu})).
Therefore, the compatibility of a given case 
with the considered hypothetical data 
on $\sin^2\theta_{23}$ clearly depends 
on the assumed best 
fit value of $\sin^2\theta_{23}$ determined from the data.  
%%%%%%%%%%%%%%%%%%%%%%%%%%%%%%%%%
\begin{figure}[h!]
  \begin{center}
   \subfigure
 {\includegraphics[width=7cm]{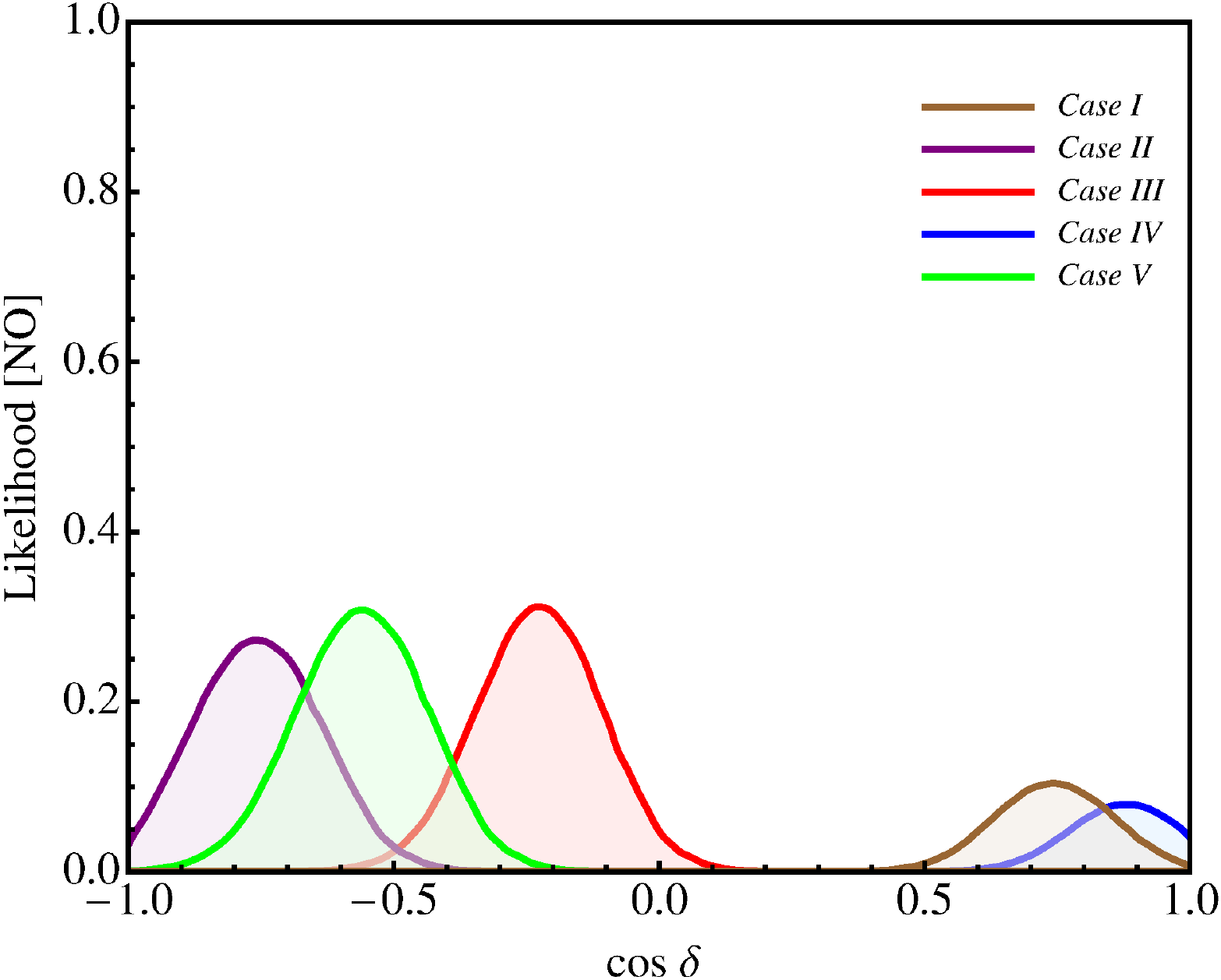}}
  {\includegraphics[width=7cm]{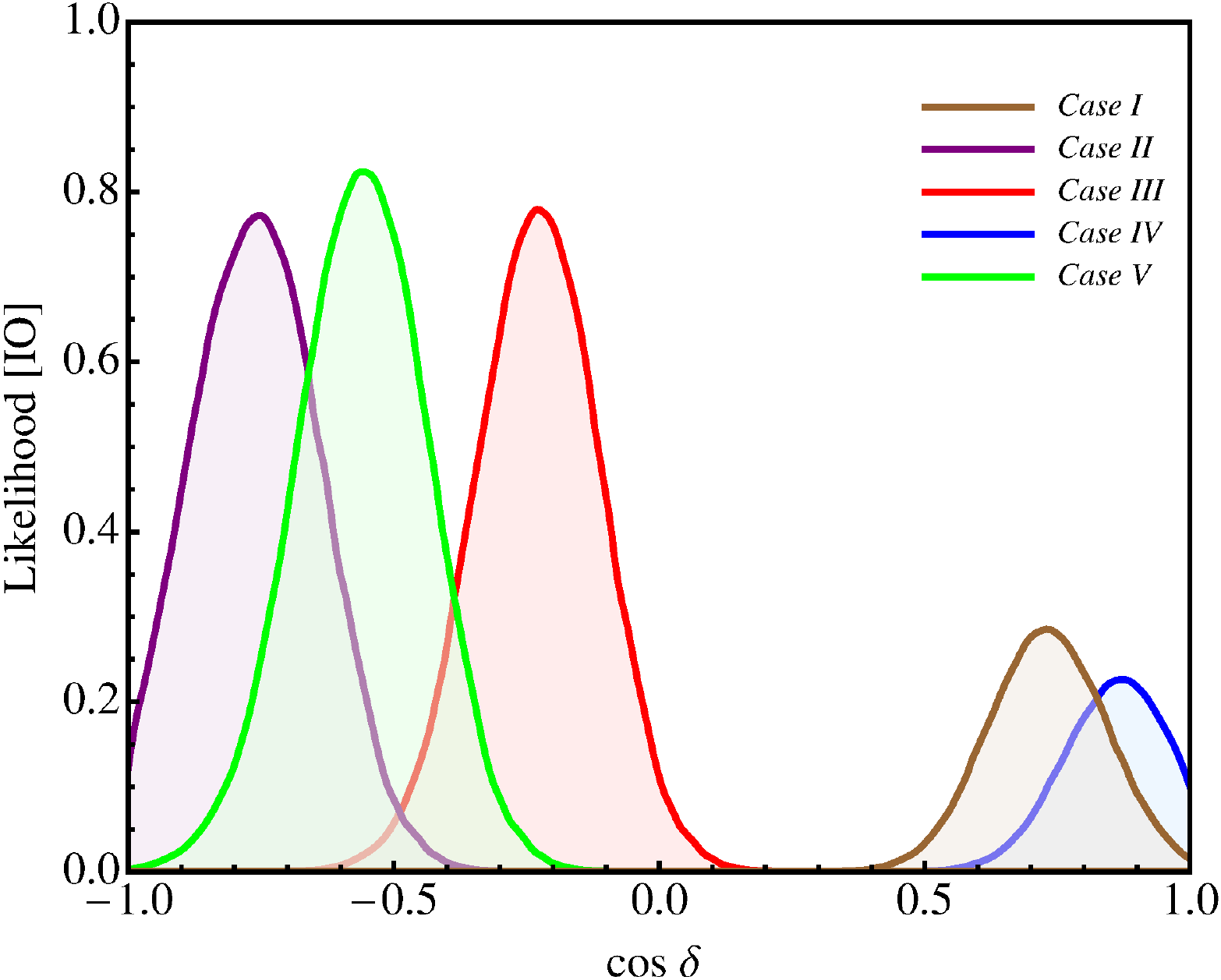}}
   \vspace{5mm}
     \end{center}
\vspace{-1.0cm} \caption{\label{Fig:cosdeltaNO_12e}
The likelihood function 
versus $\cos \delta$ for the NO (IO) neutrino mass spectrum in the
left (right) panel after marginalising over
$\sin^2\theta_{13}$ for the 
scheme $\theta^e_{12} - (\theta^\nu_{23}, \theta^\nu_{13}, \theta^\nu_{12})$
with $[\theta^\nu_{13}, \theta^\nu_{12}]$ fixed as
$[\pi/10,-\pi/4]$ (Case I), $[\pi/20,b]$ (Case II),
$[\pi/20,-\pi/4]$ (Case III), 
$[a,-\pi/4]$ (Case IV), $[\pi/20,\pi/6]$ (Case V),
where 
$a = \sin^{-1} (1 / 3)$ and $b = \sin^{-1} (1 / \sqrt{2 + r})$,
$r$ being the golden ratio.
The figure is obtained using the sum rule 
in eq.~(\ref{eq:cosdelta12e23nu13nu12nu}) and
the latest results on $\sin^2\theta_{12}$, $\sin^2\theta_{13}$, 
$\sin^2\theta_{23}$ and $\delta$ 
from the global analysis of the neutrino oscillation 
data \cite{Capozzi:2013csa}. 
}
\end{figure}
%%%%%%%%%%%%%%%%%%%%%%%%%%%%%%%%%
%

%%%%%%%%%%%%%%%%%%%%%%%%%%%%%%%%%
\begin{figure}[h!]
  \begin{center}
   \subfigure
 {\includegraphics[width=7cm]{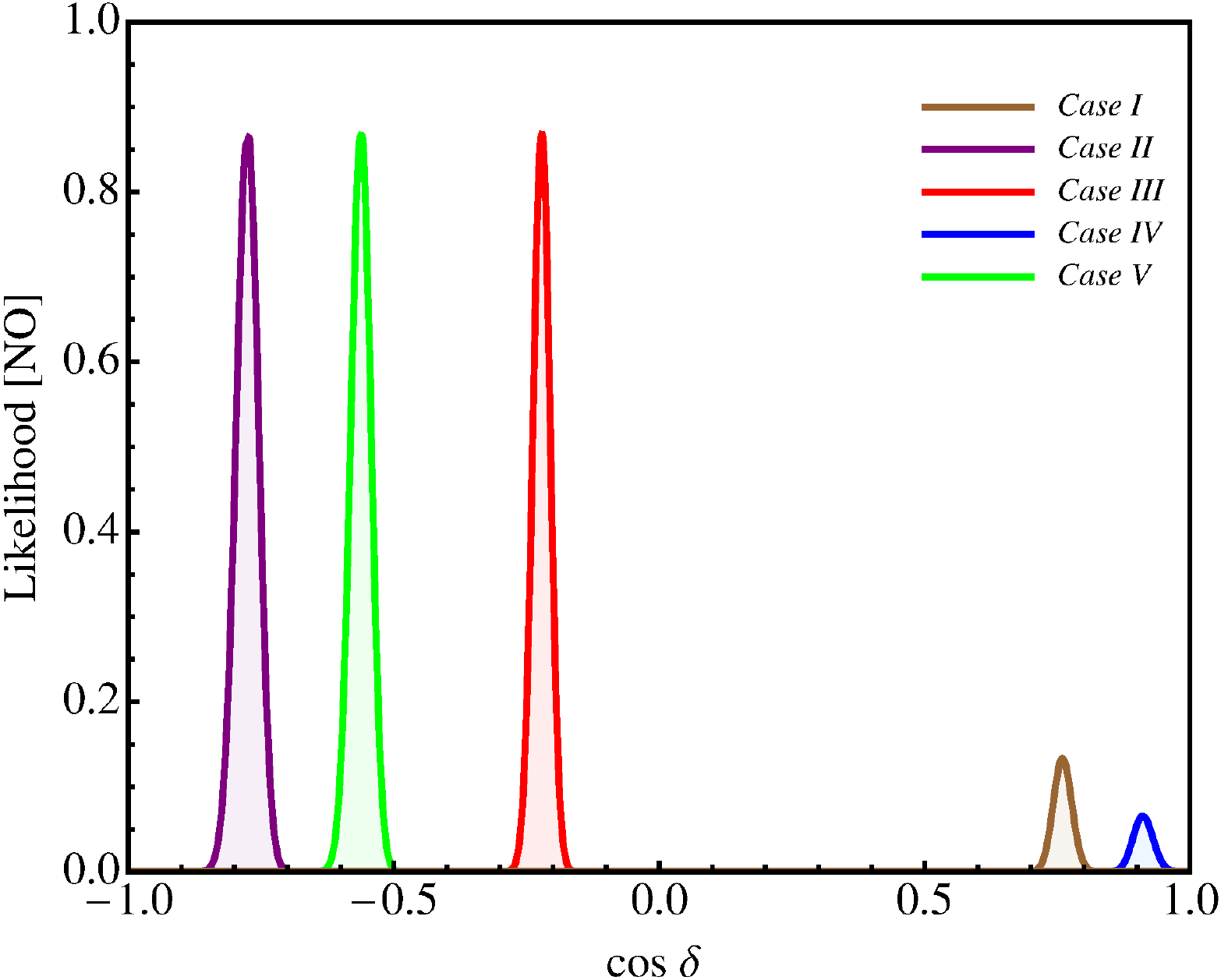}}
  {\includegraphics[width=7cm]{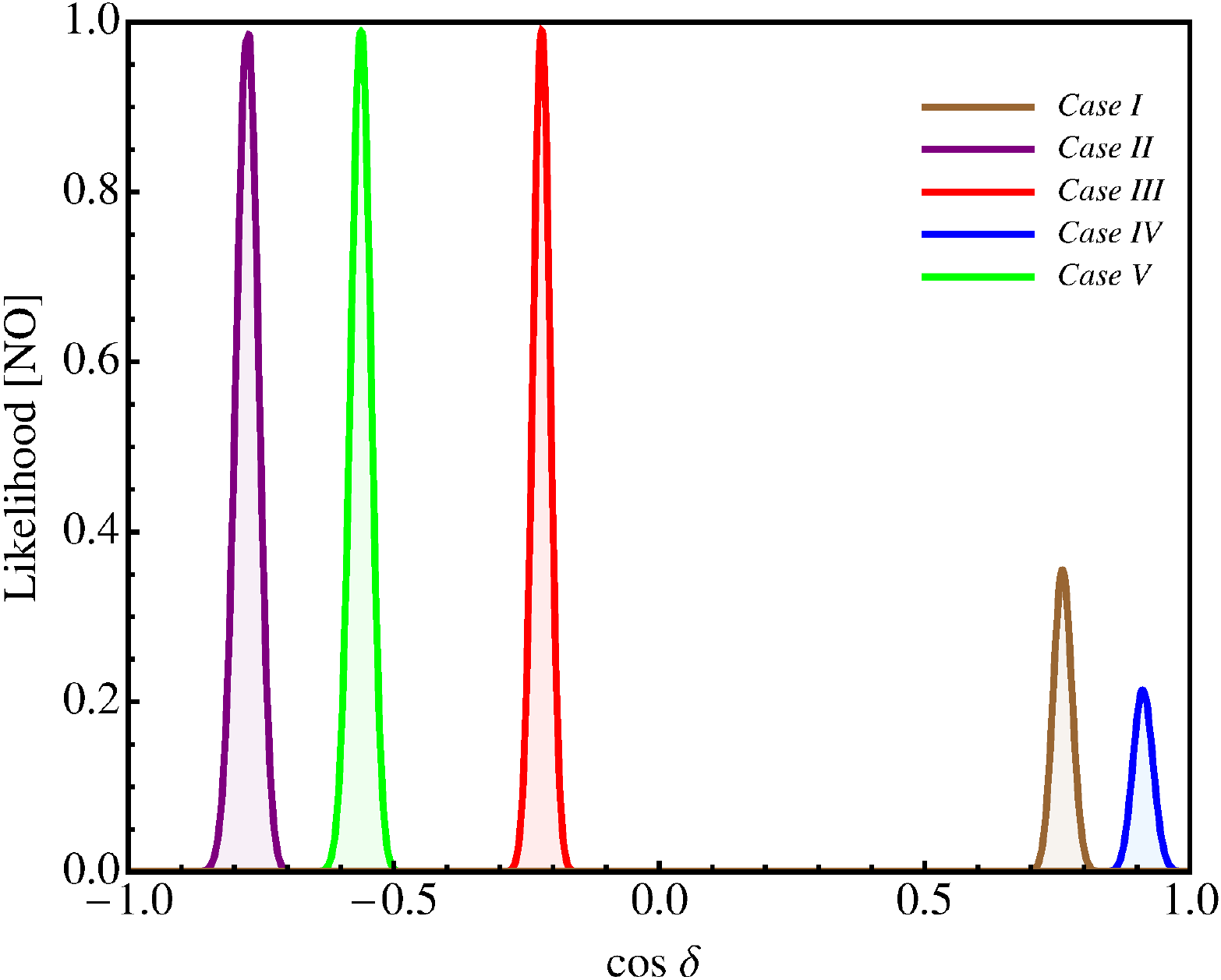}
       \vspace{2mm}}
   {\includegraphics[width=7cm]{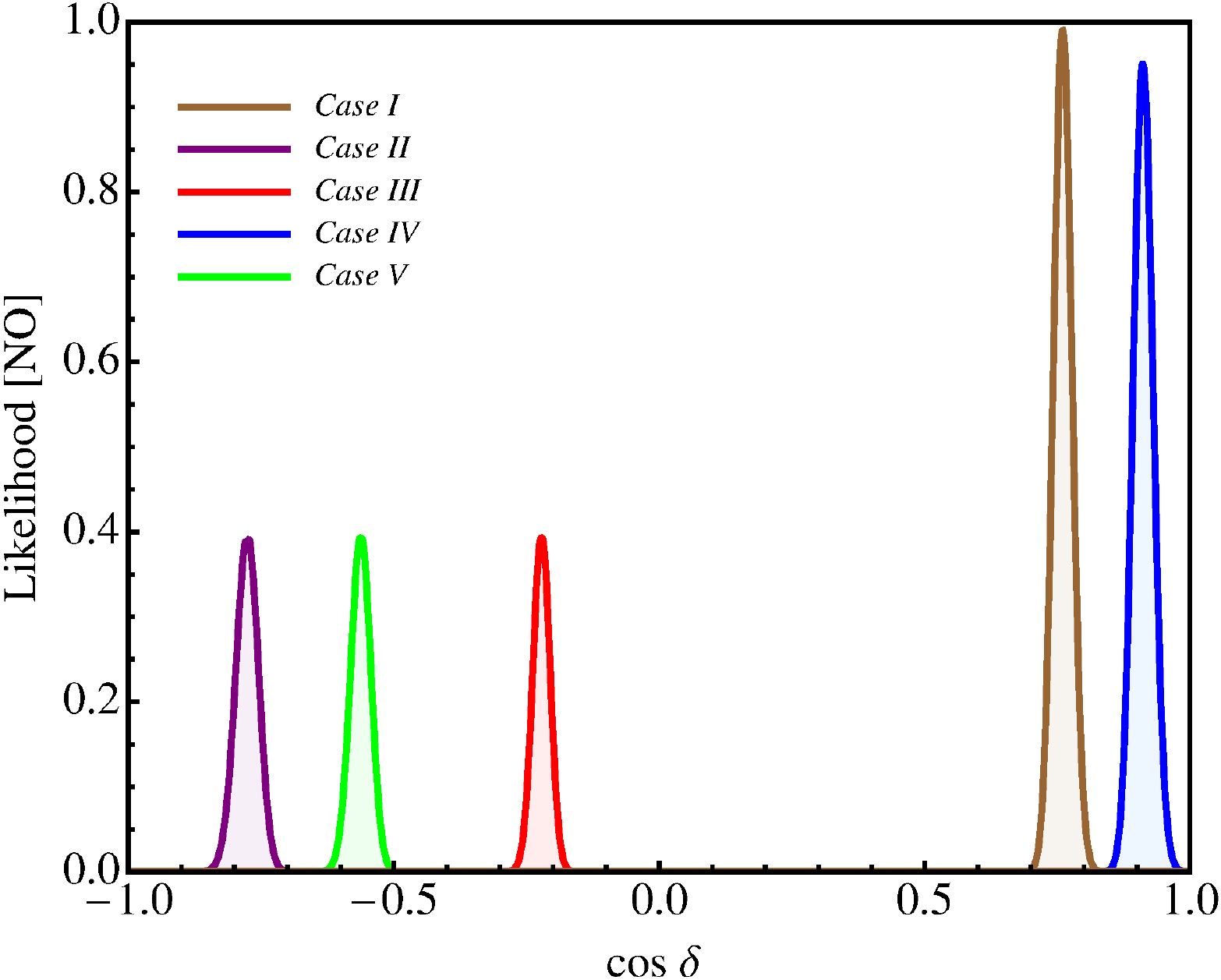}}
  {\includegraphics[width=7cm]{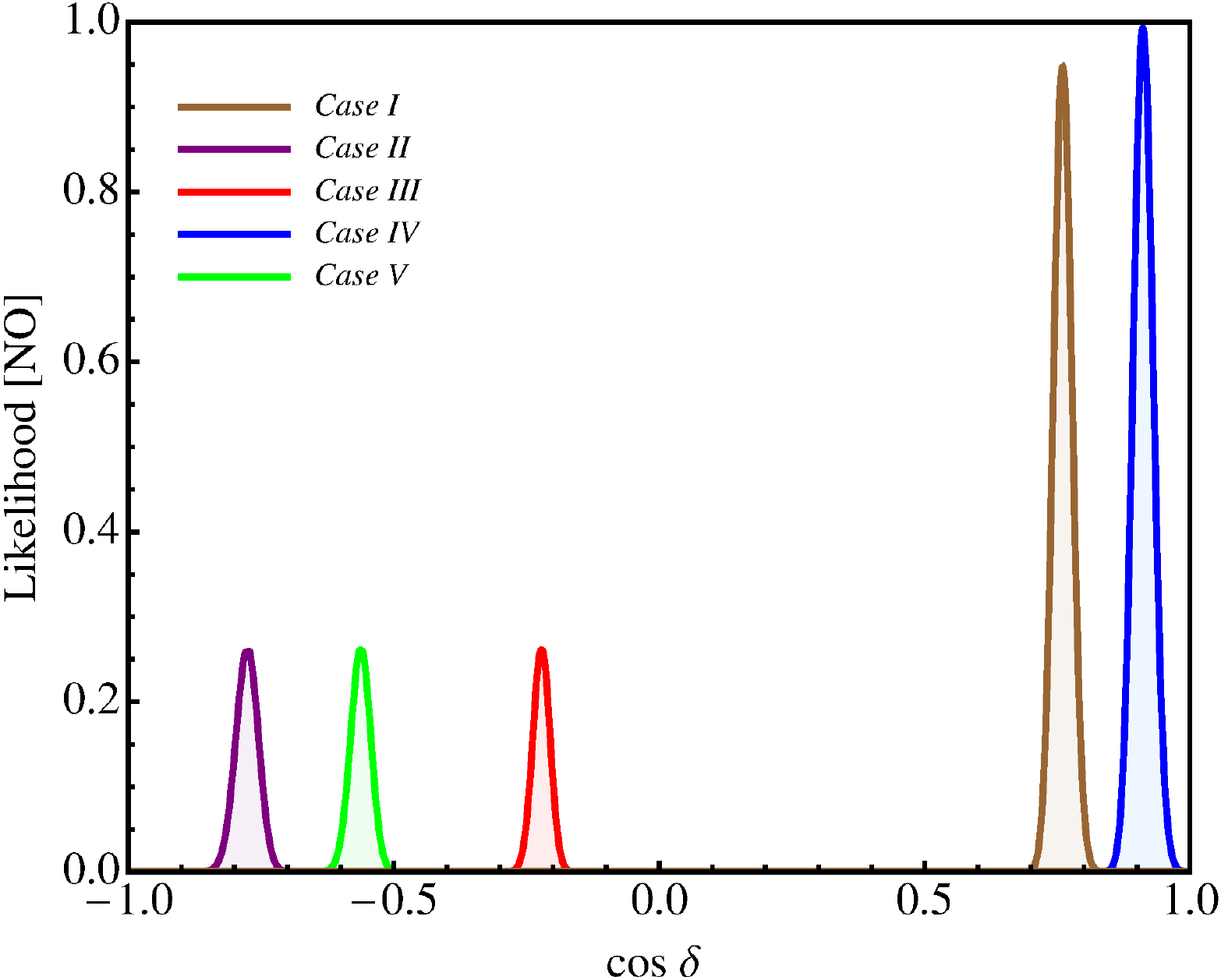}}
   \vspace{5mm}
     \end{center}
\vspace{-1.0cm} \caption{\label{Fig:cosdeltaNO_12e_fut}
The likelihood function 
versus $\cos \delta$ for the NO neutrino mass spectrum 
in the same cases as in Fig.~\ref{Fig:cosdeltaNO_12e}, but 
using  the Gaussian approximation with
the prospective uncertainties
in the measurement 
of $\sin^2\theta_{12}$, $\sin^2\theta_{13}$, 
$\sin^2\theta_{23}$,
the best fit values for $\sin^2 \theta_{12}$ and $\sin^2 \theta_{13}$
as in eqs.~(\ref{th12values}) and (\ref{th13values})
and the potential best fit values of
$\sin^2 \theta_{23} = 0.488$, $0.501$, $0.537$, $0.545$.
Upper left (right) panel: $\sin^2 \theta_{23} = 0.488$ ($0.501$); 
lower left (right) panel: $\sin^2 \theta_{23} = 0.537$ ($0.545$).
}
\end{figure}
%%%%%%%%%%%%%%%%%%%%%%%%%%%%%%%%%
%
As the results shown in Fig.~\ref{Fig:cosdeltaNO_12e_fut} indicate, 
distinguishing between the Cases I/IV and the other three cases would not 
require exceedingly high precision measurement of $\cos\delta$. 
Distinguishing between the Cases II, III and V would be more challenging 
in terms of the requisite precision on $\cos\delta$. 
In both cases the precision required will depend, in particular, 
 on the experimentally determined best fit value of 
$\cos\delta$. As Fig.~\ref{Fig:cosdeltaNO_12e_fut} also indicates, 
one of the discussed two groups of Cases might be strongly disfavoured 
by the best fit value of $\sin^2\theta_{23}$ determined in the future 
high precision experiments.

We have performed also a statistical analysis of the predictions
for the rephasing invariant $J_{\rm CP}$, minimising $\chi^2$
for fixed values of $J_{\rm CP}$.
We give $N_{\sigma} \equiv \sqrt{\chi^2}$ as a function of $J_{\rm CP}$
in Fig.~\ref{Fig:JCP12_231312}. 
The dashed lines represent the results of the global fit
\cite{Capozzi:2013csa}, while the solid ones represent the
results we obtain for each of the considered cases,
minimising the value of $\chi^2$
in $\theta^e_{12}$ for a fixed value of $J_{\rm CP}$
using eq.~(\ref{eq:JCP12e}). The blue lines correspond to the NO
neutrino mass spectrum, while the red ones are for the IO spectrum.
The value of $\chi^2$ in the minimum, which corresponds to
the best fit value of $J_{\rm CP}$ predicted in the model,
allows one to conclude about compatibility of this model
with the global neutrino oscillation data.
As it can be observed from Fig.~\ref{Fig:JCP12_231312}, 
the zero value of $J_{\rm CP}$ in the Cases III and V  is excluded at
more than $3\sigma$ with respect to the 
confidence level
of the corresponding minimum. 
Although in the other three cases 
the best fit values of $J_{\rm CP}$ are 
relatively large, as their numerical values 
quoted below show, $J_{\rm CP}=0$ is 
only weakly disfavoured statistically.
The best fit values and the $3\sigma$ ranges 
of the rephasing invariant $J_{\rm CP}$,  
obtained for the NO neutrino mass spectrum
using the current global neutrino oscillation data, 
in the five cases considered by us are given by:
%%%%%%%%%%%%%%%%%%%%%%%%%%%%%%%
\begin{align}
& \mbox{$J_{\rm CP}=-\,0.023\,,\phantom{0}$ $-0.032 \leq J_{\rm CP}\leq 0.029$ for Case I;} \\
& \mbox{$J_{\rm CP}=-\,0.022\,,\phantom{0}$ $-0.035 \leq J_{\rm CP}\leq 0.031$ for Case II;} \\
& \mbox{$J_{\rm CP}=-\,0.033\,,\phantom{0}$ $-0.039 \leq J_{\rm CP}\leq -0.025$, $0.030 \leq J_{\rm CP}\leq 0.036$ for Case III;} \\
& \mbox{$J_{\rm CP}=-\,0.016\,,\phantom{0}$ $-0.028 \leq J_{\rm CP}\leq 0.026$ for Case IV;} \\
& \mbox{$J_{\rm CP}=-\,0.028\,,\phantom{0}$ $-0.037 \leq J_{\rm CP}\leq -0.010$, $0.018 \leq J_{\rm CP}\leq 0.034$ for Case V.}
\end{align}
%%%%%%%%%%%%%%%%%%%%%%%%%%%%%%%%%%%
%

%%%%%%%%%%%%%%%%%%%%%%%%%%%%%%%%%%%%%%%%%%%%%%%%%%%%%%%%%
\begin{figure}[h!]
  \begin{center}
     \hspace{-1.6cm}
   \subfigure
 {\includegraphics[width=14cm]{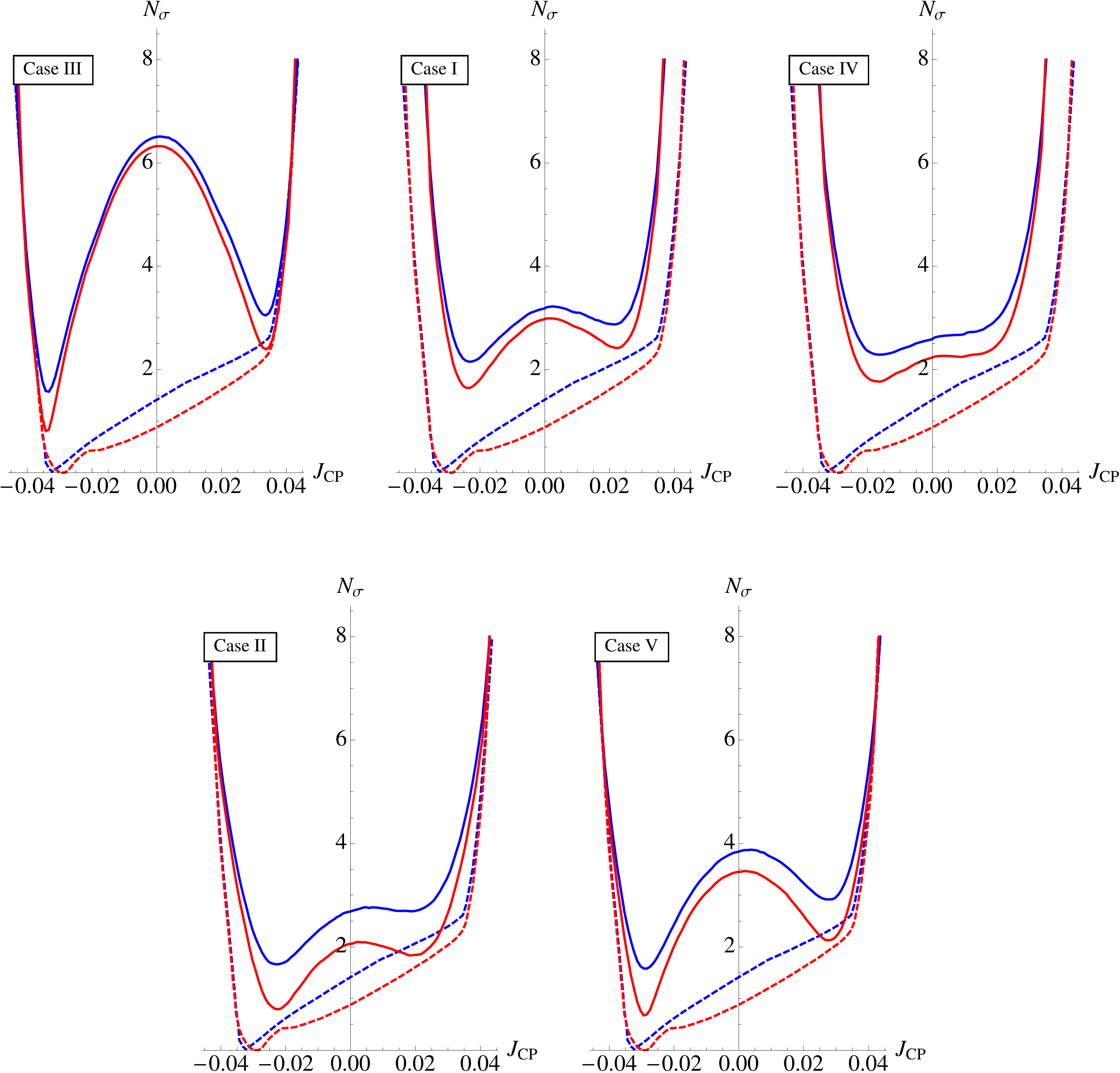}}
  \hspace{-0.8cm}
  \vspace{5mm}
     \end{center}
\vspace{-1.0cm} \caption{\label{Fig:JCP12_231312}
$N_{\sigma} \equiv \sqrt{\chi^2}$ as a function of $J_{\rm CP}$
in the scheme $\theta^e_{12} - (\theta^\nu_{23}, \theta^\nu_{13}, \theta^\nu_{12})$
with $[\theta^\nu_{13}, \theta^\nu_{12}]$ fixed as
$[\pi/10,-\pi/4]$ (Case I), $[\pi/20,b]$ (Case II),
$[\pi/20,-\pi/4]$ (Case III), 
$[a,-\pi/4]$ (Case IV), $[\pi/20,\pi/6]$ (Case V),
where $a = \sin^{-1} (1 / 3)$ and $b = \sin^{-1} (1 / \sqrt{2 + r})$,
$r$ being the golden ratio.
The dashed lines represent the results of the global fit
\cite{Capozzi:2013csa}, while the solid ones represent the
results we obtain in our set-up. The blue (red) lines are for the NO (IO)
neutrino mass spectrum.
}
\end{figure}
%%%%%%%%%%%%%%%%%%%%%%%%%%%%%%%%%%%%%%%%%%%%%%%%%%%%%%%

%%%%%%%%%%%%%%%%%%%%%
%
\subsection{The Scheme with 
$\theta^e_{13} - (\theta^\nu_{23}, \theta^\nu_{13}, \theta^\nu_{12})$
Rotations }
%
%%%%%%%%%%%%%%%%%%%%%%%%
%
\label{sec:pred13e}

 As in the set-up discussed in the subsection \ref{sec:pred12e}, 
we find for the scheme with 
$\theta^e_{13} - (\theta^\nu_{23}, \theta^\nu_{13}, \theta^\nu_{12})$
rotations that only particular values 
of $\theta^{\nu}_{12}$ and $\theta^{\nu}_{13}$
allow one to obtain the current best fit value 
of $\sin^2 \theta_{12}$.
Combining the requirements $0 < \sin^2 \theta^e_{13} < 1$
and $|\cos \omega| < 1$, 
where $\sin^2 \theta^e_{13}$ and $\cos \omega$ are given in 
eqs.~(\ref{eq:the13}) and (\ref{eq:omega}), respectively, 
and allowing $\sin^2 \theta_{13}$ to vary in its $3\sigma$ allowed 
range corresponding to the NO spectrum,
we get restrictions on the value of $\sin^2 \theta_{12}$, 
presented in Table~\ref{tab:mod1cosomega}. 
It follows from the results in Table~\ref{tab:mod1cosomega} 
that only for five out of 18 combinations
of the angles $\theta^{\nu}_{12}$ and $\theta^{\nu}_{13}$,
the best fit value of $\sin^2 \theta_{12} = 0.308$ 
and the $1.5\sigma$ experimentally allowed 
interval of values of  $\sin^2 \theta_{12}$ are
inside the allowed ranges. In Table~\ref{tab:mod1cosomega} 
these cases are marked
with the subscripts I, II, III, IV, V,
while in the case marked 
with an asterisk, the allowed range contains 
values of $\sin^2 \theta_{12}$ allowed at $2\sigma$ \cite{Capozzi:2013csa}.
%%%%%%%%%%%%%%%%%%%%%%%%%%
\begin{table}[h]
\centering
\renewcommand*{\arraystretch}{1.3}
\begin{tabular}{|c|l|l|l|l|}
\hline
$\theta^{\nu}_{12}$ & $\theta^{\nu}_{13} = \pi/20$ & $\theta^{\nu}_{13} = \pi/10$ & $\theta^{\nu}_{13} =  \sin^{-1} (1 / 3)$ \\
\hline
$\sin^{-1} (1 / \sqrt{3})$ & $(0.081,0.348)_{\rm III}$ & $(0.024,0.209)$ & $(0.019,0.189)$ \\
$\pi/4$ & $(0.197,0.516)_{\rm I}$ & $(0.103,0.361)_{\rm IV}$& $(0.091,0.338)_{\rm II}$ \\
$-\pi/4$ & $(0.484,0.803)$ & $(0.639,0.897)$ & $(0.662,0.909)$ \\
$\sin^{-1} (1 / \sqrt{2 + r})$ & $(0.051,0.291)^*$ & $(0.009,0.161)$ & $(0.006,0.143)$ \\
$\sin^{-1} (\sqrt{3 - r} / 2)$ & $(0.089,0.361)_{\rm V}$ & $(0.028,0.220)$ & $(0.022,0.200)$ \\
$\pi/6$ & $(0.038,0.264)$ & $(0.004,0.140)$ & $(0.002,0.123)$ \\
\hline
\end{tabular}
\caption{Ranges of $\sin^2 \theta_{12}$ obtained from the requirements $(0 < \sin^2 \theta^e_{13} < 1) \land (-1 < \cos \omega <1)$ 
allowing $\sin^2 \theta_{13}$ to vary in the 3$\sigma$ allowed range
for the NO neutrino mass spectrum, 
 quoted in eq.~(\ref{th13values}).
The cases for which 
the best fit value of $\sin^2 \theta_{12} = 0.308$ is within the corresponding 
allowed ranges are marked with the subscripts I, II, III, IV, V.
The case marked 
with an asterisk contains values of $\sin^2 \theta_{12}$
allowed at $2\sigma$ \cite{Capozzi:2013csa}.
}
\label{tab:mod1cosomega}
\end{table}
%%%%%%%%%%%%%%%%%%%%%%%%%%

The values of $\sin^2 \theta_{23}$ in this model depend
on the reactor angle $\theta_{13}$ and $\theta^{\nu}_{13}$
through eq.~(\ref{eq:th23B}). Using the best fit value 
of $\sin^2 \theta_{13}$ for the NO spectrum and eq.~(\ref{eq:th23B}),
we find  $\sin^2 \theta_{23} = 0.512$, $0.499$, $0.463$, $0.455$
for $\theta^{\nu}_{13} = 0$, $\pi/20$, $\pi/10$, $\sin^{-1} (1 / 3)$,
respectively. 
Thus, in the scheme under discussion  
$\sin^2 \theta_{23}$ decreases with the increase
of $\theta^{\nu}_{13}$, which is in contrast to the behaviour of 
$\sin^2 \theta_{23}$ in the 
set-up discussed in the preceding subsection.
 As we have already remarked, a measurement
of $\sin^2 \theta_{23}$ with a sufficiently high precision, 
or at least the determination of the octant of $\theta_{23}$, would 
allow one to exclude some of the values of $\theta^{\nu}_{13}$ 
considered in the literature.

%%%%%%%%%%%%%%%%%%%%%%%%%%%%%%%%%
\begin{figure}[h!]
  \begin{center}
   \subfigure
 {\includegraphics[width=7cm]{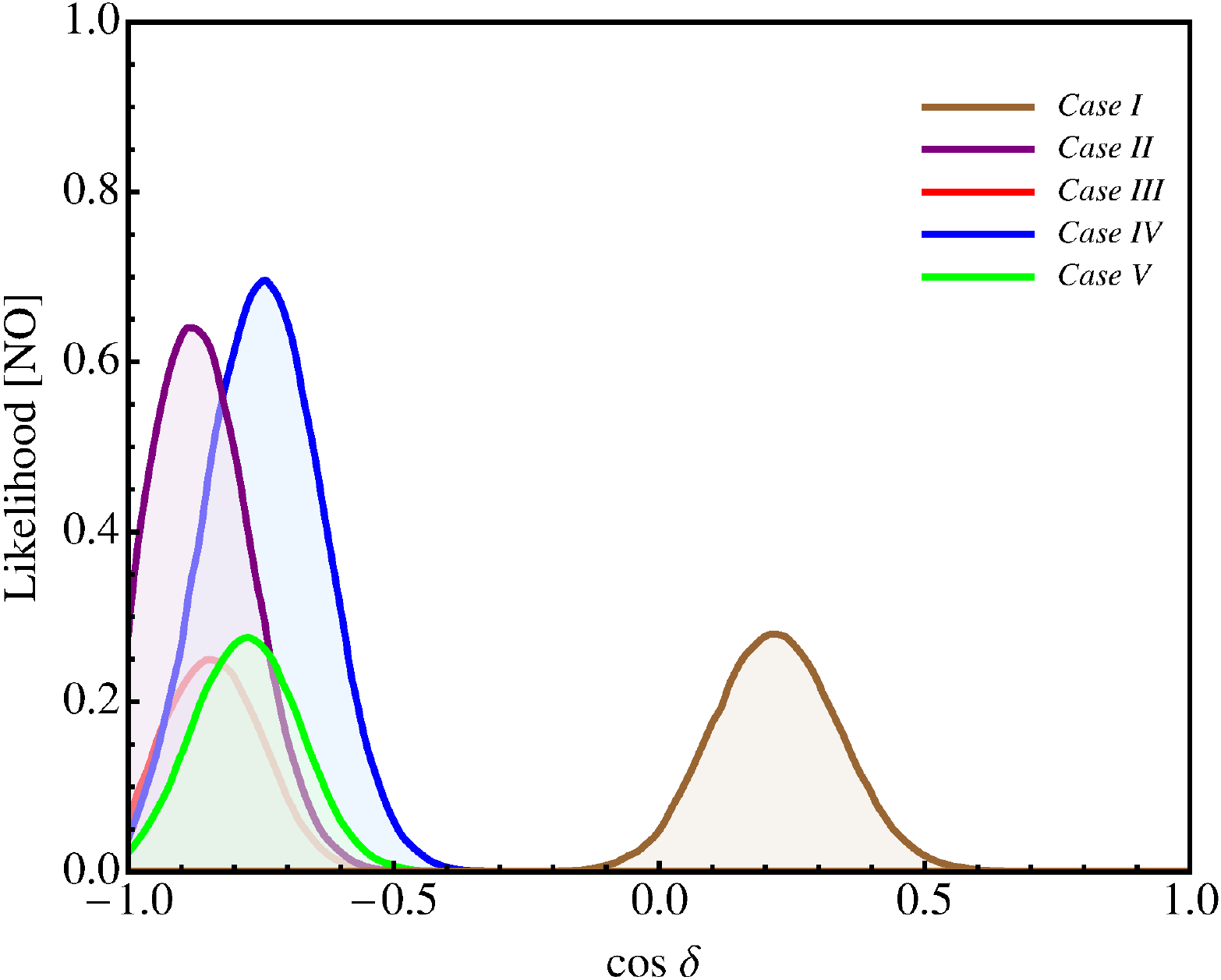}}
  {\includegraphics[width=7cm]{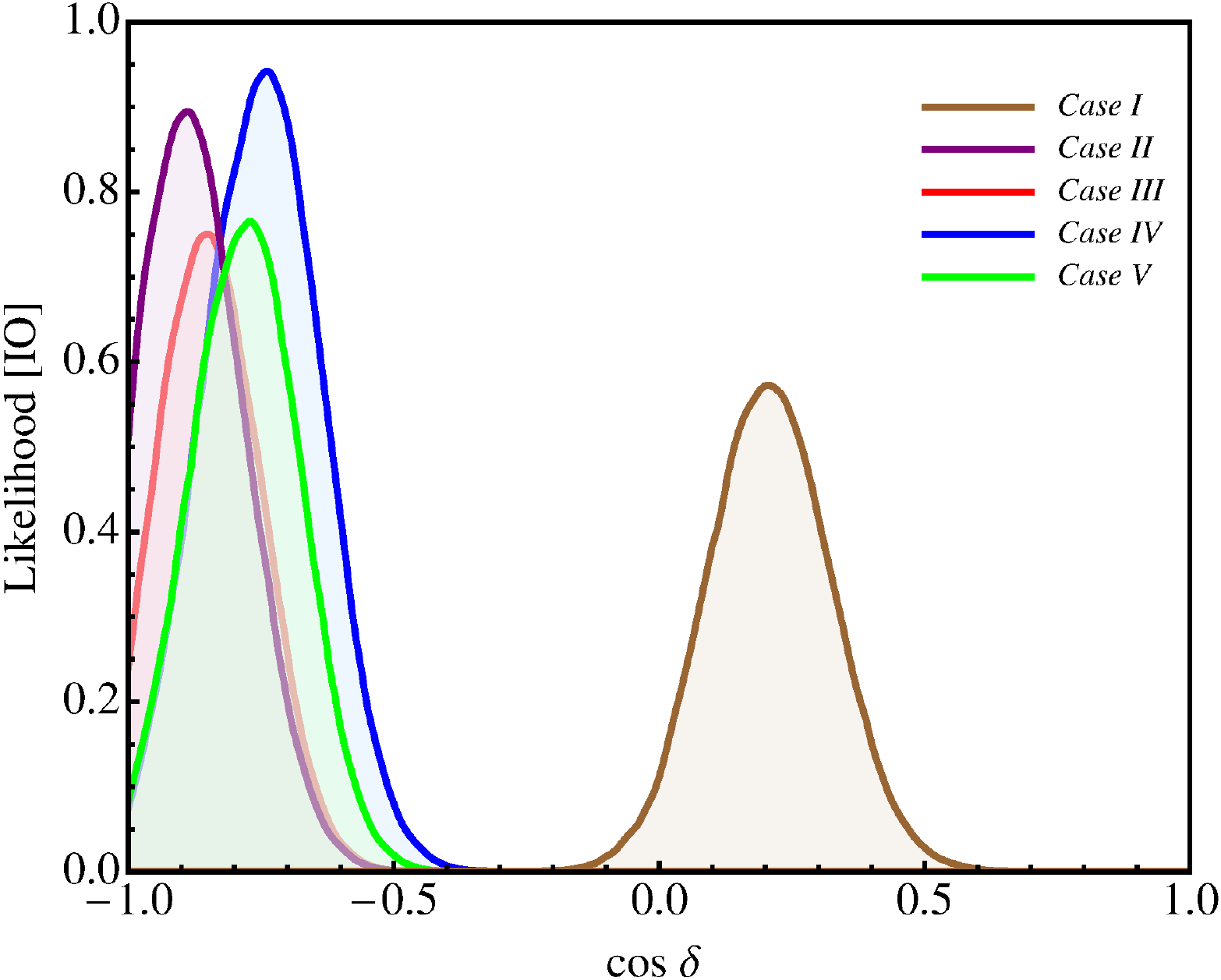}}
   \vspace{5mm}
     \end{center}
\vspace{-1.0cm} \caption{\label{Fig:cosdeltaNO_13e}
The likelihood function 
versus $\cos \delta$ for the NO (IO) neutrino mass spectrum in the
left (right) panel after marginalising over
$\sin^2\theta_{13}$ for the scheme
$\theta^e_{13} - (\theta^\nu_{23}, \theta^\nu_{13}, \theta^\nu_{12})$
with $[\theta^\nu_{13}, \theta^\nu_{12}]$ fixed as
$[\pi/20,\pi/4]$ (Case I), $[a,\pi/4]$ (Case II),
$[\pi/20,c]$ (Case III), 
$[\pi/10,\pi/4]$ (Case IV), $[\pi/20,d]$ (Case V).
We have defined $a = \sin^{-1} (1 / 3)$, 
$c = \sin^{-1} (1 / \sqrt{3})$ and $d = \sin^{-1} (\sqrt{3 - r}/2)$,
$r$ being the golden ratio.
The figure is obtained using the sum rule 
in eq.~(\ref{eq:cosdelta13e23nu13nu12nu}) and
the latest results on $\sin^2\theta_{12}$, $\sin^2\theta_{13}$, 
$\sin^2\theta_{23}$ and $\delta$
from the global analysis of the neutrino oscillation 
data \cite{Capozzi:2013csa}.
}
\end{figure}
%%%%%%%%%%%%%%%%%%%%%%%%%%%%%%%%%

The statistical analyses for $\delta$ and $J_{\rm CP}$ performed
in the present subsection are similar to those performed 
in the previous subsections.
In particular, we show in Fig.~\ref{Fig:cosdeltaNO_13e} the 
dependence of the likelihood function on $\cos \delta$
using the current knowledge on the PMNS mixing angles
and the Dirac CPV phase
 from the latest global fit results. Due to the very narrow
prediction for $\sin^2 \theta_{23}$ in this set-up, 
the prospective sensitivity likelihood curve 
depends strongly on the assumed 
best fit value of $\sin^2 \theta_{23}$.
For this reason
we present in Fig.~\ref{Fig:cosdeltaNO_13e_fut}
the predictions for $\cos \delta$
using the prospective sensitivities on the mixing angles,
the best fit values for $\sin^2 \theta_{12}$ and $\sin^2 \theta_{13}$
as in eqs.~(\ref{th12values}) and (\ref{th13values})
and the potential best fit values of
$\sin^2 \theta_{23} = 0.512$, $0.499$, $0.463$, $0.455$.
We use the value of $\sin^2 \theta_{23} = 0.512$, corresponding 
to  $\theta^{\nu}_{13} = 0$, for the same reason 
we used the value of $\sin^2 \theta_{23} = 0.488$
in the analysis in the preceding subsection, 
where we gave also a detailed explanation.

As Fig.~\ref{Fig:cosdeltaNO_13e_fut} clearly shows,
the position of the peaks
does not depend on the assumed best fit value
of $\sin^2 \theta_{23}$.
However, the height of the peaks reflects to what degree 
the model is disfavoured due to the difference between the 
assumed best fit value of $\sin^2 \theta_{23}$ and the value
predicted in the corresponding set-up.
%%%%%%%%%%%%%%%%%%%%%%%%%%%%%%%%%
\begin{figure}[h!]
  \begin{center}
   \subfigure
 {\includegraphics[width=7cm]{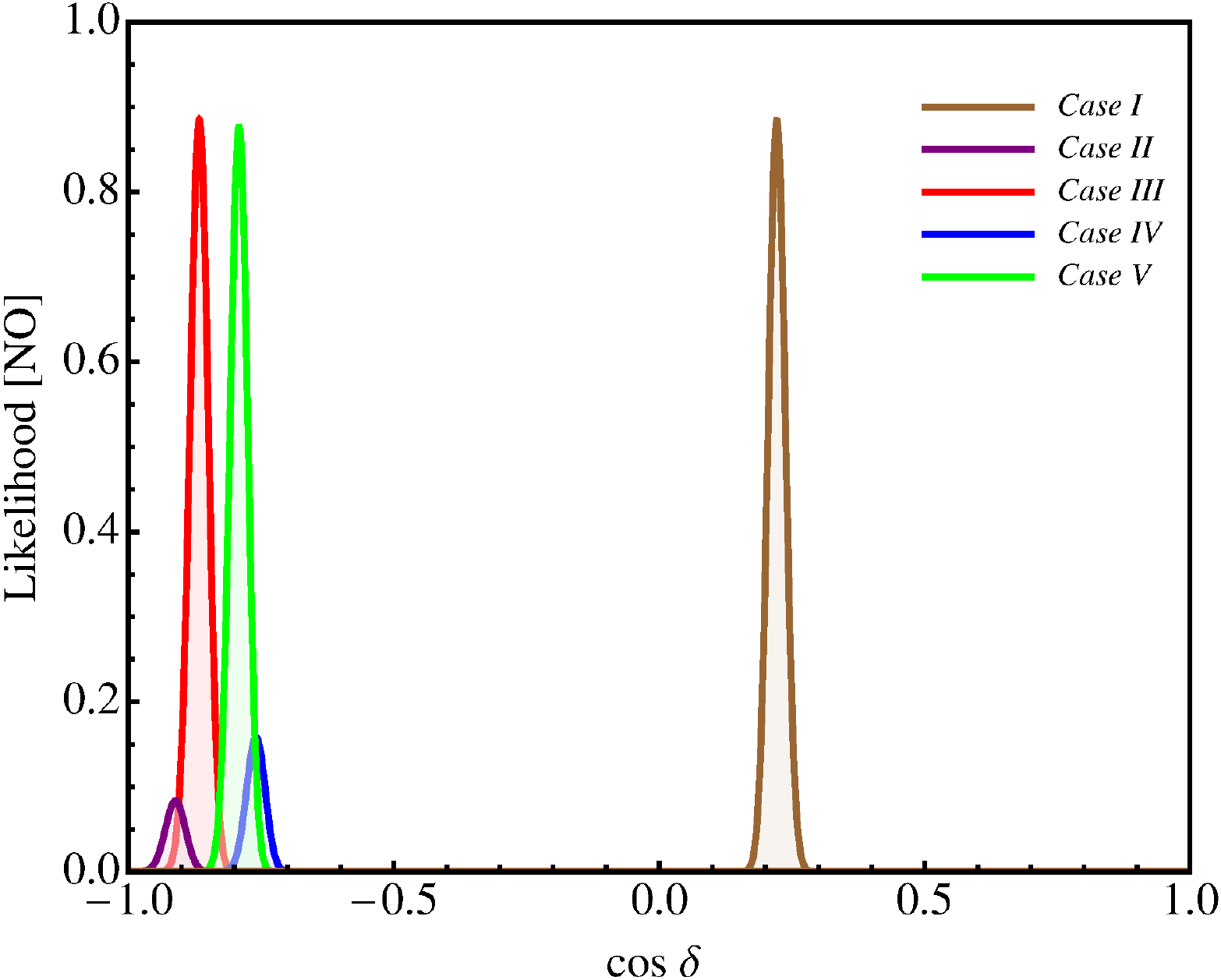}}
  {\includegraphics[width=7cm]{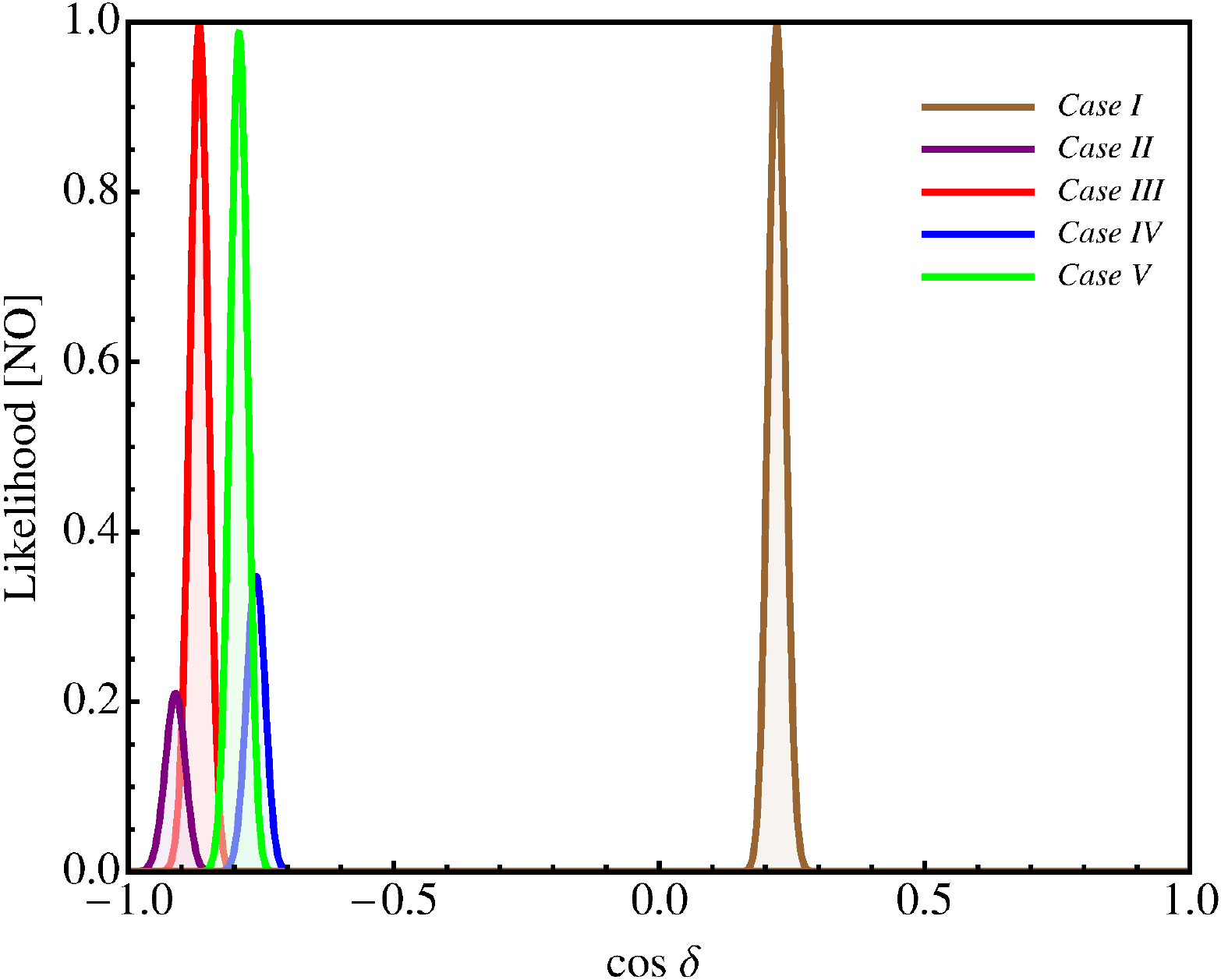}
       \vspace{2mm}}
   {\includegraphics[width=7cm]{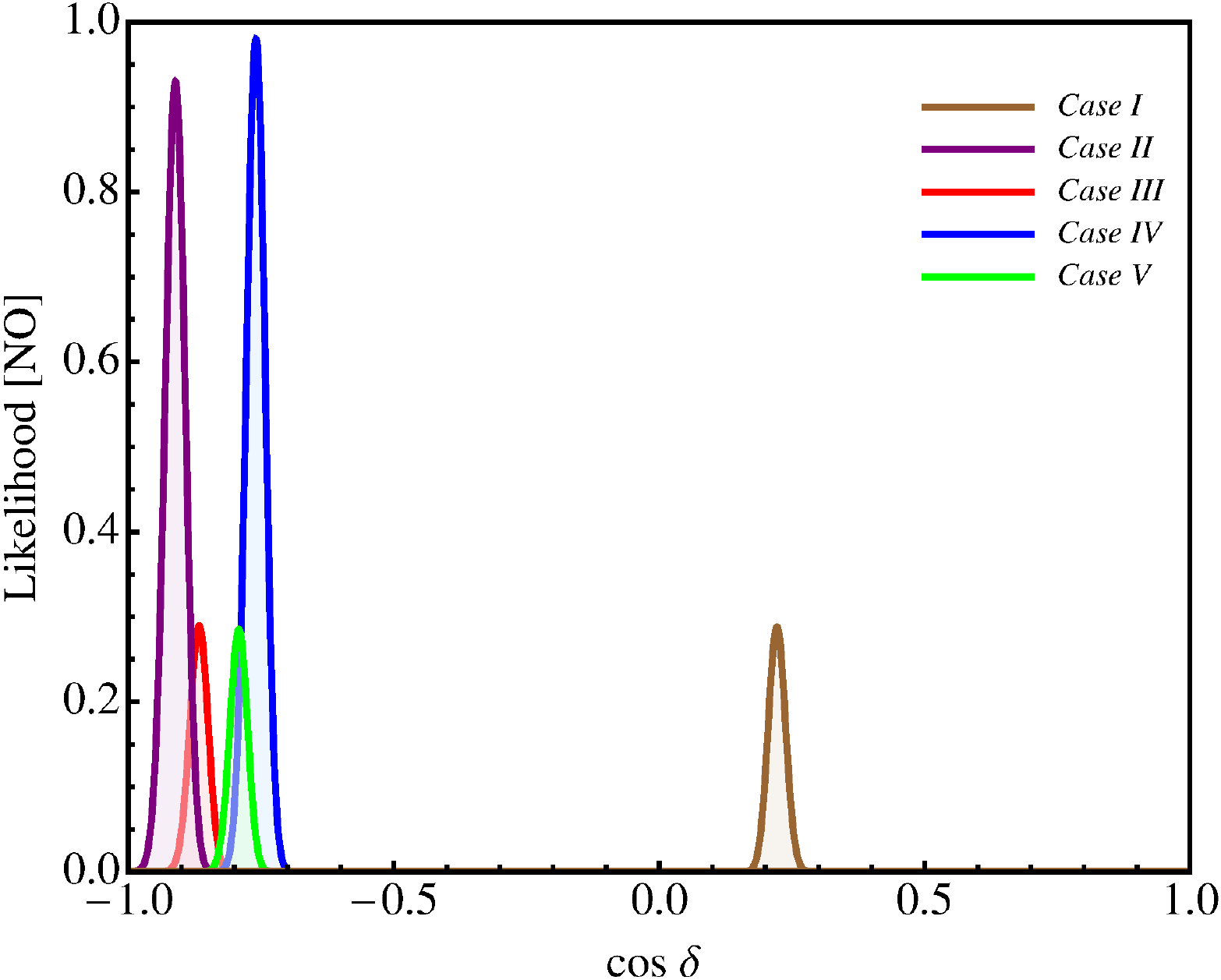}}
  {\includegraphics[width=7cm]{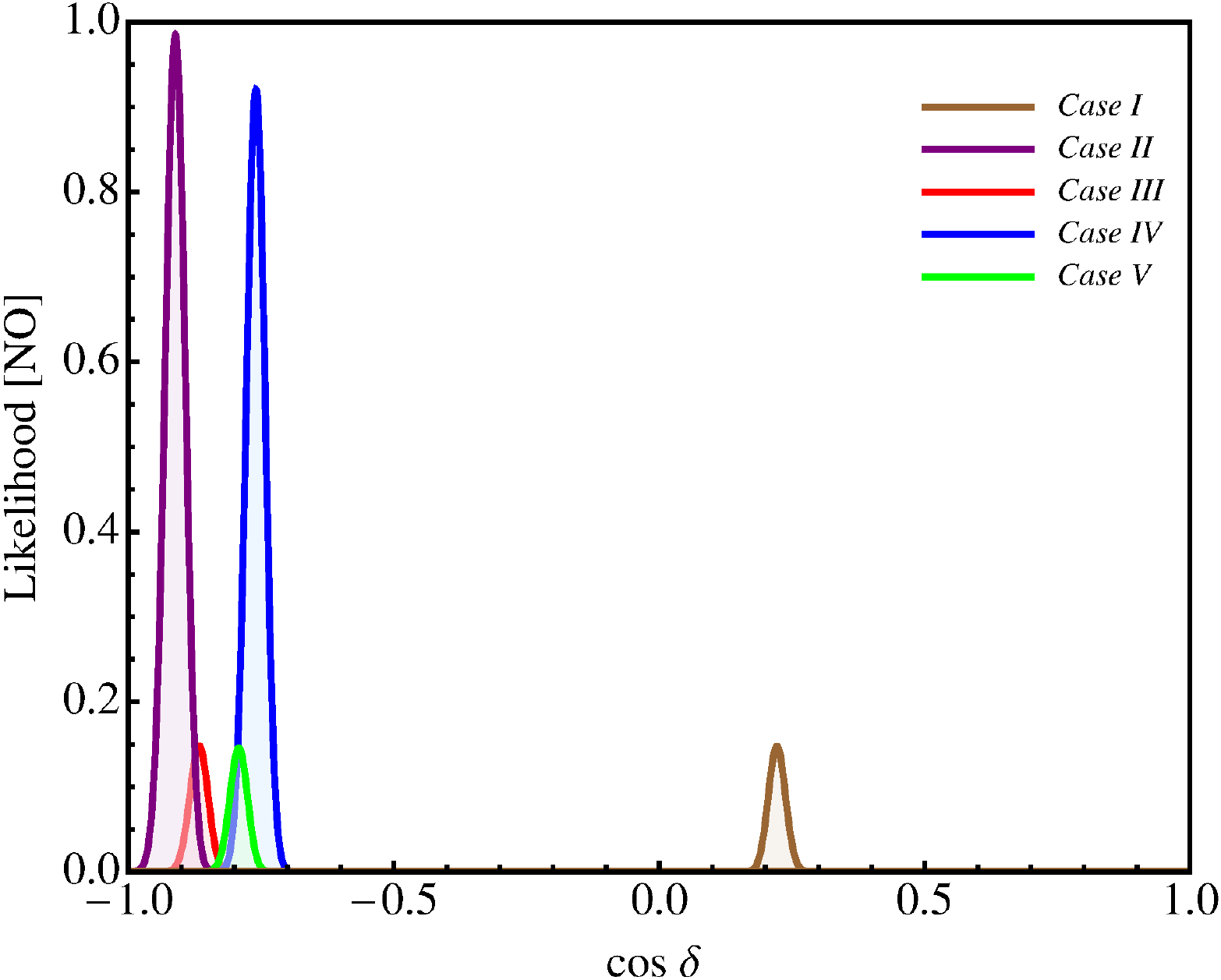}}
   \vspace{5mm}
     \end{center}
\vspace{-1.0cm} \caption{\label{Fig:cosdeltaNO_13e_fut}
The likelihood function 
versus $\cos \delta$ for the NO neutrino mass spectrum 
in the cases described in Fig.~\ref{Fig:cosdeltaNO_13e},  
but within the Gaussian approximation.
The upper left (right) panel corresponds to  
the potential best fit value of 
$\sin^2 \theta_{23} = 0.512$ ($0.499$),
while the lower left (right) panel is 
obtained for the potential best fit value of  
$\sin^2\theta_{23} = 0.463$ ($0.455$); 
the best fit values of $\sin^2\theta_{12}$ and $\sin^2\theta_{13}$ 
correspond to those 
quoted in eqs.~(\ref{th12values}) and (\ref{th13values}).
The figure is obtained using 
the prospective uncertainties in the values of $\sin^2 \theta_{12}$,
$\sin^2 \theta_{13}$ and $\sin^2 \theta_{23}$.
}
\end{figure}
%%%%%%%%%%%%%%%%%%%%%%%%%%%%%%%%%
%

The results shown in  Fig.~\ref{Fig:cosdeltaNO_13e_fut} 
clearly indicate that  
i) the measurement of $\cos\delta$
can allow one to distinguish between the Case I and the other 
four cases; 
ii) distinguishing between the Cases II/III and 
the Cases IV/V might be possible, but is very challenging in terms 
of the precision on $\cos\delta$ required to achieve that; and 
iii) distinguishing between the Cases II and III (the Cases IV and V) 
seems practically impossible. 
Some of, or even all, these cases  
would be strongly disfavoured if the best fit value 
of $\sin^2\theta_{23}$ determined with the assumed high precision 
in the future experiments were relatively large, say, 
$\sin^2\theta_{23} \gtap 0.54$.

 The results on the predictions for the rephasing invariant 
$J_{\rm CP}$ are presented in Fig.~\ref{Fig:JCP13_231312},
where we show the dependence of $N_{\sigma} \equiv \sqrt{\chi^2}$ on
$J_{\rm CP}$. 
It follows from the results presented in 
Fig.~\ref{Fig:JCP13_231312}, in particular, that 
$J_{\rm CP} =0 $ is excluded
at more than $3\sigma$ with respect to
the confidence level of the corresponding minimum
only in the Case I.
For the rephasing invariant $J_{\rm CP}$, using the current global
neutrino oscillation data, we find for the different cases
considered the 
following best fit values and $3\sigma$ ranges 
for the NO neutrino mass spectrum:
%%%%%%%%%%%%%%%%%%%%%%%%%%%%%%%%
\begin{align}
& \mbox{$J_{\rm CP}=-\,0.033\,,\phantom{0}$ $-0.039 \leq J_{\rm CP}\leq -0.025$,  $0.029 \leq J_{\rm CP}\leq 0.037$ for Case I;} \\
& \mbox{$J_{\rm CP}=-\,0.016\,,\phantom{0}$ $-0.028 \leq J_{\rm CP}\leq 0.025$ for Case II;} \\
& \mbox{$J_{\rm CP}=-\,0.018\,,\phantom{0}$ $-0.029 \leq J_{\rm CP}\leq 0.026$ for Case III;} \\
& \mbox{$J_{\rm CP}=-\,0.023\,,\phantom{0}$ $-0.031 \leq J_{\rm CP}\leq 0.029$ for Case IV;} \\
& \mbox{$J_{\rm CP}=-\,0.022\,,\phantom{0}$ $-0.030 \leq J_{\rm CP}\leq 0.028$ for Case V.}
\end{align}

%%%%%%%%%%%%%%%%%%%%%%%%%%%%%%%%%%%%%%%%%%%%%%%%%%%%%%%%%
\begin{figure}[h!]
  \begin{center}
     \hspace{-1.6cm}
   \subfigure
 {\includegraphics[width=14cm]{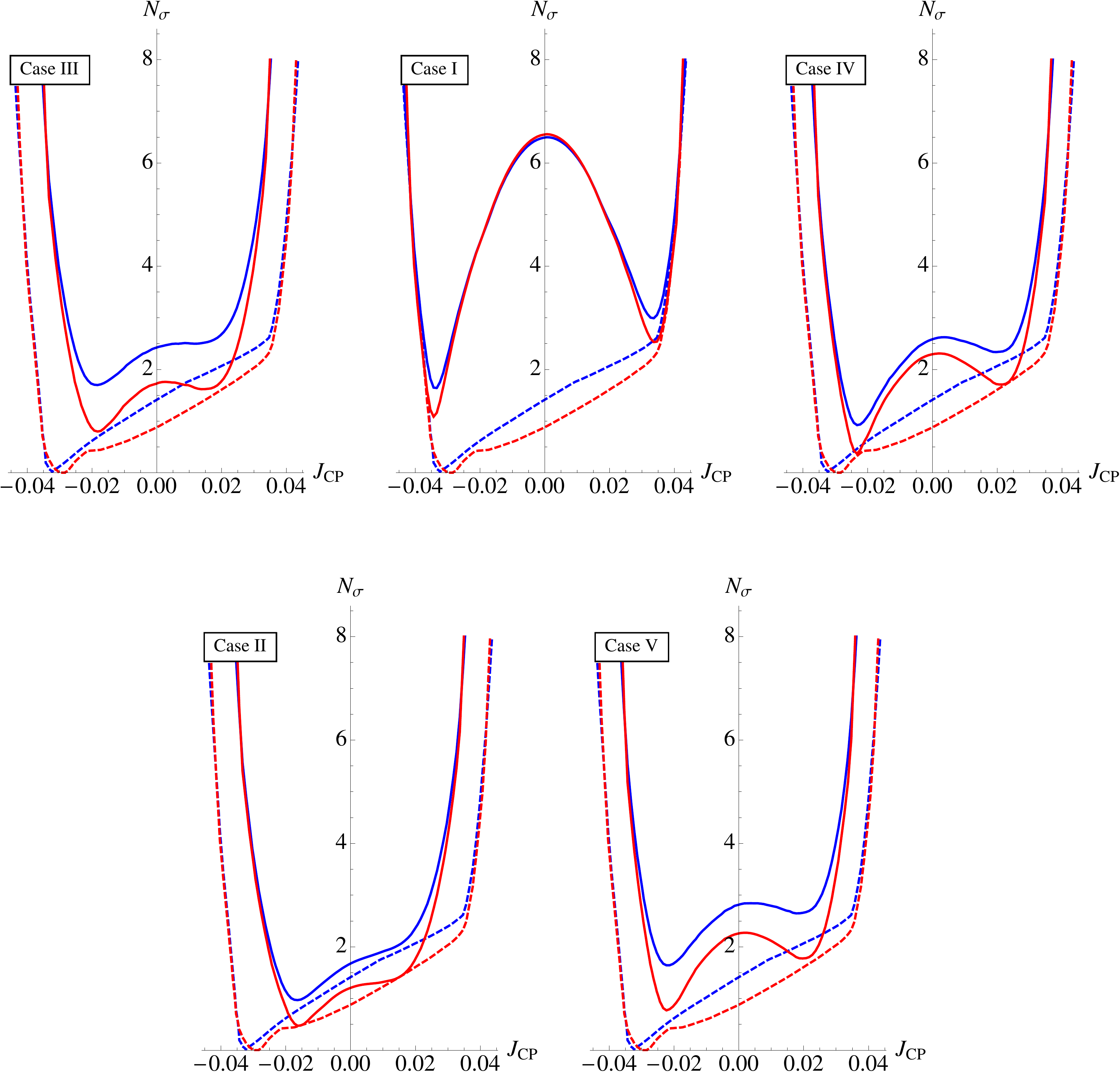}}
 \hspace{-0.9cm}
  \vspace{5mm}
     \end{center}
\vspace{-1.0cm} \caption{\label{Fig:JCP13_231312}
The same as in Fig.~\ref{Fig:JCP12_231312}, but for the
scheme $\theta^e_{13} - (\theta^\nu_{23}, \theta^\nu_{13}, \theta^\nu_{12})$
with $[\theta^\nu_{13}, \theta^\nu_{12}]$ given by 
$[\pi/20,\pi/4]$ (Case I), $[a,\pi/4]$ (Case II),
$[\pi/20,c]$ (Case III), 
$[\pi/10,\pi/4]$ (Case IV), $[\pi/20,d]$ (Case V),
where $a = \sin^{-1} (1 / 3)$, 
$c = \sin^{-1} (1 / \sqrt{3})$ and $d = \sin^{-1} (\sqrt{3 - r}/2)$,
$r$ being the golden ratio.
The dashed lines represent the results of the global fit
\cite{Capozzi:2013csa}, while the solid ones represent the
results we obtain in our set-up. The blue (red) lines are for the NO (IO)
neutrino mass spectrum.
}
\end{figure}
%%%%%%%%%%%%%%%%%%%%%%%%%%%%%%%%%%%%%%%%%%%%%%%%%%%%%%%

%%%%%%%%%%%%%%%%%%%%%%%%%%%%%%%%%
\section{Summary and Conclusions}
\label{sec:summary}
%%%%%%%%%%%%%%%%%%%%%%%%%%%%%%%%%

In the present article we have derived 
predictions for the Dirac phase $\delta$ present
in the $3\times 3$ unitary neutrino mixing
matrix $U = U_e^{\dagger} \, U_{\nu} =
(\tilde{U}_{e})^\dagger\, \Psi \tilde{U}_{\nu} \, Q_0$,
where $U_e$ ($\tilde{U}_e$)
and $U_{\nu}$  ($\tilde{U}_\nu$)
are $3\times 3$ unitary (CKM-like) 
matrices which arise from the diagonalisation, 
respectively, of the charged lepton and the neutrino mass matrices, 
and  $\Psi$ and $Q_0$ are diagonal phase matrices 
each containing in the general case two physical CPV phases. 
The phases in the matrix $Q_0$ contribute to the Majorana 
phases in the PMNS matrix. 
After performing a systematic search, 
we have considered forms of $\tilde{U}_e$ and 
$\tilde{U}_{\nu}$ allowing us to express
$\delta$ as a function of the 
PMNS mixing angles,
$\theta_{12}$, $\theta_{13}$ and $\theta_{23}$, present in $U$, 
and the angles contained in $\tilde{U}_{\nu}$. 
We have derived such sum rules for $\cos\delta$ in the cases of forms 
for which the sum rules of interest do not exist 
in the literature. More specifically,
we have derived new sum rules for $\cos\delta$ 
in the following cases:\\
i) $U = R_{12}(\theta^e_{12})\Psi R_{23}(\theta^{\nu}_{23}) 
R_{12}(\theta^{\nu}_{12}) Q_0$ 
($\theta^e_{12} - (\theta^\nu_{23}, \theta^\nu_{12})$ scheme),\\
ii) $U = R_{13}(\theta^e_{13})\Psi R_{23}(\theta^{\nu}_{23}) 
R_{12}(\theta^{\nu}_{12}) Q_0$ 
($\theta^e_{13} - (\theta^\nu_{23}, \theta^\nu_{12})$ scheme),\\
iii) $U =R_{13}(\theta^e_{13})R_{23}(\theta^e_{23})\Psi
R_{23}(\theta^{\nu}_{23}) R_{12}(\theta^{\nu}_{12}) Q_0$
($(\theta^e_{13},\theta^e_{23}) - (\theta^\nu_{23}, \theta^\nu_{12})$ 
scheme),\\
iv) $U =R_{12}(\theta^e_{12}) R_{13}(\theta^e_{13}) \Psi 
R_{23}(\theta^{\nu}_{23}) R_{12}(\theta^{\nu}_{12}) Q_0$
($(\theta^e_{12},\theta^e_{13}) - (\theta^\nu_{23}, \theta^\nu_{12})$ 
scheme),\\
v) $U =R_{12}(\theta^e_{12})\Psi R_{23}(\theta^{\nu}_{23})
R_{13}(\theta^{\nu}_{13}) R_{12}(\theta^{\nu}_{12}) Q_0$
($\theta^e_{12} - (\theta^\nu_{23}, \theta^\nu_{13}, \theta^\nu_{12})$ 
scheme), and \\ 
vi)  $U = R_{13}(\theta^e_{13}) \Psi R_{23}(\theta^{\nu}_{23})
R_{13}(\theta^{\nu}_{13}) R_{12}(\theta^{\nu}_{12}) Q_0$
($\theta^e_{13} - (\theta^\nu_{23}, \theta^\nu_{13}, \theta^\nu_{12})$ 
scheme), \\
where $R_{ij}$ are real orthogonal matrices describing rotations 
in the $i$-$j$ plane, and  $\theta^e_{ij}$ and $\theta^{\nu}_{ij}$ 
stand for the rotation angles contained in 
$\tilde{U}_e$ and $\tilde{U}_{\nu}$, respectively.  
In the sum rules $\cos\delta$ is expressed, in general, 
in terms of the three angles of the PMNS matrix, $\theta_{12}$, 
$\theta_{13}$ and $\theta_{23}$, measured, e.g.,
in the neutrino oscillation experiments, 
and the angles in $\tilde{U}_{\nu}$, which are 
assumed to have fixed known values.
In the case of the scheme iv), 
$\cos\delta$ depends in addition on an a priori unknown 
phase $\omega$, whose value can only be fixed in a 
self-consistent model of neutrino mass generation.
A summary of the sum rules derived in the present article 
is given in Table \ref{tab:summarysumrules}. 

 To obtain predictions for $\cos\delta$, $\delta$ and 
the $J_{\rm CP}$ factor, which controls the magnitude 
of the CP-violating effects in neutrino oscillations, 
we have considered several forms of $\tilde{U}_{\nu}$ 
determined by, or associated with, symmetries, 
for which the angles in $\tilde{U}_{\nu}$ 
have specific values. 
More concretely, in the cases i) - iv), 
we have performed analyses for  the  
TBM, BM (LC), GRA, GRB,
and HG forms of $\tilde{U}_{\nu}$.  
For all these forms we have $\theta^{\nu}_{23} = -\pi/4$ 
and $\theta^{\nu}_{13} = 0$. The forms differ by the value of 
the angle $\theta^{\nu}_{12}$, which for the different forms of interest 
was given in the Introduction.
In the schemes v) and vi) 
with non-zero fixed values of $\theta^{\nu}_{13}$, 
which are also inspired by certain types of flavour symmetries, 
we have considered three representative values of 
$\theta^{\nu}_{13}$ discussed in the literature, 
$\theta^{\nu}_{13} = \pi/20,~\pi/10$ and $a = \sin^{-1} (1 / 3)$, 
in combination with specific values of $\theta^{\nu}_{12}$ ---  
altogether five sets of different pairs of values 
of  $[\theta^{\nu}_{13},\theta^{\nu}_{12}]$ in each of the 
two schemes. They are given in Table \ref{tab:12}. 

We first obtained predictions for $\cos\delta$ and  $\delta$ using 
the current best fit values of $\sin^2\theta_{12}$,  $\sin^2\theta_{13}$ 
and  $\sin^2\theta_{23}$, given in eqs.~(\ref{th12values})~--~(\ref{th13values}). 
They are summarised in Tables \ref{tab:12} and \ref{tab:1}.
The quoted values  of $\cos\delta$ and  $\delta$
for the scheme iv) are for $\omega = 0$.
For completeness, in Tables \ref{tab:12} and \ref{tab:1} 
we have presented results also for\\
vii) the $(\theta^e_{12} - (\theta^\nu_{23}, \theta^\nu_{12})$ 
scheme (in which $(\tilde{U}_{e})^\dagger   = R_{12}(\theta^e_{12})$,
 $\tilde{U}_{\nu}= R_{23}(\theta^{\nu}_{23}) R_{12}(\theta^{\nu}_{12})$), and\\
viii) the $(\theta^e_{12},\theta^e_{23}) - (\theta^\nu_{23}, \theta^\nu_{12})$ 
scheme (in which $(\tilde{U}_{e})^\dagger = 
R_{12}(\theta^e_{12})R_{23}(\theta^e_{23})$, 
 $\tilde{U}_{\nu}= R_{23}(\theta^{\nu}_{23}) R_{12}(\theta^{\nu}_{12})$).\\ 
For these two schemes results were given earlier in \cite{Petcov:2014laa}.
We have updated the predictions obtained in  \cite{Petcov:2014laa} 
using the best fit values of $\sin^2\theta_{12}$,  $\sin^2\theta_{13}$ 
and  $\sin^2\theta_{23}$, found in the most recent 
analyses of the neutrino oscillation data.

We have not presented predictions for the BM (LC) symmetry form 
of $\tilde{U}_{\nu}$ in Tables \ref{tab:12} and \ref{tab:1},
because for the current best fit values of 
$\sin^2 \theta_{12}$, $\sin^2 \theta_{23}$, $\sin^2 \theta_{13}$ 
the corresponding sum rules  were found to give unphysical values of 
$\cos\delta$ (see, however, ref. \cite{Girardi:2014faa}). 

 We have found that the predictions for $\cos\delta$ of the 
 $\theta^e_{12} - (\theta^\nu_{23}, \theta^\nu_{12})$
and  $\theta^e_{13} - (\theta^\nu_{23}, \theta^\nu_{12})$
schemes for each of the symmetry forms of $\tilde{U}_{\nu}$ 
considered differ only by sign. The 
$\theta^e_{12} - (\theta^\nu_{23}, \theta^\nu_{12})$ 
scheme and the  $(\theta^e_{12},\theta^e_{13}) - (\theta^\nu_{23}, \theta^\nu_{12})$ 
scheme with $\omega = 0$ provide 
very similar  predictions for $\cos\delta$.

 In the schemes with three rotations in $\tilde{U}_{\nu}$ 
we consider, $\cos\delta$ is predicted to have values 
which typically differ significantly (being larger in absolute value) from 
the values predicted by the schemes 
with two rotations in  $\tilde{U}_{\nu}$ discussed by us,
 the only exceptions being two cases (see Table \ref{tab:12}).

We have found also that the predictions for $\cos\delta$ of the set-ups  
denoted as $(\theta^e_{12},\theta^e_{23}) - (\theta^\nu_{23}, \theta^\nu_{12})$ 
and $(\theta^e_{13},\theta^e_{23}) - (\theta^\nu_{23}, \theta^\nu_{12})$
differ for each of the symmetry forms of  $\tilde{U}_{\nu}$
considered both by sign and magnitude. 
If the best fit value of $\theta_{23}$ were $\pi/4$, these 
predictions would differ only by sign. 
 In the case of the 
 $(\theta^e_{12},\theta^e_{13}) - (\theta^\nu_{23}, \theta^\nu_{12})$ scheme,  
the predictions for $\cos \delta$ 
depend on the value chosen of the phase $\omega$.

 We have performed next a 
statistical analysis of the predictions 
a) for $\cos\delta$ and $J_{\rm CP}$ 
using the latest results of the global
fit analysis of neutrino oscillation data, 
and b) for $\cos\delta$ using prospective sensitivities
on the PMNS mixing angles. This was done 
by constructing likelihood functions in the two cases.

For the reasons related to the dependence of 
$\cos\delta$ on $\omega$ we did not present results 
of the statistical analysis for the 
$(\theta^e_{12},\theta^e_{13}) - (\theta^\nu_{23}, \theta^\nu_{12})$ 
scheme. This can be done in self-consistent models of neutrino mixing, 
in which the value of the phase $\omega$ is fixed by the model.

We have found also that in the case of the 
$\theta^e_{12} - (\theta^\nu_{23}, \theta^\nu_{12})$ scheme, 
the results for $\chi^2$ as a function of $\delta$ or $J_{\rm CP}$ 
are rather similar to those obtained in \cite{Girardi:2014faa} 
in the  $(\theta^e_{12},\theta^e_{23}) - (\theta^\nu_{23}, \theta^\nu_{12})$ set-up.
The main difference between these two schemes 
is the predictions for $\sin^2 \theta_{23}$, which 
can deviate only by approximately $0.5 \sin^2 \theta_{13}$
from $0.5$ in the first scheme, and by a significantly 
larger amount in the second. 
Similar conclusions hold comparing the results for 
the $\theta^e_{13} - (\theta^\nu_{23},\theta^\nu_{12})$ scheme
and in the  
$(\theta^e_{13},\theta^e_{23}) - (\theta^\nu_{23},\theta^\nu_{12})$ scheme. 
Therefore, in what concerns these 
four schemes, given the above 
conclusions and the fact that for the 
$(\theta^e_{12},\theta^e_{23}) - (\theta^\nu_{23},\theta^\nu_{12})$ 
scheme detailed results already exist 
in the literature (see \cite{Girardi:2014faa}),
 we have presented results of statistical 
analysis of the predictions for $\cos\delta$ 
and the $J_{\rm CP}$ factor only for the 
$(\theta^e_{13},\theta^e_{23}) - (\theta^\nu_{23},\theta^\nu_{12})$ 
scheme. This was done for the five symmetry forms 
considered --- TBM, BM (LC), GRA, GRB and HG.
We have found, in particular, that for a given symmetry form,  
$\cos\delta$ is predicted to have opposite 
sign to that predicted in the 
$(\theta^e_{12},\theta^e_{23}) - (\theta^\nu_{23},\theta^\nu_{12})$ 
scheme. Thus, in the 
$(\theta^e_{13},\theta^e_{23}) - (\theta^\nu_{23}, \theta^\nu_{12})$  
scheme analysed in the present article, one has $\cos\delta > 0$ in the
TBM, GRB and BM (LC) cases, and $\cos\delta <0$ in the
cases of GRA and HG symmetry forms of  $\tilde U_{\nu}$. 
As in the $(\theta^e_{12},\theta^e_{23}) - (\theta^\nu_{23}, \theta^\nu_{12})$
set-up, there are significant overlaps between the 
predictions for $\cos\delta$ for the TBM  and GRB forms, and for  
the GRA and HG forms, respectively.
The BM (LC) case is disfavoured 
at more than $2\sigma$ confidence level.
Due to the fact that the best fit value of $\sin^2\theta_{23} < 0.5$, 
the predictions for $\cos\delta$ for
each symmetry form, obtained in the discussed two set-ups
differ not only by sign but also in absolute value. 
We found also that in the  
$(\theta^e_{13},\theta^e_{23}) - (\theta^\nu_{23}, \theta^\nu_{12})$  
scheme relatively large 
CP-violating effects in neutrino oscillations are predicted 
for all symmetry forms considered, the only 
exception being the case of the BM symmetry form.

In the case of the 
$\theta^e_{12} - (\theta^\nu_{23}, \theta^\nu_{13}, \theta^\nu_{12})$ 
and
$\theta^e_{13} - (\theta^\nu_{23}, \theta^\nu_{13}, \theta^\nu_{12})$ 
schemes we have performed statistical analyses 
of the predictions for $\cos\delta$ and the $J_{\rm CP}$ factor 
for the five sets of values of the angles $[\theta^\nu_{13}, \theta^\nu_{12}]$ 
listed in Tables \ref{tab:12} and  \ref{tab:1}. These sets 
differ for the two schemes. For the values of 
$[\theta^\nu_{13},\theta^\nu_{12}]$  given in 
Tables \ref{tab:12} and  \ref{tab:1}, 
the allowed intervals of values of 
$\sin^2\theta_{12}$ in the two schemes, in particular, 
satisfy the requirement that they contain the 
best fit value and the $1.5\sigma$ experimentally
allowed range of $\sin^2\theta_{12}$.  
In the discussed two schemes the value of 
$\sin^2\theta_{23}$ is determined by the values of 
$\theta_{13}$, $\theta^\nu_{13}$ and $\theta^\nu_{23}$ 
(see Table \ref{tab:summarysin2th23}). In the statistical analyses we have 
performed  $\theta^\nu_{23}$ was set to $(-\pi/4)$.
Setting $\sin^2\theta_{13}$ to its best fit value, 
in the scheme 
$\theta^e_{12} - (\theta^\nu_{23}, \theta^\nu_{13}, \theta^\nu_{12})$ 
and for $\theta^{\nu}_{13} = 0$, $\pi/20$, $\pi/10$ and $\sin^{-1}(1/3)$ 
we found, respectively:  
$\sin^2 \theta_{23} = 0.488$, $0.501$, $0.537$ and $0.545$. 
For the same values of $\sin^2\theta_{13}$ and  
$\theta^{\nu}_{13}$ we obtained in the scheme 
$\theta^e_{13} - (\theta^\nu_{23}, \theta^\nu_{13}, \theta^\nu_{12})$: 
$\sin^2 \theta_{23} = 0.512$, $0.499$, $0.463$, $0.455$.

 Further, the statistical analyses we have performed 
showed that for each of the two schemes, 
the five cases considered form two groups for which 
$\cos \delta$ differs in sign and in magnitude 
(Figs.~\ref{Fig:cosdeltaNO_12e} and \ref{Fig:cosdeltaNO_13e}).
This suggests that distinguishing between
the two groups for each of the two schemes considered 
could be achieved with a not very demanding 
(in terms of  precision) measurement of $\cos \delta$.
In the analyses performed using the prospective 
sensitivities on $\sin^2\theta_{12}$, $\sin^2\theta_{13}$ 
and $\sin^2\theta_{23}$, assuming the current best fit 
values of $\sin^2\theta_{12}$, $\sin^2\theta_{13}$ 
will not change, we have chosen as potential best fit values of 
$\sin^2 \theta_{23}$ those predicted by the two schemes 
in the five cases considered 
(the values are listed in the preceding paragraph).
These analyses have revealed, in particular, 
that for each of the two schemes, 
distinguishing between the 
cases inside the two groups which 
provide opposite sign predictions for $\cos\delta$
would be more challenging 
in terms of the requisite precision on $\cos \delta$; 
for certain pairs of cases predicting $\cos \delta < -0.5$ 
in the scheme 
$\theta^e_{13} - (\theta^\nu_{23}, \theta^\nu_{13}, \theta^\nu_{12})$, 
this seems impossible to achieve in practice. These conclusions are 
well illustrated by Figs.~\ref{Fig:cosdeltaNO_12e_fut} and 
\ref{Fig:cosdeltaNO_13e_fut}. 
However, we have found that, 
depending on the chosen potential best fit value
of $\sin^2 \theta_{23}$, some of the cases
are strongly disfavoured.
Thus, a high precision measurement of $\sin^2 \theta_{23}$ 
would certainly rule out some of (if not all) the cases 
of the two schemes we have considered.

The analysis performed of the predictions 
for the $J_{\rm CP}$ factor
showed that in the 
$\theta^e_{12} - (\theta^\nu_{23}, \theta^\nu_{13}, \theta^\nu_{12})$
set-up, the CP-conserving value of $J_{\rm CP} = 0$ is excluded at more than 
$3\sigma$ with respect to the confidence level 
of the corresponding minimum, 
in two cases, namely, for 
$[\theta^{\nu}_{13}, \theta^{\nu}_{12}] = [\pi/20, - \pi/4]$, 
$[\pi/20,\pi/6]$ 
(denoted in the text as Cases III and V).
In the other three cases in spite of the relatively large 
predicted best fit values of $J_{\rm CP}$, $J_{\rm CP} = 0$
is only weakly disfavored (Fig.~\ref{Fig:JCP12_231312}). 
For the $\theta^e_{13} - (\theta^\nu_{23}, \theta^\nu_{13}, \theta^\nu_{12})$ 
scheme, $J_{\rm CP} = 0$ is excluded at more than 
$3\sigma$ (with respect to the confidence level                
of the corresponding minimum), only in one case 
(denoted as Case I in the text),
namely, for $[\theta^{\nu}_{13}, \theta^{\nu}_{12}] = [\pi/20, \pi/4]$ 
(Fig.~\ref{Fig:JCP13_231312}).

 The results obtained in the present article
confirm the conclusion 
reached in earlier similar studies  
that the measurement of the Dirac phase
in the PMNS mixing matrix, together with an improvement
of the precision on the mixing angles 
$\theta_{12}$, $\theta_{13}$ and $\theta_{23}$,
can provide unique information as regards the 
possible existence of symmetry 
in the lepton sector. 
These measurements could also 
provide an indication about
the structure of the matrix $\tilde U_e$ 
originating from the charged lepton sector, 
and thus about the charged lepton mass matrix.
 
%\vspace{0.2cm}
 \section*{Acknowledgements}
%{\bf Acknowledgements.}

 This work was supported in part by the European Union FP7
ITN INVISIBLES (Marie Curie Actions, PITN-GA-2011-289442-INVISIBLES),
by the INFN program on Theoretical Astroparticle Physics (TASP),
by the research Grant  2012CPPYP7 ({\sl  Theoretical Astroparticle Physics})
under the program  PRIN 2012 funded by the Italian Ministry 
of Education, University and Research (MIUR)
and by the World Premier International Research Center
Initiative (WPI Initiative, MEXT), Japan (STP).

\appendix

\section{Appendix: Statistical Details}
\label{app:statdetails}

In order to perform a statistical analysis 
of the schemes considered we use as input
the latest results on $\sin^2\theta_{12}$, $\sin^2\theta_{13}$,
$\sin^2\theta_{23}$ and $\delta$, obtained in the global analysis of the
neutrino oscillation data performed in \cite{Capozzi:2013csa}.
The aim is to derive the allowed ranges
for $\cos\delta$ and $J_{\rm CP}$,
predicted on the basis of the current data on
the neutrino mixing parameters for each scheme considered.
For this purpose we construct
the $\chi^2$ function in the following way:
$\chi^2(\{x_i\}) = \sum_i \chi_i^2(x_i)$,
with $x_i = \{\sin^2 \theta_{12},\sin^2 \theta_{13},\sin^2 \theta_{23},\delta\}$.
The functions $\chi^2_i$ have been extracted from 
the 1-dimensional projections given in \cite{Capozzi:2013csa} and, thus, 
the correlations between the oscillation parameters have been neglected.
This approximation is sufficiently precise since it allows one
to reproduce the contours in the planes 
$(\sin^2 \theta_{23} , \delta)$, $(\sin^2 \theta_{13} , \delta)$ 
and $(\sin^2 \theta_{23} , \sin^2\theta_{13})$, given in \cite{Capozzi:2013csa}, 
with a rather high accuracy (see \cite{Girardi:2014faa}).
We construct, e.g., $\chi^2 (\cos \delta)$ by marginalising $\chi^2(\{x_i\})$
over the free parameters, e.g., $\sin^2 \theta_{13}$ and $\sin^2 \theta_{23}$, for a fixed value of $\cos \delta$.
Given the global fit results, the likelihood function,
\be
L(\cos \delta) \propto \exp \left( - \frac{\chi^2(\cos \delta)}{2} \right) \,,
\ee
represents the most probable values of $\cos \delta$ 
in each considered case.
When we present the likelihood function
versus $\cos \delta$ within the Gaussian approximation
we use $\chi^2_{\rm G} = \sum_i  (y_i - \overline y_i)^2 / \sigma^2_{y_i}$, 
with $y_i = \{\sin^2 \theta_{12},\sin^2 \theta_{13},\sin^2 \theta_{23}\}$, 
$\overline y_i$ are the potential best fit values 
of the indicated mixing parameters and $\sigma_{y_i}$ are the prospective 
$1\sigma$ uncertainties 
in the determination of these mixing parameters.
More specifically, we use as $1\sigma$ 
uncertainties i) 0.7\% for $\sin^2 \theta_{12}$, 
which is the prospective sensitivity 
of the JUNO experiment \cite{Wang:2014iod}, 
ii) 5\% for $\sin^2 \theta_{23}$, 
obtained from the prospective uncertainty 
of 2\% \cite{deGouvea:2013onf}
on $\sin^2 2 \theta_{23}$ expected to be reached in 
the NOvA and T2K experiments, and
iii) 3\% for $\sin^2 \theta_{13}$, deduced from the error of 3\% on 
$\sin^2 2 \theta_{13}$ planned to be reached in the  Daya Bay 
experiment \cite{deGouvea:2013onf,Zhang:2015fya}.

\section{Appendix: $\sin^2 \theta_{23}$ in the $(\theta^e_{13},\theta^e_{23}) - (\theta^\nu_{23}, \theta^\nu_{12})$ Set-up}
\label{app:th2313e23e}

For completeness in Fig.~\ref{Fig:chi2th2313e23e} we give $N_{\sigma} \equiv \sqrt{\chi^2}$ 
as a function of $\sin^2 \theta_{23}$ for the scheme $(\theta^e_{13},\theta^e_{23}) - (\theta^\nu_{23}, \theta^\nu_{12})$.
%
%
%
%
%
%%%%%%%%%%%%%%%%%%%%%%%%%%%%%%%%%%%%%%%%%%%%%%%%%%%%%%%%
\begin{figure}[h]
  \begin{center}
\vspace{0.8cm}
   \hspace{-1.2cm}
%   \subfigure
 \includegraphics[width=14cm]{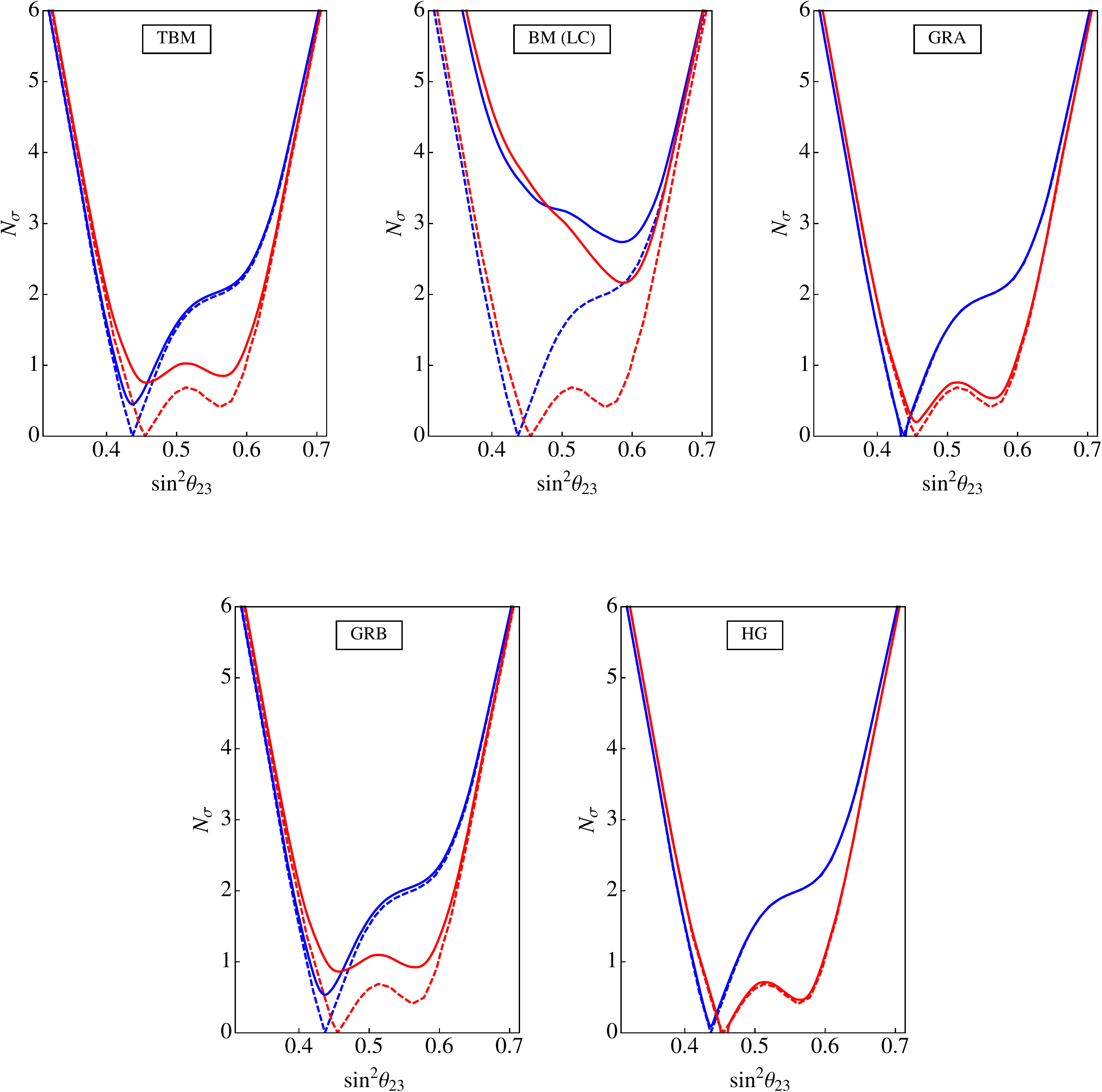}
  \hspace{-0.85cm}
  \vspace{5mm}
     \end{center}
\vspace{-1.0cm} \caption{\label{Fig:chi2th2313e23e}
$N_{\sigma} \equiv \sqrt{\chi^2}$ as a function of $\sin^2 \theta_{23}$
in the scheme $(\theta^e_{13},\theta^e_{23}) - (\theta^\nu_{23}, \theta^\nu_{12})$.
The dashed lines represent the results of the global fit
\cite{Capozzi:2013csa}, while the solid ones represent the
results we obtain for the TBM, BM (LC), GRA
(upper left, central, right panels), GRB and HG
(lower left and right panels) neutrino mixing symmetry forms.
The blue (red) lines are for the NO (IO) neutrino mass spectrum.}
\end{figure}
%%%%%%%%%%%%%%%%%%%%%%%%%%%%%%%%%%%%%%%%%%%%%%%%%%%%%%%
%
%
The dashed lines represent the results of the global fit
\cite{Capozzi:2013csa}, while the solid ones represent the
results we obtain for each of the considered symmetry 
forms of the matrix $\tilde U_{\nu}$,
minimising the value of $\chi^2$
for a fixed value of $\sin^2 \theta_{23}$.
The blue lines correspond to the NO
neutrino mass spectrum, while the red ones are for the IO one.

%%%%%%%%%%%%%%%%%%%%%%%%%%%%%%%%%%%%%%%%%%%%%

\end{document}